\documentclass[12pt]{article}
\usepackage{graphicx} 
\usepackage{bm}
\usepackage{amsmath}
\usepackage{float} 
\usepackage{booktabs} 
\usepackage{multirow} 
\usepackage{subfig} 
\usepackage[left=1in, right=1in, top=1in, bottom=1in]{geometry}
\usepackage{appendix} 
\usepackage[numbers]{natbib} 
\usepackage{amssymb} 

\usepackage[affil-it]{authblk}

\usepackage{moreverb,url}
\usepackage[colorlinks,bookmarksopen,bookmarksnumbered,citecolor=red,urlcolor=red, linkcolor=black]{hyperref}
\newcommand\BibTeX{{\rmfamily B\kern-.05em \textsc{i\kern-.025em b}\kern-.08em
T\kern-.1667em\lower.7ex\hbox{E}\kern-.125emX}}

\usepackage[dvipsnames]{xcolor}
\definecolor{gray}{rgb}{0.5,0.5,0.5}
\usepackage{listings}
\lstdefinelanguage{rstan}{
morekeywords={data,int,matrix,vector,real,lower,transformed,parameters,for,in,target,generated,quantities,model,rock,around,the,tonight},
morecomment=[l]{//}
}
\title{A Bayesian hierarchical model for disease mapping that accounts for scaling and heavy-tailed latent effects}

\author[1]{Victoire Michal\thanks{Corresponding author: victoire.michal@mail.mcgill.ca}}
\author[1]{Alexandra M. Schmidt}
\author[2]{La\'is Picinini Freitas}
\author[3]{Oswaldo Gonçalves Cruz}
\affil[1]{\small Department of Epidemiology, Biostatistics and Occupational Health, McGill University, Montreal, Canada}
\affil[2]{School of Public Health, University of Montreal, Montreal, Canada \& Centre de Recherche en Santé Publique, Montreal, Canada}
\affil[3]{Programa de Computação Científica (PROCC), Oswaldo Cruz Foundation, Rio de Janeiro, Brazil}
\date{}

\begin{document}
\fontsize{10pt}{20pt}\selectfont

\maketitle

\begin{abstract}
In disease mapping, the relative risk of a disease is commonly estimated across different areas within a region of interest. The number of cases in an area is often assumed to follow a Poisson distribution whose mean is decomposed as the product between an offset and the logarithm of the disease’s relative risk. The log risk may be written as the sum of fixed effects and latent random effects. The BYM2 model decomposes each latent effect into a weighted sum of independent and spatial effects. We build on the BYM2 model to allow for heavy-tailed latent effects and accommodate potentially outlying risks, after accounting for the fixed effects. We assume a scale mixture structure wherein the variance of the latent process changes across areas and allows for outlier identification. We \textcolor{black}{propose} two prior specifications \textcolor{black}{for} this scale mixture \textcolor{black}{parameter. These are compared through} simulation studies and in the analysis of Zika cases from the first (2015-2016) epidemic in Rio de Janeiro city, Brazil. The simulation studies show that, in terms of the model assessment criterion WAIC and outlier detection, the two proposed parametrisations perform better than the model proposed by Congdon (2017) to capture outliers. In particular, the proposed parametrisations are more efficient, in terms of outlier detection, than Congdon's when \textcolor{black}{outliers are neighbours.}
Our analysis of Zika cases finds 19 out of 160 districts of Rio as potential outliers, after accounting for the socio-development index. Our proposed model may help prioritise interventions and identify potential issues in the recording of cases.
\end{abstract}


\section{Motivation}
\label{sec:Motivation}

The first Zika cases in the Americas were identified in 2015, when it was considered a benign disease. However, in October 2015 an unprecedented increase in the number of microcephaly cases in neonates was reported in the Northeast of Brazil and was later associated with the Zika virus infection during pregnancy \citep{Lowe2018}. The Zika virus is transmitted to humans by the bite of infected \textit{Aedes} mosquitoes, the same vectors that transmit  dengue, chikungunya and yellow fever. Dengue is the most prevalent \textit{Aedes}-borne disease in the world and around 3.9 billion people in 129 countries are at risk of acquiring the disease \cite{WHOvector}. Because of climate change, the global distribution of \textit{Aedes} mosquitoes is expanding, increasing the number of people exposed to \textit{Aedes}-borne diseases. 

In the city of Rio de Janeiro, Brazil, the first Zika epidemic occurred between 2015 and 2016, with more than 35 thousand confirmed cases \cite{Freitas2019}. The city is the second-largest in Brazil, with approximately 6.3 million inhabitants, and its main tourist destination. Rio de Janeiro has a tropical climate and a favourable environment for the \textit{Ae. aegypti} mosquitoes, which are highly adapted to urban settings. Despite efforts to control the vector population, the city has suffered from dengue epidemics every three to four years, in general \cite{Nogueira1999,Honrio2009,Santos2019}. The widespread presence of the mosquito also allowed the entry and rapid dispersion of Zika and chikungunya viruses \cite{Freitas2019}. This epidemiological scenario highlights the need for novel strategies to help design interventions that are more effective in decreasing the burden of established \textit{Aedes}-borne diseases and preventing emerging and re-emerging arbovirus diseases from causing new outbreaks. In this sense, we propose a model that has the potential to help prioritise interventions by identifying areas with outlying risks with respect to the entire region and with respect to their neighbours, while accounting for covariates.

\begin{figure}[ht!]
	\centering
	\subfloat[\centering Map of the SMR]{\includegraphics[page=1, width=0.6\textwidth]{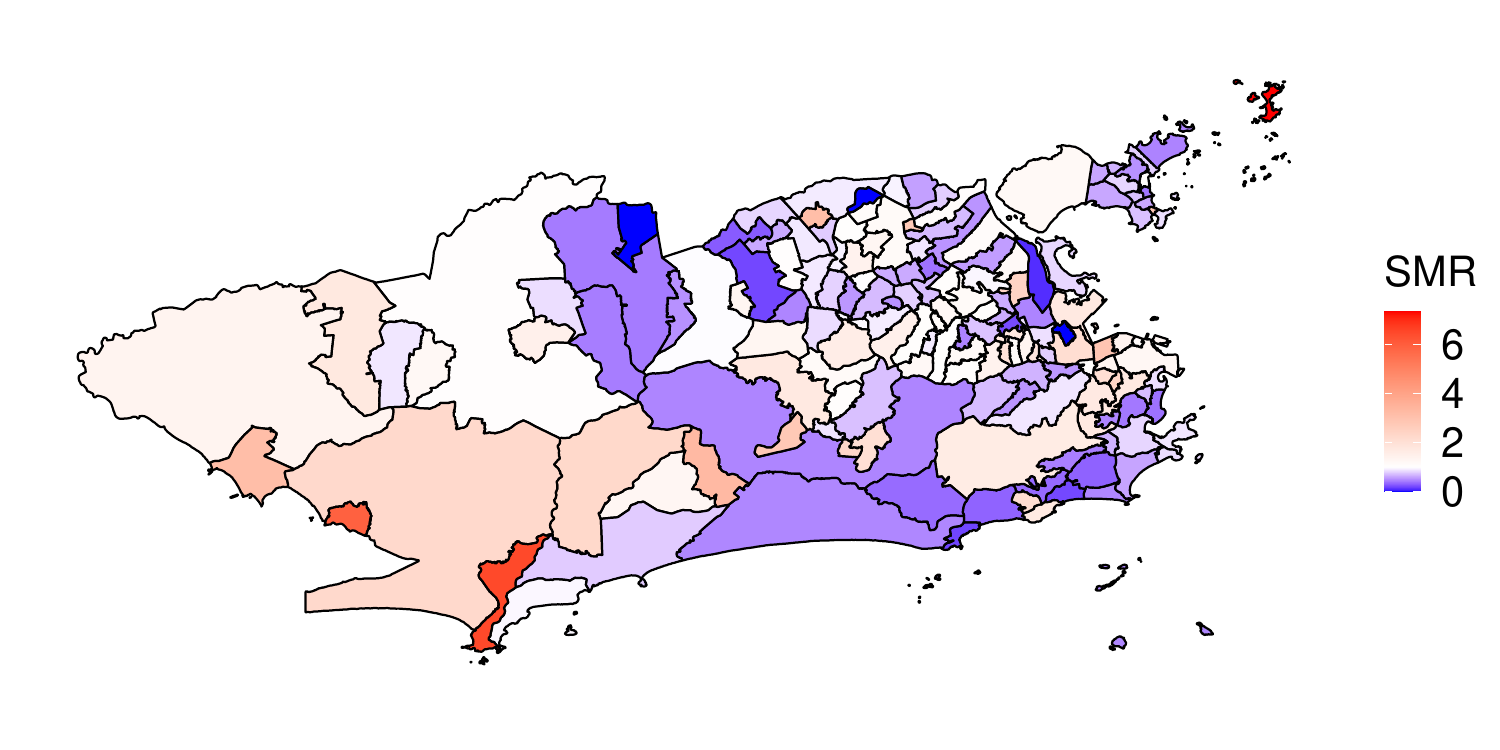}}
	\subfloat[\centering Histogram of the SMR]{\includegraphics[page=1,width=0.4\textwidth]{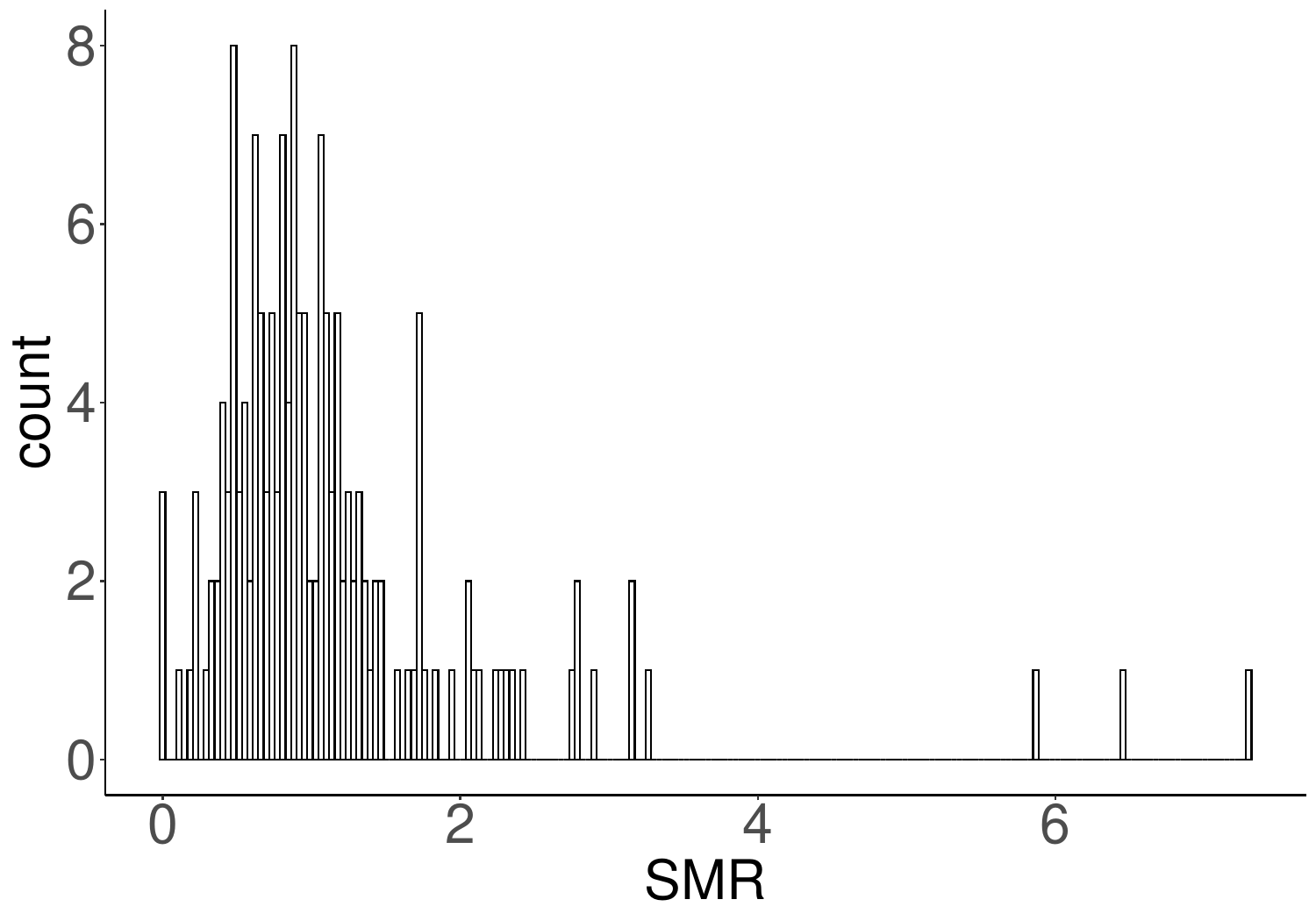}}
	\caption{\footnotesize\centering {\color{black} Map and histogram} of the SMR distribution for the Zika counts across the 160 neighbourhoods of Rio de Janeiro, between  2015 and 2016.}
	\label{fig:Zika_SMR}
\end{figure}

Motivating the proposed model, we have available the Zika cases counts aggregated by neighbourhood for the period of the first Zika epidemic in the city of Rio de Janeiro. The data come from the Brazilian Notifiable Diseases Information System (SINAN – \textit{Sistema de Informação de Agravos de Notificação}). In Brazil, cases attending healthcare facilities with a suspected diagnosis of Zika are reported to this system, usually by the physician. The standardised morbidity ratios (SMR) for the Zika counts by neighbourhood during the study period are presented in Figure \ref{fig:Zika_SMR}. Although the epidemic affected most of the city, some neighbourhoods seem to have been hit harder than others and some, not at all. The diversity of the territory of Rio de Janeiro is possibly an important factor influencing this. Regarding the city's geography, for instance, there are mountains that separate different areas. Additionally, Rio's territory is heterogeneous in terms of demographic, socio-economic, and environmental characteristics that are involved in the distribution of \textit{Aedes}-borne diseases \cite{Freitas2021}. 

For this analysis, we have available the socio-development index, an index that includes indicators related to sanitation, education and income, and for which higher values represent better socio-economic conditions. In places with inadequate sanitary conditions, the female \textit{Ae. aegypti} can more easily find any type of container filled with water to deposit her eggs. In Rio de Janeiro, \textcolor{black}{a city with great social disparities}, the socio-development index ranges from 0.282 (in Grumari, a neighbourhood in the West region) to 0.819 (in Lagoa, South region) \cite{prefeitura_ids}. 

\subsection{Literature review}
\label{sec:Litt}
 
In the last 30 years, the area of disease mapping has experienced an enormous growth. This is because it is an important tool for decision makers to obtain reliable areal estimates of disease rates over a region of interest. 
Disease mapping methods further help understanding the underlying associations between covariates and the disease risk. Commonly, Bayesian hierarchical models are used to model the disease cases observed \textcolor{black}{across} the different areas that form a region of interest. The number of cases in an area is assumed to follow a Poisson distribution whose mean is decomposed as the product of an offset by the relative risk of the disease. Further, in the log scale, the relative risk is decomposed as the sum of covariates and latent (unobserved) areal effects. The latent components accommodate overdispersion as this decomposition of the log-relative risk can be seen as a Poisson-lognormal mixture model, if the latent effects follow a normal prior distribution.

Usually, these latent effects follow a spatial structure, \textit{a priori}, such that neighbouring locations will adjust similarly after accounting for the available covariates. Indeed, it seems natural to expect that areas that are close to each other are more correlated than areas that are further apart. \textcolor{black}{Let $\bm{b}=[b_1, \dots, b_n]^\top$ be the vector of latent effects for the $n$ areas of the region of interest. Different models have been proposed in the literature for the $b$'s. First,} a commonly used spatial model for the latent effects that does not accommodate outliers is the intrinsic conditional auto-regressive (ICAR) prior \cite{CAR}. Under the ICAR prior distribution, it is assumed that  
$b_i \mid \bm{b}_{(-i)} ,\sigma_b^2 \sim \mathcal{N}\left((1/d_i)\sum_{j=1}^n w_{ij}b_j, \sigma_b^2/d_i\right), \ i=1, \dots, n,$
where $\bm{b}_{(-i)} = [b_1, \dots, b_{i-1}, b_{i+1}, \dots, b_n]^\top$, $\bm{W}=[w_{ij}]$ is a $n \times n$ matrix of weights, $w_{ij}$, that defines the neighbourhood structure and where $d_i=\sum_{j=1}^n w_{ij}$. Note that $\sigma_b^2$ is the variance of the \textit{conditional distribution} of $b_i$ given its neighbours. It can be shown \cite{Banerjee} that the joint distribution of $\bm{b}$ is proportional to $\exp \left[-(1/2\sigma^2_b)\bm{b}^\top \bm{Q} \bm{b}\right]$, with $\bm{Q}=\bm{D}-\bm{W}$, where $\bm{D}=diag(d_i)$. 
\textcolor{black}{The spatial weights are often set as $w_{ij}=1$ if areas $i$ and $j$ share a border and $w_{ij}=0,$ otherwise.} To ease the notation, let $\bm{b}\sim \textcolor{black}{\mathrm{ICAR}(\sigma_b, \bm{Q})}$ \textcolor{black}{denote the multivariate ICAR distribution. Using this common adjacency matrix,} the joint ICAR distribution is not a proper multivariate normal distribution as the \textcolor{black}{``}precision" matrix, $\bm{Q}$, is not positive definite. 
One issue with the ICAR model is that it does not perform well when there is no underlying spatial structure in the data \textcolor{black}{\cite{BYM2}}.

To accommodate the presence of independent latent effects, Besag et al. \cite{BYM} proposed the so-called BYM model, where each areal latent effect is decomposed as the sum of an unstructured component and a spatially structured component. As pointed out by MacNab \cite{BYM_IdentifiabilityIssue}, this model presents an identifiability issue as the two variance components cannot be distinguished. To avoid the introduction of two random effects for each area, \textcolor{black}{like in the BYM model}, Leroux et al. \cite{Leroux} proposed an alternative distribution for the latent spatial effects that includes a spatial dependence parameter, $\lambda$. The latter is a mixing parameter in the unit interval that allows the variance of the latent effects to be decomposed into a weighted sum between an unstructured and a spatially structured variance components. 
\textcolor{black}{On the other hand, regarding the BYM model,} S{\o}rbye and Rue \cite{SorbyeRue} argued that scaling the spatially structured effects is essential to ease interpretation and prior assignment of the variance parameter of the latent effects, independently of the neighbourhood structure. 
Hence, Riebler et al. \cite{BYM2} proposed the BYM2 model, that decomposes the latent effects into a weighted sum of unstructured random noises with unit variance and scaled structured components. The vector of latent spatial effects is scaled according to the neighbourhood structure. \textcolor{black}{This BYM2 model is a modification of the Dean model \cite{DeanPrior}, which is itself a modification of the BYM model.} \textcolor{black}{In the BYM2 model,} the decomposition of the $i$th latent effect is as follows: 
\begin{equation}
    b_i = \sigma_B\left(\sqrt{1-\lambda}\theta_i + \sqrt{\lambda}u_i^\star\right), \ i=1, \dots, n,
    \label{eq:BYM2}
\end{equation}
where $\lambda \in [0,1]$ and $\bm{\theta} \sim \mathcal{N}(\bm{0}, \bm{I})$ is independent of the scaled spatially structured components, $\bm{u}^\star=[u_1^\star, \dots, u_n^\star]^\top \sim  \textcolor{black}{\mathrm{ICAR}(1, \bm{Q}_\star)}.$ \textcolor{black}{Let} the matrix $\bm{Q}_\star^-$ be the generalised inverse of $\bm{Q}_\star$, which is a scaled version of the ICAR \textcolor{black}{``}precision" matrix, $\bm{Q}$: $\bm{Q}_\star = h \bm{Q}$. The scaling factor, $h$, is proportional to the generalised variance that arises from an ICAR model, $ h= \exp\left[(1/n)\sum_{i=1}^n \ln\left(\bm{Q}_{ii}^-\right)\right]$.
Note that the scaling factor only depends on the \textcolor{black}{graph of the region under study}. This scaled ICAR prior corresponds to $\bm{u}^\star = \left[u_1/\sqrt{h}, \dots, u_n/\sqrt{h}\right]^\top,$ for $\bm{u} \sim \textcolor{black}{\mathrm{ICAR}(1, \bm{Q})}.$ As stated in S{\o}rbye and Rue \cite{SorbyeRue}, this scaling process allows each structured component to have a variance of approximately 1. \textcolor{black}{For further discussion on the scaling process, refer to section 3.2 of Riebler et al.\cite{BYM2}.} It results that $\mathbb{V}(b_i \mid \sigma_B) \textcolor{black}{= \sigma_B^2\left[(1-\lambda) \mathbb{V}(\theta_i) + \lambda \mathbb{V}(u_i^\star)\right]} \simeq \sigma_B^2 \left[(1-\lambda)\times 1 + \lambda \times 1\right] = \sigma^2_B.$ Hence, a \textit{marginal} variance, $\sigma^2_B$, is defined for the latent effects and all the parameters can be interpreted for all neighbourhood structures.

\textcolor{black}{S}patial heteroscedasticity is not explicitly considered in the previous models. However, it is reasonable to imagine that some areas may have abnormally high or low disease risks. \textcolor{black}{Richardson et al. \cite{richardson2004interpreting} emphasised the importance for disease mapping models to be able to differentiate and adapt between smoothing the risk surface and capture abrupt changes in relative risks.} 
\textcolor{black}{This issue of spatial heteroscedasticity has been increasingly considered over the recent years. For instance, regarding geostatistical data, Palacios and Steel \cite{PalaciosSteel} proposed a log-normal scale mixture of a Gaussian process to accommodate heavy tails.} 

To allow for disparities, Congdon \cite{Congdon} proposed a modification of the Leroux prior by including scale mixture parameters. More specifically, Congdon \cite{Congdon} assumes
\begin{equation}
    b_i \mid \bm{b}_{(-i)}, \bm{\kappa}, \lambda, \sigma^2_C \sim \mathcal{N}\left(\frac{\lambda}{1-\lambda + \lambda d_i} \sum_{j=1}^n w_{ij}\kappa_j b_j, \frac{\sigma_C^2}{\kappa_i(1-\lambda + \lambda d_i)}\right), \ i=1, \dots, n,
    \label{eq:Congdon}
\end{equation}
with $\kappa_i \overset{i.i.d.}{\sim} \mathrm{Gamma}(\nu/2, \nu/2), \ i=1, \dots, n$ and $\nu \sim \mathrm{Exp}(1/\mu_\nu),$ for some value of $\mu_\nu$  fixed by the analyst.
These positive parameters, $\bm{\kappa}=[\kappa_1, \dots, \kappa_n]^\top$, allow for discrepancies in the neighbouring estimated risks, while the usual CAR-type priors aim to locally smooth the risk surface. The scale mixture parameters are termed outlier indicators as $\kappa<1$ captures outliers. Again, \textcolor{black}{for $\lambda \in (0,1)$,} $\sigma_C^2$ is the variance of the \textit{conditional distribution} of $b_i$ given its neighbours. This implies that the interpretation of $\sigma_C^2$ differs with every spatial structure, \textcolor{black}{which renders its prior assignment not straightforward and makes interpretation difficult.}
It can be shown \cite{Congdon} that the joint distribution of the latent effects is $ \bm{b} \mid \sigma_C^2, \lambda, \bm{\kappa} \sim \mathcal{N}\left(\bm{0}, \sigma_C^2 \bm{Q}_C^-\right), $ where the \textcolor{black}{``}precision" matrix has diagonal elements $\bm{Q}_{C_{ii}}=\kappa_i(1-\lambda+\lambda d_i)$ and off-diagonal elements $\bm{Q}_{C_{ij}}=-\lambda w_{ij}\kappa_i \kappa_j$. The diagonal dominance condition \cite{RueHeld} states that a sufficient condition for a symmetric matrix $\bm{Q}_C$ to be symmetric positive definite is $Q_{C_{ii}} > \sum_{j \neq i} |Q_{C_{ij}}|, \ \forall i.$ Hence, it is sufficient that $\lambda \in [0,1)$ and $\lambda < \underset{i}{\min} \left\{1/\left(1-d_i + \sum_{j \neq i} w_{ij}\kappa_j\right)\right\},$ for $\bm{Q}_C$ to be a valid precision matrix. Note that if $\bm{\kappa}=\bm{1}_n$, then Congdon's prior is the Leroux prior, which is proper for $\lambda \in [0,1)$. This mixture differs from the commonly used normal-gamma model, as the scale mixture components appear both in the mean and in the variance of the conditional distribution. \textcolor{black}{Because the scale mixture components appear in the conditional mean, areas that share a border with an outlying area give this outlier a lower weight. \textcolor{black}{Let neighbouring areas $i$ and $j$ be outliers, and let area $k$ be a neighbour of $i$ and not an outlier. Then, $b_j$ contributes by a weight of $\kappa_j<1$ to the conditional mean of $b_i$, whereas $b_k$ contributes by a factor of $\kappa_k > \kappa_j$.} This is a drawback when there are multiple outlying areas that are neighbours, as they will not borrow strength from each other.}

\textcolor{black}{Different from Congdon \cite{Congdon}, 
Dean et al. \cite{dean2019frontiers} addressed local discrepancies 
by changing the neighbouring structure according to the observed data. This approach differs from Congdon's as it is a two-step procedure that implies changing the neighbourhood structure. Other models have been proposed to allow \textcolor{black}{the strength of} the spatial autocorrelation to vary over a region of interest. Corpas-Burgos and Martinez-Beneito\cite{corpas2020use} proposed the so-called adaptive ICAR and adaptive Leroux models, which are modifications of the ICAR and Leroux models, by estimating the weights in the matrix $\bm{W}$. The adaptive Leroux model they proposed (CB-MB model) can be tied to Congdon's model (\ref{eq:Congdon}). \textcolor{black}{ For $\lambda=0$, Congdon's model yields independent latent effects with variance divided by the scaling mixture component. Similarly, when $\lambda=0$, the CB-MB model yields independent latent effects with variance divided by the spatial weight.} However, Corpas-Burgos and Martinez-Beneito point out that a single dataset is not enough to learn about those weights; so they suggest that their method is more suitable when modelling a multivariate outcome, where the neighbourhood structure is the same for the different outcomes. On the other hand, MacNab\cite{macnab2023revisiting}  recently proposed a model that allows the spatial mixing parameter, $\lambda$, to change across space. This approach allows the underlying structure of the latent effects of the areas to differ from their neighbours, when necessary. The model proposed by MacNab differs from our proposal because it points out which structure, between the independent and spatially structured included in the BYM2 model, is more important for each region. The method proposed by MacNab does not allow for different variances across the region of interest, nor the identification of outlying areas. }

The \textcolor{black}{main aim} of this paper is to propose a method to accommodate and identify outlying areas, \textcolor{black}{following a single step inference procedure. We propose a modification of the BYM2 prior \textcolor{black}{(\ref{eq:BYM2})} that is able to identify outlying areas, after accounting for the effect of covariates. A scale mixture is introduced in the BYM2 model. The proposed model keeps the appealing property of parameter interpretation while capturing potentially outlying areas and allowing the neighbouring outlying areas to borrow strength from each other. Areas may be outliers with respect to the whole region of interest, namely areas with extreme disease risks; or with respect to their neighbours, termed spatial outliers. 
Throughout, the term ``outlier" refers to both types of outliers: extremes and spatial outliers. }
 This paper is organised as follows: Section \ref{sec:Model} describes the proposed model, \textcolor{black}{then a} simulation study showcases the performance of the proposed model in section \ref{sec:Analyses}. Additionally, the application of the proposed model to the data presented in section \ref{sec:Motivation} from the 2015-2016 Zika epidemic in the 160 neighbourhoods of Rio de Janeiro is shown in section \ref{sec:Analyses}. \textcolor{black}{Section \ref{sec:Discussion} concludes with a discussion.}

\section{Proposed model}
\label{sec:Model}

Let a region of interest be partitioned into $n$ non-intersecting areas.
Let $Y_i$ be the number of cases in area $i, \ i=1, \dots, n,$ and $E_i$, the expected number at risk in that area. The counts are modelled through the following Poisson model: 
$$
Y_i \mid E_i,\mu_i \sim \mathcal{P}(E_i \mu_i),
$$
where $\mu_i$ denotes the relative risk in area $i$ and $E_i$ is an offset. Commonly, the risk is decomposed in the log scale as follows: $$\ln(\mu_i) = \beta_0 + \bm{x}_i\bm{\beta} + b_i,$$ where $\beta_0$ is the overall log risk, $\bm{x}_i$ is a $p$-dimensional vector with the explanatory variables in area $i$, associated with the $p$ coefficients $\bm{\beta}$, and $b_i$ is a random effect for area $i$. \textcolor{black}{This latent effect} is included in order to allow for overdispersion in the Poisson model that would otherwise assume equal mean and variance for area $i$. The latent areal effects can also accommodate an assumed underlying spatial structure in the data. To that end, a spatial structure is defined through the matrix $\bm{W}=[w_{ij}]$. Throughout this paper, we assume that two areas are said to be neighbours if they share a border. This implies that $w_{ij}=1$ if areas $i$ and $j$ are neighbours and $w_{ij}=0,$ otherwise. In this setting, $d_i=\sum_{j=1}^n w_{ij}$ corresponds to the number of neighbours of area $i$. To model the latent areal effects accounting for such 0-1 spatial structure, we propose a modification of the \textcolor{black}{BYM2} prior (\ref{eq:BYM2}), that is, we assume
\begin{equation}
    b_i= \frac{\sigma}{\sqrt{\kappa_i}} \left( \sqrt{1-\lambda} \theta_i + \sqrt{\lambda} u_i^\star \right), \quad i=1, \dots, n,
    \label{eq:Proposed}
\end{equation}
where  $\sigma>0$ is scaled by $\kappa_i >0$, and where $\lambda \in [0,1]$. The component $\theta_i$ is assumed independent of $u_i^\star$. In particular, $\bm{\theta} \equiv [\theta_1, \dots, \theta_n]^\top \sim \mathcal{N}(\bm{0}, \bm{I})$, and $\bm{u}^\star \equiv [u^\star_1, \dots, u^\star_n]^\top \sim \textcolor{black}{\mathrm{ICAR}(1, \bm{Q}_\star)}$. \textcolor{black}{Components $\theta_i$ and $u_i^\star$} are termed the unstructured and the scaled structured components, respectively. Like in the BYM2 model \cite{BYM2} (\ref{eq:BYM2}), the \textcolor{black}{``}precision" matrix is such that $\bm{Q}_\star = h \bm{Q}$, where the scaling factor, $h$, is computed from the neighbourhood structure (see section \ref{sec:Litt}). It results that, $\mathbb{V}(b_i \mid \sigma, \kappa_i) {\color{black} = (\sigma^2/\kappa_i) \left[(1-\lambda)\mathbb{V}(\theta_i) + \lambda \mathbb{V}(u_i^\star)\right]} \simeq (\sigma^2/\kappa_i)\left[(1-\lambda)\times 1 + \lambda \times 1\right] = \sigma^2/\kappa_i$.
Hence, $\sigma^2/\kappa_i$ represents the approximate \textit{marginal} variance of the $i$th area's latent effect. Moreover, the variance-covariance matrix, $\mathbb{V}V$, of the proposed latent effects, $\bm{b}$, is given by $\mathbb{V}V=\sigma^2 \bm{K}^{-1}\left[(1-\lambda)\bm{I} + \lambda \bm{Q}_\star^-\right]$, where $\bm{K}=diag(\kappa_i)$. Thus, the parameter $\lambda$ represents the weight of the spatial effect in the variance of the latent process. Note that this distribution is a proper multivariate normal for small values of $\lambda$, depending on the neighbourhood structure. Indeed, the diagonal dominance condition \cite{RueHeld} implies that it is sufficient that $\lambda \in [0,1)$ and $\lambda < \underset{i}{\min} \left\{1/\left(1-\bm{Q}_{\star_{ii}}^- +\sum_{j \neq i}|\bm{Q}_{\star_{ij}}^-|\right)\right\}$ for the covariance matrix, $\mathbb{V}V$, to be valid.\\
 
 In a nutshell, the proposed model uses interpretable parameters to accommodate outlying areas while identifying them. The proposed model points at neighbourhoods that need heavy-tailed latent effects, through the introduction of the scale mixture components, $\bm{\kappa}=[\kappa_1, \dots, \kappa_n]^\top$. \textcolor{black}{Area $i$ is identified as an outlier when $\kappa_i<1$.} Different from Congdon (\ref{eq:Congdon}), the proposed model makes use of parameters that intervene on the \textit{marginal} distribution of the latent effects. Therefore, their prior assignment is simplified as their interpretation remains the same regardless of the neighbourhood structure. This concerns the  weight of the spatial structure $\lambda$, the marginal variance $\sigma^2$, as well as the scaling mixture parameters \textcolor{black}{$\kappa_1, \ \dots, \kappa_n$ when the $\kappa$'s are assumed independent across the region}.
 
\textcolor{black}{We now compare the interpretation and roles of the scale mixture components $\bm{\kappa}$ in the proposed model and in Congdon's model.} To interpret the scale mixture components $\bm{\kappa}$, the importance of the spatial structure in the data, measured by $\lambda$, must be taken into account. When $\lambda=0$, both \textcolor{black}{models} reduce to independent latent effects without spatial structure. In that case, $\kappa_i<1$ only impacts the \textit{marginal} variance of the $i$th latent effect and identifies an outlying area that showcases an extreme disease risk, after accounting for covariates. When $\lambda=1$, \textcolor{black}{the proposed latent effects become $b_i = (\sigma/\sqrt{\kappa_i})(u_i/\sqrt{h}), \ i=1, \dots, n$. The $\kappa$'s intervene on the \textit{marginal} variances and $\kappa_i<1$ acts as an outlier indicator by inflating the $i$th \textit{marginal} variance and hence allowing the $i$th effect to differ from the overall mean structure. Additionally, when $\lambda=1$, the \textit{conditional} distribution of the latent effects may be written as follows:}
\begin{equation}
    b_i \mid \bm{b}_{(-i)}, \sigma^2, \bm{\kappa} \sim \mathcal{N}\left(\frac{1}{d_i}\sum_{j=1}^n w_{ij}\sqrt{\frac{\kappa_j}{\kappa_i}}b_j, \frac{\sigma^2/h}{\kappa_id_i}\right), \quad i=1, \dots, n.
    \label{eq:Proposed_Conditional}
\end{equation}
\textcolor{black}{We} compare the \textit{conditional} distributions (\ref{eq:Congdon}) and (\ref{eq:Proposed_Conditional}) consider\textcolor{black}{ing} the case where neighbouring areas $i$ and $j$ are both outliers with $\kappa_i, \kappa_j<1$ and $i \sim j$. In both \textcolor{black}{distributions (\ref{eq:Congdon}) and (\ref{eq:Proposed_Conditional})}, the $i$th and $j$th \textit{conditional} variances are inflated by $\kappa_i$ and $\kappa_j$, respectively. Regarding the \textit{conditional} means, \textcolor{black}{in the proposed model, $\kappa_j/\kappa_i \simeq \kappa_i/\kappa_j \simeq 1$ and outlying effects are allowed to borrow strength from neighbouring outliers. However, in Congdon's model, the mutual weights of $b_i$ and $b_j$ are deflated and areas $i$ and $j$ contribute less to their mutual latent effects. This feature of borrowing strength in the proposed model is attractive in the case where neighbouring areas have extreme disease risks.}

In the next subsection, different prior distributions are discussed for the scale mixture components.
 
\subsection{Prior specification of the scale mixture component}
\label{sec:kappas}
 
A natural choice, and used by Congdon \cite{Congdon}, is to assume:
 \begin{equation}
 \kappa_i \overset{i.i.d.}{\sim} \mathrm{Gamma}\left(\nu/2, \nu/2\right), \ i=1, \dots, n, \quad \mbox{and} \quad \nu \sim \mathrm{Exp}(1/\mu_\nu),
 \label{eq:Kappa_Congdon}
 \end{equation}
 where the hyperparameter's mean $\mu_\nu$ controls the magnitude of $\nu$.
When $\lambda=0$, marginalising the proposed distribution (\ref{eq:Proposed}) of the latent effect, $b_i$, with respect to $\kappa_i$ yields a Student-$t$ distribution with $\mu_\nu$ degrees of freedom, that is $t_{\mu_\nu}$. The introduction of $\kappa_i$ hence allows for heavier tails than a Gaussian distribution for the latent effects. In this case,  $\mu_\nu$ corresponds to choosing the degrees of freedom of the resulting $t$ distribution, which impact the moments of the distribution as well as its tails. A large $\mu_\nu$ results in a distribution close to being normal, which is inadequate to capture outliers. On the other hand, $\mu_{\nu} < 3$ implies a $t$ distribution whose variance is not defined. Some simulation studies showed that setting $\mu_\nu=4$ performed well, which \textcolor{black}{is the value suggested by} Gelman et al. \cite{Gelman_t4}.
 
 Another possible prior specification for the $\kappa$'s is to borrow ideas from Palacios and Steel \cite{PalaciosSteel} who proposed the inclusion of a scale mixture component in the variance of a Gaussian process. The authors suggest the usual gamma mixing is not always appropriate, as not all positive moments exist. Additionally, they point out that the $t$ distribution that results from marginalising over the gamma scaling mixture parameters may still overestimate the overall variance and struggle to detect specific outlying areas.
 In particular, they assume that the scale mixture component follows a log-Gaussian process with the same spatial structure as the one defined for the main Gaussian process. Here, we propose a scaled log proper CAR prior distribution for the $\kappa$'s. 
 This form of discretisation of the method proposed by Palacios and Steel \cite{PalaciosSteel} is applied to the latent effects, $b_i, \ i=1, \dots, n$, which include both the structured and unstructured components, in order to keep the interpretative property of the parameters. This contrasts with the method proposed by Palacios and Steel \cite{PalaciosSteel} as they introduced a scale mixture only for the spatially dependent components, leaving the unstructured components untouched. Let the scale mixture components be modelled as follows: 
 \begin{align}
\label{eq:kappa_logCAR}
     \begin{split}
         \ln(\kappa_i) &\equiv - \frac{\nu_\kappa}{2} + z_i, \ i=1, \dots, n, \\
         \mbox{where} \quad \bm{z} \equiv [z_1, \dots, z_n]^\top &\mid \nu_\kappa \sim \mathcal{N}\left(\bm{0}, \nu_\kappa \bm{Q}_{\alpha,\star}^{-1}\right) \quad  \mbox{and} \quad \nu_\kappa \sim \mathrm{Exp}(1/\mu_{\nu_\kappa}),
     \end{split}
 \end{align}
 where $\bm{Q}_{\alpha, \star}=h\bm{Q}_\alpha = h_\alpha[\bm{D}-\alpha\bm{W}]$ is again \textcolor{black}{a} precision matrix that is scaled by $h_\alpha$, which is computed based on $\bm{D}-\alpha\bm{W}$. The parameter $\alpha$ guarantees $\bm{Q}_\alpha$ to be a valid precision matrix for $\alpha \in [0,1)$ \cite{Banerjee}. \textcolor{black}{For this proper distribution to be close to an ICAR prior}, we impose $\alpha=0.99$. The proper CAR distribution is scaled in order to approximately have that $\mathbb{V}[\ln(\kappa_i) \mid \nu_\kappa] \simeq \nu_\kappa \times 1$. 
 Similarly to Palacios and Steel \cite{PalaciosSteel}, this prior implies $\mathbb{E}(\kappa_i\mid \nu_\kappa) \simeq 1$, which corresponds to a constant marginal variance across the areal latent effects, and $\mathbb{V}(\kappa_i\mid \nu_\kappa) { \color{black}= \left[\exp\left(\mathbb{V}(\ln(\kappa_i \mid \nu_\kappa)) -1 \right)\right] \exp\left(2\mathbb{E}(\ln(\kappa_i \mid \nu_\kappa)) + \mathbb{V}(\ln(\kappa_i \mid \nu_\kappa)) \right) \simeq \left[\exp(\nu_\kappa)-1\right]\exp\left(-\nu_\kappa + \nu_\kappa\right)} = \exp\left(\nu_\kappa\right)-1, \ \forall i$. For $\nu_\kappa$ close to 0, $\kappa$ is close to 1 with a small variance. A bigger $\nu_\kappa$ allows the $\kappa$'s to differ greatly from 1 and to be closer to 0, when necessary. 
 Palacios and Steel \cite{PalaciosSteel} suggest that a reasonable prior mean for $\nu_\kappa$ is $\mu_{\nu_\kappa}=0.2$. The simulation studies we conducted suggest that a sensible choice for $\mu_{\nu_\kappa}$ is $\mu_{\nu_\kappa}=0.3$, which yields $[0.2, 2.4]$ as the 95\% prior credible interval for the $\kappa$'s. This includes, $\kappa_i=1$ while allowing for departure from $\kappa_i=1$, to accommodate the potentially outlying random effect of area $i$. This prior specification for the $\kappa$'s allows the mixture components to borrow strength from neighbouring $\kappa$'s. This may be of particular interest when outlying areas are neighbours.

\subsection{Inference procedure}
\label{sec:inference}

Following the specifications discussed in the previous section, the resultant posterior distributions, regardless of the prior specification for $\kappa_i$, do not have a closed analytical form.
Therefore, the posterior distributions are approximated through computational methods. In particular, Markov Chain Monte Carlo (MCMC) methods are considered. The Hamiltonian Monte Carlo method implemented in the R package {\tt rstan} \cite{Stan} is used for the simulation studies and real data application that follow. Morris et al. \cite{MorrisBYM2} note that the No U-Turn Sampler implemented in {\tt rstan} is more efficient than other MCMC samplers to obtain reliable estimates of the posterior distributions induced by the complex auto-regressive type of models that are of interest in this paper.

One way to approximate a proper posterior distribution when assigning an ICAR prior, is to add a sum-to-zero constraint on the parameters in order to distinguish them from any added constant. This is necessary due to the invariance of the ICAR distribution to the addition of a constant \cite{RueHeld}. The sum-to-zero constraint is applied to the spatial components of the proposed model, $\bm{u}$, that need to be distinguished from the global intercept, $\beta_0$.
More precisely, we add a soft sum-to-zero constraint, that is $\sum_{i=1}^n u_i \sim \mathcal{N}\left(0, (n/1000)^2\right)$. 
The \texttt{rstan} implementation of the BYM2 model is discussed by Morris et al. \cite{MorrisBYM2} and the code for the proposed model, which is a modification of the BYM2, is available in Appendix \ref{sec:Code}.

The scaling factor, $h$, needed in the BYM2 and in the proposed model, is computed through the R package {\tt R-INLA} (Integrated Laplace Approximation, Rue et al. \cite{INLA}, \url{www.r-inla.org}) as explained by Riebler et al. \cite{BYM2}.

\section{Data analyses}
\label{sec:Analyses}

In this section, we present the results of \textcolor{black}{a} simulation stud\textcolor{black}{y} that \textcolor{black}{was} conducted to assess the performance of the proposed model. The results from fitting the proposed model to data obtained from the first Zika epidemic that took place \textcolor{black}{between} 2015 and 2016 in Rio de Janeiro are also shown. In both cases, we consider the two parametrisations of the proposed model, which correspond to the two prior specifications of the scaling mixture components described in section \ref{sec:kappas}. In the simulation stud\textcolor{black}{y} and in the data application, the proposed model is compared to Congdon's model \cite{Congdon}. Out of completeness, we also consider the two prior specifications for the $\kappa$'s for Congdon's model. Namely, Congdon's model is fitted with the $\kappa$'s following the original independent prior Gamma distributions (\ref{eq:Kappa_Congdon}), as well as with spatially structured $\kappa$'s (\ref{eq:kappa_logCAR}). 

In the simulation stud\textcolor{black}{y}, we generated data for \textcolor{black}{the 96 French departments and contaminate some areas.} 
The goal 
is to check whether our proposed model is able to identify \textcolor{black}{the} generated outliers. \textcolor{black}{Then}, in the Zika data \textcolor{black}{analysis} \textcolor{black}{in Rio de Janeiro}, we compare the results of our proposed model to Congdon's as well as the BYM2 \cite{BYM2} and Leroux \cite{Leroux} models. We identify some potentially outlying districts which might be of interest to decision makers.

\subsection{\textcolor{black}{Simulation study: neighbouring outliers in France}}
\label{sec:sim_clust}

In this section, we present the results from \textcolor{black}{a} simulation study wherein some arbitrary \textcolor{black}{neighbouring} areas \textcolor{black}{in France} are contaminated into outlying areas, to assess the performance of the proposed model in comparison to the one proposed by Congdon. The design of \textcolor{black}{the} simulation stud\textcolor{black}{y} is inspired by Richardson et al. \cite{richardson2004interpreting}\textcolor{black}{, where the goal is to assess the ability of the proposed model to both smooth over non-contaminated areas while capturing and identifying the contaminated ones. Richardson et al. \cite{richardson2004interpreting} emphasised the importance for disease mapping models to adapt to these abrupt changes in the risk surface.}

\begin{figure}[H]
	\centering
	\includegraphics[page=1, width=\textwidth]{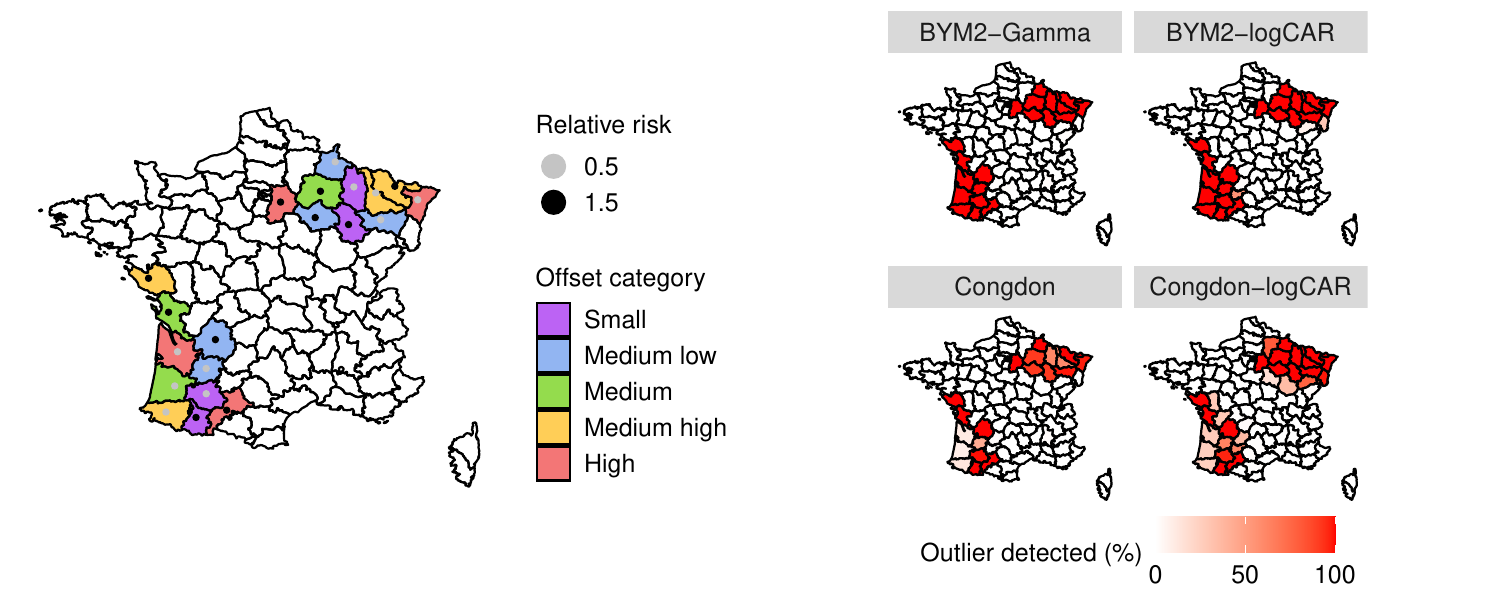}
	\caption{\footnotesize\textcolor{black}{Left panel: French departments arbitrarily chosen to be outliers in the second simulation study. Colours depict the offset category based on the empirical offset quantiles. The points represent the relative risk set to each outlying district. Right panel: Percentage of times among 100 replicates that the outliers were identified by each model, in the simulation study. The outliers are pointed out when $\kappa_u<1$, where $\kappa_u$ is the upper bound of the posterior 95\% credible interval of $\kappa$.}}
	\label{fig:Out_Clust_France_&ID}
\end{figure}

In this simulation study, 20 \textcolor{black}{departments} are contaminated such that 2 groups of 10 neighbouring outliers are created. Out of simplicity, there are no covariates included in the generating process nor when fitting the models. First, all \textcolor{black}{$n=96$} latent effects, which correspond to relative risks in this covariate-free simulation study, are set to 1: $b_i=1, \ i=1, \dots, n$. Then, the offsets $[E_1, \dots, E_n]^\top$ are \textcolor{black}{computed based on the 2019 department size estimates available on the Institut National de la Statistique et des Études Économiques (INSEE) website (\url{https://statistiques-locales.insee.fr/\#c=indicator}).} We define five offset categories based on the empirical offset quantiles. The first category corresponds to the smallest offsets and the fifth category, to the largest ones. The categories are termed \textcolor{black}{``}Small" for $E \leq \textcolor{black}{568}$, \textcolor{black}{``}Medium low" for $E \in \textcolor{black}{(568, 906]}$, \textcolor{black}{``}Medium" for $E \in \textcolor{black}{(906, 1428]}$, \textcolor{black}{``}Medium high" for $E \in \textcolor{black}{(1428, 2399]}$ and \textcolor{black}{``}High" for $E>\textcolor{black}{2399}$. Based on these categories, we select 20 \textcolor{black}{departments} to be outliers, such that each group of 10 neighbouring outliers contains 2 areas of each offset category. Within each such pair of districts, the relative risks are contaminated into outliers by setting $b_i=0.5$ and $b_{i'}=1.5$. The resulting outliers are mapped in \textcolor{black}{the left panel of} Figure \ref{fig:Out_Clust_France_&ID}, highlighting the offset sizes and imposed relative risks. Finally, $R=100$ populations of size $n=96$ are created according to a hierarchical Poisson model, that is, $Y_i \sim \mathcal{P}\left(E_i\exp[b_i]\right)$. The only source of randomness across the 100 replicates comes from the repeated sampling from a Poisson distribution. 

Using the two scale mixtures described in section \ref{sec:kappas}, the Congdon model is compared to the proposed model. The first version of the proposed model is denoted BYM2-Gamma and the second, BYM2-logCAR. The original Congdon model is termed Congdon, whereas the one with spatially structured scale mixture components is denoted Congdon-logCAR. For the four models, the intercept is given a quite vague prior: $\beta_0 \sim \mathcal{N}(0, 10^2)$ and the mixing parameter, $\lambda$, is assigned a uniform, $\mathcal{U}(0,1)$, prior distribution. The same $\mathcal{N}_+(0,1)$ prior is considered for $\sigma$, which is a \textit{marginal} standard deviation in the proposed model, while it is a \textit{conditional} standard deviation in Congdon's. Finally, in the BYM2-Gamma and Congdon models, the prior distribution for the $\kappa$'s is described in (\ref{eq:Kappa_Congdon}) with $\nu \sim \mathrm{Exp}(1/4)$. For the BYM2-logCAR and Congdon-logCAR parametrisations, the $\kappa$'s follow \textit{a priori} the distribution in (\ref{eq:kappa_logCAR}) and we set $\nu \sim \mathrm{Exp}(1/0.3)$. 

The models are fitted through the R package {\tt rstan} (Stan Development Team, 2020). For each dataset, the MCMC procedure consists of 2 chains of 20,000 iterations with a 10,000 burn-in period and a thinning factor of 10. Convergence of the chains is assessed through trace plots, effective sample sizes and the $\widehat{R}$ statistic (Gelman et al. \cite{Gelman_Diagnostic}, Vehtari et al. \cite{Rhat_Stan}).

\begin{figure}[H]
	\centering
	\includegraphics[page=1, width=\textwidth]{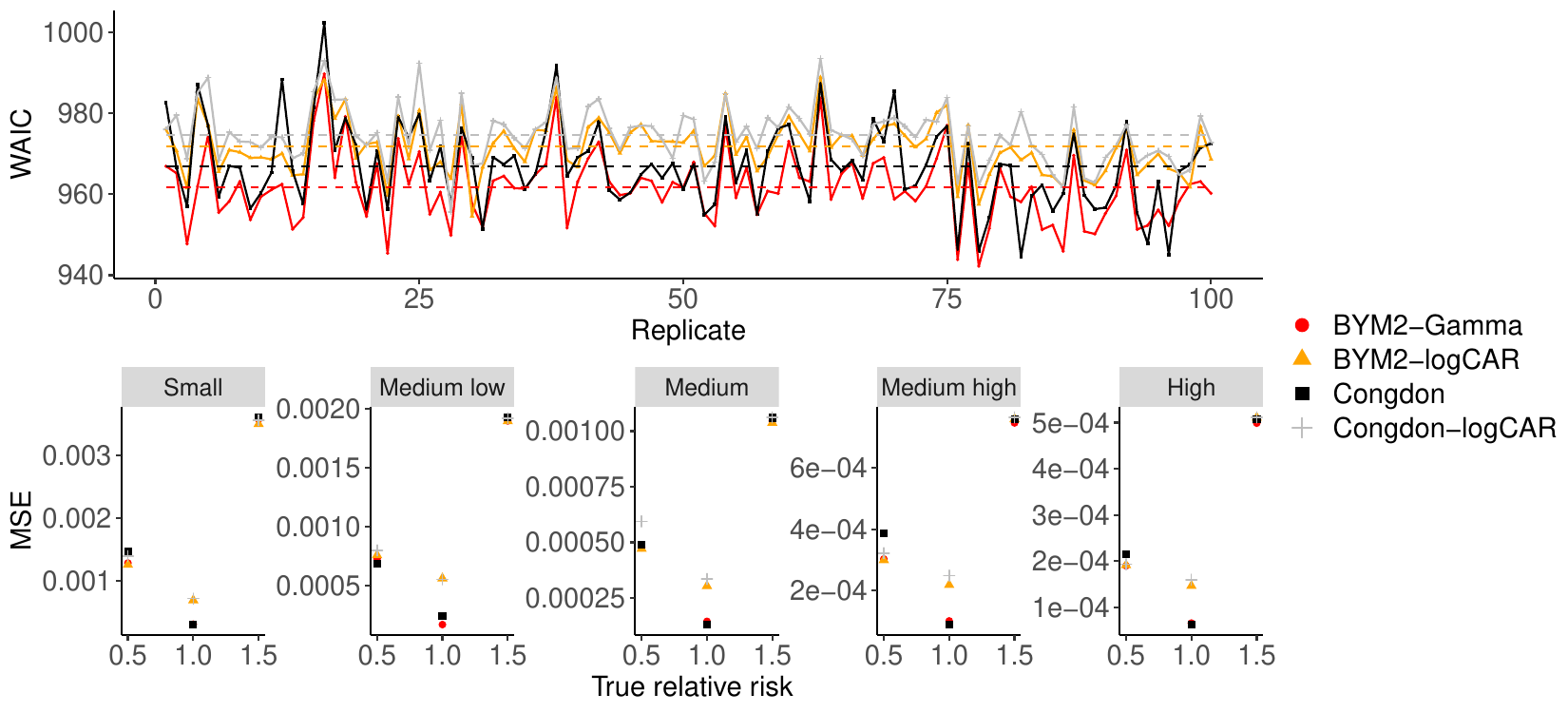}
	\caption{\footnotesize\textcolor{black}{Top panel: WAIC across the 100 replicates for the proposed models and Congdon's, in the simulation study. Dashed lines: mean WAIC for each model. Bottom panel: MSE over the 100 replicates for the proposed models and Congdon's according to the true relative risk and the offset size, in the second simulation study.}}
	\label{fig:WAIC_MSE_Clust_France}
\end{figure}

In terms of WAIC, \textcolor{black}{the proposed BYM2-Gamma model yields the smallest value among the four models, as shown in Figure \ref{fig:WAIC_MSE_Clust_France}, with an average WAIC of 962 versus 967, 972 and 975 for Congdon, BYM2-logCAR and Congdon-logCAR, respectively.} 
\textcolor{black}{In terms of MSE, Figure \ref{fig:WAIC_MSE_Clust_France} shows that all models perform similarly: on average over the 100 replicates and all areas, the BYM2-Gamma's MSE is 0.0003, versus 0.0004 for Congdon and 0.0005 for both models with the logCAR parametrisation.} 

Regarding the detection of outliers, the results are summarised in Table \ref{tab:Sens_Spe_ClustOut_France} and \textcolor{black}{the right panel of } Figure \ref{fig:Out_Clust_France_&ID}. \textcolor{black}{Area $i$ is detected as an outlier when $\kappa_{u,i} < 1$, where $\kappa_{u,i}$ is the upper bound of the 95\% posterior credible interval of $\kappa_i$.} \textcolor{black}{Congdon's} model with spatially structured $\kappa$'s tends to identify more outliers than truly present in the data (overall specificity of \textcolor{black}{93}\%, versus 99.9\% for both BYM2-Gamma and Congdon, \textcolor{black}{and 98.7 for BYM2-logCAR}). \textcolor{black}{More importantly, while both parametrisations of the proposed model always identify all the contaminated areas, overall, the two versions of Congdon's model miss 22\% and 13\% of the outliers. That is, the proposed spatially structured prior for the $\kappa$'s allows Congdon's model to identify 10\% more outliers than the model with independent mixture components.} 

\begin{table}[H]
\footnotesize
    \centering
    \begin{tabular}{cc cccc}
\toprule
& Offset category & BYM2-Gamma & BYM2-logCAR & Congdon & Congdon-logCAR \\
\midrule
\multirow{6}{*}{Sensitivity} & Small & 100.0 & 100.0 & 87.7 & 99.0 \\
& Medium low & 100.0 & 100.0 & 86.4 & 92.6 \\
& Medium & 100.0 & 100.0 & 66.7 & 75.0 \\
& Medium high & 100.0 & 100.0 & 68.0 & 81.2 \\
& High & 100.0 & 100.0 & 77.0 & 81.7 \\
& Overall & 100.0 & 100.0 & 78.1 & 86.8 \\
\addlinespace
\multirow{6}{*}{Specificity} & Small & 100.0 & 99.2 & 99.9 & 89.2 \\
& Medium low & 99.9 & 96.1 & 99.9 & 90.1 \\
& Medium & 99.7 & 99.9 & 99.9 & 92.6 \\
& Medium high & 99.9 & 98.1 & 100.0 & 93.5 \\
& High & 100.0 & 100.0 & 100.0 & 100.0 \\
& Overall & 99.9 & 98.7 & 99.9 & 93.1 \\
\bottomrule
    \end{tabular}
    \caption{\footnotesize\textcolor{black}{Sensitivity and specificity of the outlier detection for each model, depending on the offset size, in the simulation study.}}
    \label{tab:Sens_Spe_ClustOut_France}
\end{table}

\subsubsection{Further simulation studies}
\textcolor{black}{To further assess the performance of the proposed model, other simulation studies were conducted. In Appendices \ref{sec:SimFromModel} and \ref{sec:SimFromBLC}, two simulation studies show the ability of the two versions of the proposed model to recover the true parameters when data are generated from the model itself. This suggests that the proposed model does not suffer from identifiability issues. In particular, the proposed model is able to identify and distinguish, for each district, the outlier indicators, the spatial components and the unstructured components, individually. Appendix \ref{sec:SimNoOut} presents a simulation study without \textcolor{black}{contaminating any areas into outliers}, which results in the proposed model performing well compared to the prior by Congdon \cite{Congdon}, in terms of WAIC and in terms of outlier detection, where Congdon's model wrongly identifies non-outlying areas as outliers. Appendix \ref{sec:sim_dist} presents the results from a simulation study where arbitrary distant areas in France are contaminated into outliers. Again, the goal is to assess the ability of the proposed model to identify these outliers. As  discussed in Section \ref{sec:Model}, in that scenario where outliers are far from each other, the proposed model performs similarly to Congdon's model. To show that the performance of the proposed model is independent of the neighbourhood structure under study, we present in Appendix \ref{sec:sim_clustdist_Rio} the results from two simulation studies that use the map of Rio de Janeiro, where some districts are contaminated into outliers. A third simulation study shown in Appendix \ref{sec:sim_clustwithx} aims to resemble the data analysis presented in Section \ref{sec:zika}, wherein a covariate is included, and relative risks vary more over the region of interest. We found that the proposed model performed better in identifying the outliers, compared to Congdon's model.}

\subsection{Cases of Zika during the 2015-2016 epidemic in Rio de Janeiro}
\label{sec:zika}

The total numbers of cases of Zika were recorded across the 160 neighbourhoods of Rio de Janeiro during the first epidemic, which took place between 2015 and 2016. Let $Y_i$ be the disease count in district $i=1, \dots, 160.$ A hierarchical Poisson model is fitted to these data with offsets, $E$, computed from, $P$, the areal population sizes, $E_i=P_i\left(\sum_{j}Y_j/\sum_j P_j\right)$. We consider a socio-development index, $x$, as an explanatory variable for the number of cases. 
Identifying districts with potentially outlying risks, after accounting for the covariate, may be useful for decision makers to understand how to prevent Zika and where to start from. The distribution of Zika is described through a map and a histogram of the standardised morbidity ratio (SMR), $Y/E$, in Figure \ref{fig:Zika_SMR} in section \ref{sec:Motivation}. Some districts seem to present different SMR values than the mean surface, such as the island Paquetá, Barra de Guaratiba and Pedra de Guaratiba, with SMRs of 7.3, 6.5 and 5.9, respectively. In the lower tail of the SMR distribution, three districts did not record any cases and thus present null SMRs, namely Gericinó, Vasco da Gama and Parque Colúmbia. However, the SMR being an exploratory tool, one cannot conclude that high or low SMR values necessarily indicate outlying districts. Therefore, we are interested in comparing which districts are identified as potential outliers, after accounting for the socio-development index, by the two versions of the proposed model and Congdon's. 
The same priors are defined for the parameters as in the simulation study presented in section \ref{sec:sim_clust} and the two versions of the proposed model and Congdon's are again denoted BYM2-Gamma, BYM2-logCAR, Congdon and Congdon-logCAR. 
We further compare the performance of the four models to the BYM2 and Leroux models which do not accommodate potential outliers.

All models are fitted in {\tt rstan} (Stan Development Team, 2020) with 2 chains of 20,000 iterations thinned by 10 and of which 10,000 are burnt. As assessed by the trace plots, the effective sample sizes and the $\widehat{R}$ statistics, the two chains have mixed well for all six models and convergence is attained. \textcolor{black}{Appendix \ref{sec:MCMC_CV} presents the trace plots, effective sample sizes and $\widehat{R}$ statistics for a selection of parameters from the two parametrisations of the proposed model. The proposed BYM2-Gamma model took 15 minutes to run while the proposed BYM2-logCAR needed 11 minutes. In comparison, Congdon's model converged in 22 minutes and the Congdon-logCAR, in 11 minutes.}

\begin{figure}[H]
	\centering
	\includegraphics[page=1, width=\textwidth]{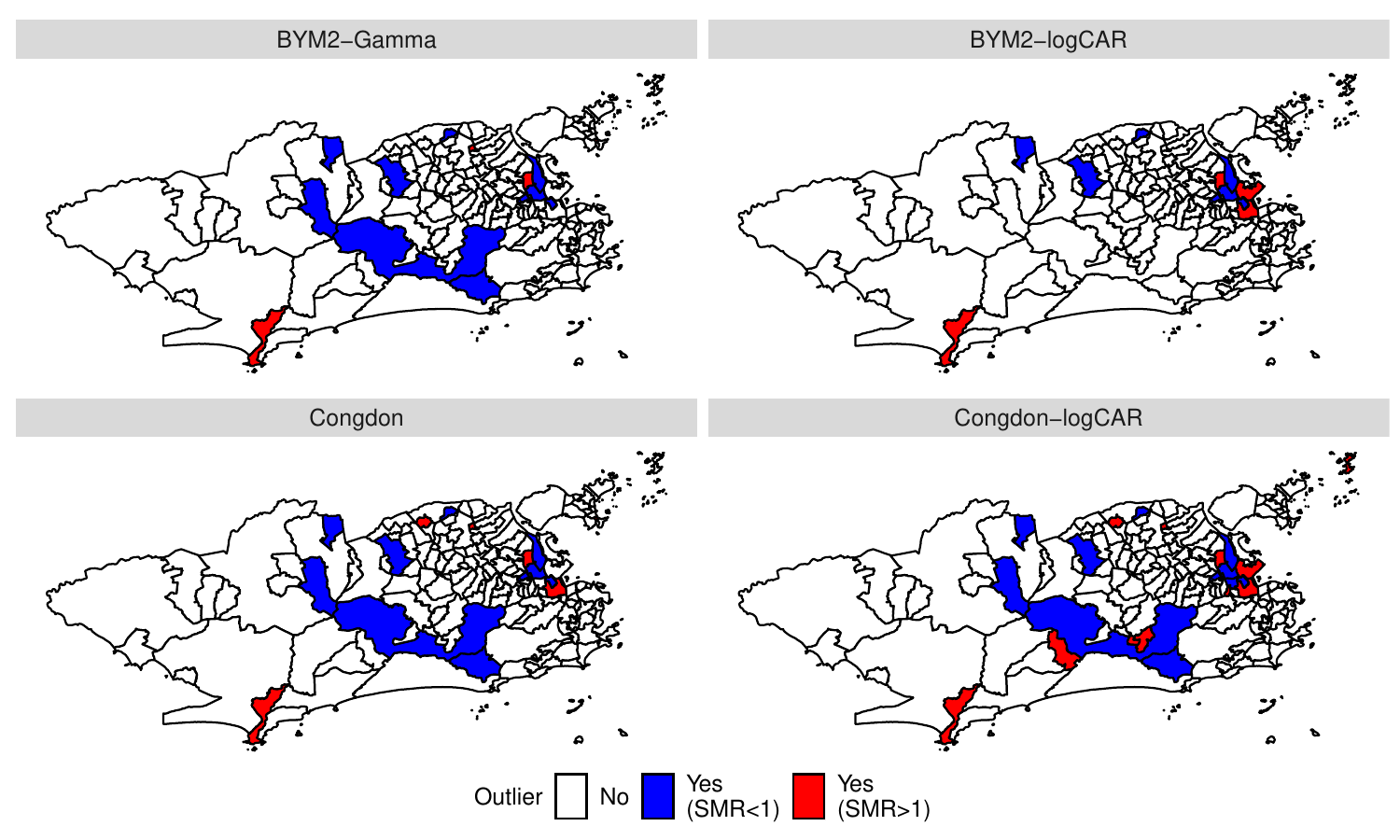}
	\caption{\footnotesize Maps of the outliers indicated by each model when analysing the Zika counts.  The outliers are pointed out when $\kappa_u<1$, where $\kappa_u$ is the upper bound of the posterior 95\% credible interval of $\kappa$. The outliers on the lower tail are distinguished from the ones on the upper tail of the SMR distribution.}
	\label{fig:Zika_Out}
\end{figure}

The results from the fitted models are presented in Table \ref{tab:Zika_Res} and Figure \ref{fig:Zika_Out}. In terms of WAIC, the proposed BYM2-Gamma model performs best among the six considered. There is an important performance gain when accommodating outliers (BYM2-Gamma, BYM2-logCAR, Congdon and Congdon-logCAR: 1335, 1342, 1337 and 1339, respectively, vs BYM2 and Leroux: 1371 and 1374, respectively). Congdon's prior does not seem to perform significantly worse than the BYM2-Gamma model. 
Interestingly, even though the proposed model has 160 more parameters than Congdon's, its effective number of parameters is similar (80 vs 81). In terms of MSE, all models yield similar values, between 243.5, for the Congdon-logCAR model, and 245.7 for the Leroux model. 

Regarding the intercept, $\beta_0$, the proposed models and Congdon's give similar results, whereas the Leroux and BYM2 models yield smaller posterior means and lower credible interval bounds. This is probably due to the difference in the spatial effects that are allowed to be more extreme in the Congdon, Congdon-logCAR, BYM2-Gamma and BYM2-logCAR models. All six models indicate a negative relationship between the development index and the risk of Zika, with negative posterior 95\% credible intervals for $\beta$ that do not include 0. We cannot directly compare the parameters $\lambda$ and $\sigma$ between the BYM2-type models and Leroux-type priors, as these lie in the \textit{marginal} and \textit{conditional} distributions of the latent effects, respectively. 
Marginally, the BYM2-type models yield similar weights of the spatially structured components on the latent effects (posterior means for $\lambda$ of 0.6 and 0.7). 
For the Leroux-type models, the point estimates for $\lambda$ show slightly more difference (e.g. 0.6 for Leroux and 0.8 for Congdon). This difference may be due to the presence of outliers in the data, which results in the Leroux model finding more random noise in the latent effects.
The same observation can be made for the marginal and conditional standard deviation, $\sigma$, regarding the BYM2-type models and the Leroux-type models, respectively. The posterior credible interval for $\sigma$ is significantly higher in the BYM2 model compared to the two parametrisations of the proposed model, and in the Leroux model compared to the two versions of Congdon's model. Indeed, the proposed models are able to estimate a smaller overall variance for the latent effects, which is then adjusted through the $\kappa$'s when needed. Finally, it can be noted that there seems to be enough information in the data to learn about the hyperparameter $\nu$. This parameter was assigned a prior mean of 4 and prior 95\% credible interval of $[0.1, 14.7]$ for the BYM2-Gamma and Congdon models and resulted in posterior means of about 2 and posterior 95\% credible intervals of about $[1,3]$. The BYM2-logCAR and Congdon-logCAR models assigned an exponential distribution with mean 0.3 for $\nu$, inducing a prior 95\% credible interval of $[0.0, 1.1]$, and yielded posterior credible intervals of $[0.7, 2.3]$ and $[0.9, 2.9]$, showing the need for some $\kappa$'s to be different from 1, \textit{a posteriori}.

\begin{table}[H]
\footnotesize
	\centering
	\begin{tabular}{l c c c c c c}
		\toprule
		& BYM2 & BYM2-logCAR & BYM2-Gamma & Congdon & Congdon-logCAR & Leroux \\
		\midrule
		\multicolumn{7}{l}{Model fit} \\
		\midrule
		WAIC & 1371.2 & 1342.3 & 1335.6 & 1337.5 & 1339.2 & 1373.9\\
		$p_W$ & 88.6 & 82.3 & 80.0 & 81.0 & 81.1 & 89.2 \\
		\addlinespace
		\addlinespace
		MSE & 244.8 & 243.7 & 244.1 & 244.3 & 243.5 & 245.7 \\
		\addlinespace
		\addlinespace
		\multicolumn{7}{l}{Parameters' posterior summaries}\\
		\midrule
	& Mean (95\% CI) & Mean (95\% CI) & Mean (95\% CI) & Mean (95\% CI)& Mean (95\% CI) & Mean (95\% CI) \\
  & & & & & & \\
		\midrule
		$\beta_0$ & 1.6 (0.4,2.8) & 2.5 (1.3,3.5) & 2.5 (1.7,3.4) & 2.4 (1.4,3.2) & 2.0 (1.0,3.0) & 1.2 (-0.1,2.4) \\ 
		$\beta$ & -2.8 (-4.8,-0.8) & -4.2 (-5.8,-2.3) & -4.3 (-5.6,-2.9) & -4.0 (-5.4,-2.6) & -3.7 (-5.1,-1.9) & -1.9 (-4.1,-0.1) \\ 
		$\lambda$ & 0.7 (0.4,0.9) & 0.6 (0.2,0.9) & 0.7 (0.3,0.9) & 0.8 (0.5,0.9) & 0.6 (0.2,0.9) & 0.6 (0.2,0.9)\\
		$\sigma$ & 0.8 (0.7,0.9) & 0.4 (0.3,0.5) & 0.4 (0.3,0.5) & 0.6 (0.4,0.8) & 0.6 (0.4,0.8) & 1.2 (0.9,1.5) \\ 
		$\nu$ & - & 1.4 (0.7,2.3) & 2.2 (1.4,3.3) & 1.9 (1.3,2.8) & 1.7 (0.9,2.9) & - \\
		\bottomrule
	\end{tabular}
	\caption{\footnotesize Results from the analysis of Zika reported cases in Rio de Janeiro in 2015-2016. Model assessment (WAIC) and parameter posterior summaries: posterior mean and 95\% credible interval (CI) for BYM2, BYM2-logCAR, BYM2-Gamma, Congdon and Leroux.}
	\label{tab:Zika_Res}
\end{table}

We now focus on the outliers detected by the proposed models and Congdon's, as shown in Figure \ref{fig:Zika_Out}. District $i$ is again found to be a potential outlier, after accounting for the socio-development index, if $\kappa_{u,i}$, the upper bound of the posterior 95\% credible interval of $\kappa_i$, is below 1. 
In Figure \ref{fig:Zika_Out}, the blue and red coloured districts help distinguish the detected outliers on the lower tail of the SMR distribution from the ones on the upper tail.
After accounting for the socio-development index, some districts are pointed out by the four models, such as Gericinó, Parque Col\'umbia, Vasco da Gama and Maré, on the lower tail of the SMR distribution, Barra de Guaratiba, Bonsucesso and Vista Alegre, on the upper tail. However, the BYM2-Gamma and Congdon's models both do not point out Paquetá in the upper tail, whereas the BYM2-logCAR and Congdon-logCAR models detect it. This may be explained by the offset size of Paquetá, which is among the smallest in the entire region of Rio. Note, however, that the BYM2-Gamma model is close to identifying Paquetá as an outlier as it results in $\kappa_u=1.04$ for this district. 
Neither of the four models identify Pedra de Guaratiba, which has a high SMR, as shown in Figure \ref{fig:Zika_SMR}. 
Interestingly, the district of São Cristóvão is detected as an outlier by all models except the BYM2-Gamma model, with $\kappa_u=1.2$. The BYM2-logCAR and Congdon's models detect few more potential outliers than the BYM2-Gamma model. 
Our simulations have shown that the BYM2-logCAR and Congdon models tend to detect non-outliers more often than the BYM2-Gamma model. We believe that this explains the differences in the outliers identified after accounting for the socio-development index.

\section{Discussion}
\label{sec:Discussion}

In this paper, we propose a disease mapping model that is able to identify areas with potentially outlying disease risks, after accounting for the effects of covariates. Outliers refer to areas with extreme risks - on the tail of the risk distribution - as well as spatial outliers, after accounting for covariates. Spatial outliers correspond to areas whose risk differs from their neighbours, after accounting for covariates. The proposed model is a scale mixture of the BYM2 model \cite{BYM2}. Two different prior specifications are proposed for the scale mixture components in order to compare independent components and spatially structured components. Our model allows for a straightforward interpretation of the parameters, that is common to every data application, while accommodating outliers. The parameters' interpretation is eased by the scaling process of the latent spatially structured components \cite{SorbyeRue}.

\textcolor{black}{A simulation study presents} the performance of the two versions of the proposed model compared to the one by Congdon\cite{Congdon}, as well as a version of Congdon's model that uses our proposed spatially structured mixture components. The neighbourhood structure of \textcolor{black}{France} is used and the latent effects of some \textcolor{black}{neighbouring departments} are contaminated to control the presence of outliers. The BYM2-Gamma version of the proposed model always performs best in terms of WAIC and in terms of MSE. Regarding the detection of outliers, the two versions of the proposed model \textcolor{black}{always} identify the contaminated \textcolor{black}{departments, compared to} the two parametrisations of Congdon's model \textcolor{black}{that miss up to 33\% of the outliers}. \textcolor{black}{Additionally}, the BYM2-Gamma version of the proposed model does not detect non-contaminated districts. Finally, in all of our simulation studies, the proposed model always performs at least as well as Congdon's, and often better, both in terms of WAIC, MSE and of outlier identification (see, e.g., \textcolor{black}{Appendices \ref{sec:sim_dist}, \ref{sec:sim_clustdist_Rio}}).

The cases of Zika that were recorded in Rio de Janeiro during the first 2015-2016 epidemic are analysed using the two parametrisations of the proposed model as well as the model by Congdon \cite{Congdon} and its version with spatially structured mixture components, the BYM2 \cite{BYM2} and the Leroux prior \cite{Leroux}. All six models find that there is a fairly strong negative association between the socio-development index and the number of cases, meaning that richer districts have lower disease risks. This finding is consistent with previous studies conducted in Rio de Janeiro, one investigating the first chikungunya epidemic in the city \cite{Freitas2021} and another also investigating Zika, but using a different methodological approach \cite{Raymundo2021}. These studies, including ours, indicate that improving sanitary conditions and reducing socio-economic disparities are of paramount importance to fight \textit{Aedes}-borne diseases. 

\begin{figure}[H]
    \centering
    \includegraphics[width=0.7\textwidth]{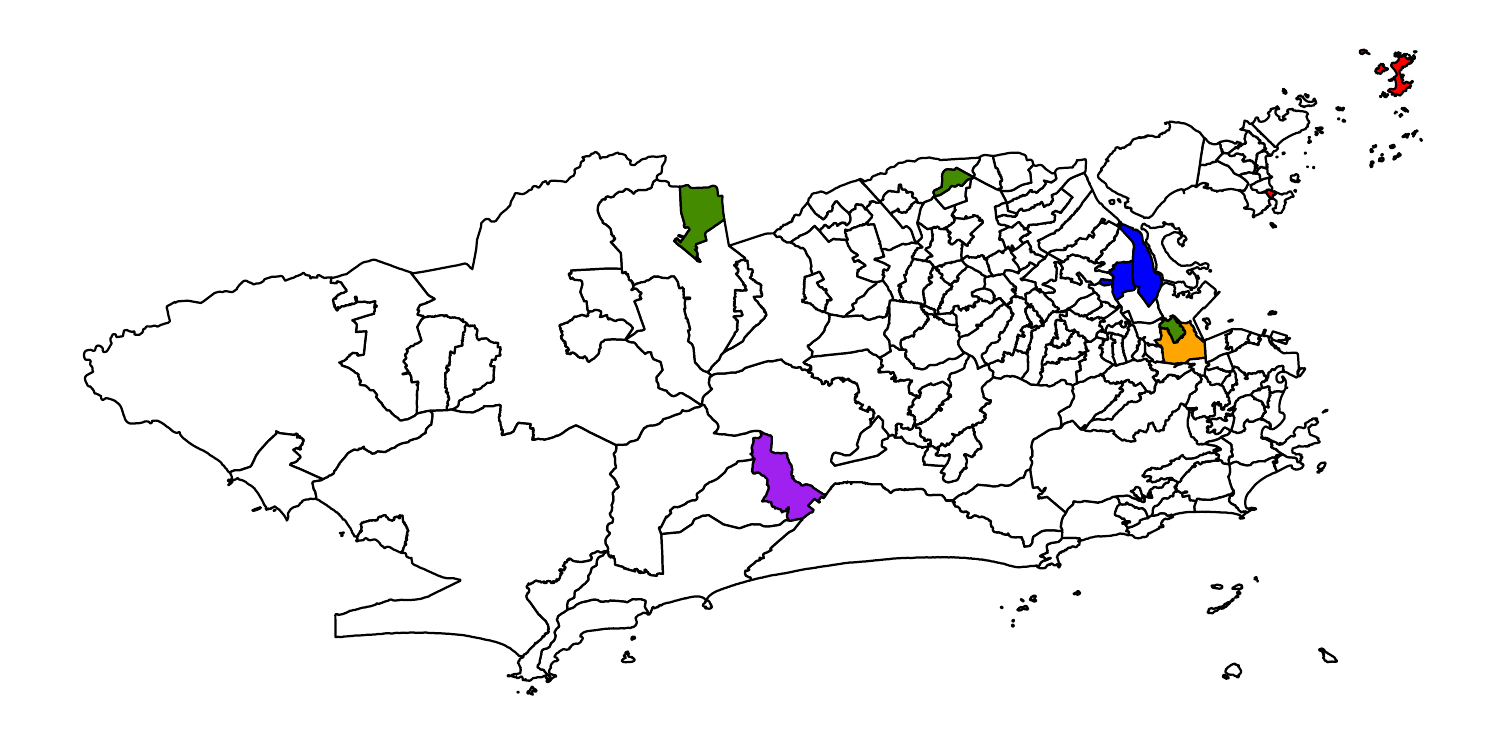}
    \caption{\footnotesize Map highlighting some districts identified as outliers by at least one model when analysing the Zika counts. Orange: São Cristóvão; Red: districts with small offsets; Blue: districts whose population sizes increased significantly after the 2010 census; Purple: districts combining both characteristics; Green: districts with zero cases recorded.}
    \label{fig:Zika_Out_Highlight}
\end{figure}

After accounting for the effect of the socio-development index, some neighbourhoods are detected as potential outliers by the proposed models and Congdon's, both in the lower and upper tails of the number of cases' distribution across the districts. Out of the 23 neighbourhoods identified as outliers, irrespective of the model, the proposed models BYM2-logCAR and BYM2-Gamma identified  19 (90.5\%) and 14 (60.9\%), respectively. 
The four models do not always point out the same districts as potential outliers. One possible explanation for that is the small offset sizes of some districts. The simulation study with neighbouring outliers showed that, when the offset is small, the models that impose a spatially structured prior on the scaling mixture components tend to accurately identify outlying areas more often than the models with \textit{a priori} independent mixture components. Regarding the analysis of Zika cases, Figure \ref{fig:Zika_Out_Highlight} shows in red and purple the districts identified as outliers by at least one of the four models and whose offsets are among the smaller 5\%. For example, based on the results from the second simulation study, it is possible that, when analysing the Zika counts, Camorim (purple) and the island Paquet\'a (red) are missed by the BYM2-Gamma and Congdon models while they are pointed out by the BYM2-logCAR and Congdon-logCAR models (Figure \ref{fig:Zika_Out}) because of their smaller offset sizes (Figure \ref{fig:Zika_Out_Highlight}).

Figure \ref{fig:Zika_Out_Highlight} highlights in green the districts with zero Zika cases recorded between 2015-2016: Parque Col\'umbia, Gericin\'o and Vasco da Gama. These 3 districts are pointed out as outliers by the four models, as shown in Figure \ref{fig:Zika_Out}. One potential explanation for these zero recorded cases is that when the disease appeared for the first time in 2015, it was not immediately identified as Zika. Further, there is evidence that epidemics in Rio de Janeiro tend to spread starting from the north-east of the city \citep{Freitas2019}. It is then possible that when the authorities began registering the Zika cases, there were no cases to record in the two northern districts highlighted in blue, Parque Col\'umbia and Gericin\'o. Another potential reason is that it is not uncommon in Rio de Janeiro for a person to report as their neighbourhood of residence a neighbourhood that actually shares a border with the one where they actually live. For instance, Parque Col\'umbia and Gericin\'o are relatively new districts and the population might not yet be used to naming them as their districts of residence. Similarly, a person living in Vasco da Gama (southern blue district) may report São Cristóvão (orange) as their district. This would artificially cause Vasco da Gama to record zero cases and be detected as a potential outlier. Further, if a given district is accounting for a proportion of the cases that are in fact from the neighbouring areas (e.g., São Cristóvão), this would artificially increase the risk of this district. In fact, Figure \ref{fig:Zika_Out} shows that São Cristóvão is pointed out as a potential outlier by all models but the BYM2-Gamma. Therefore, the inaccurate information on the district of residency may artificially create outliers.

Finally, artificial outliers may be caused by inaccurate information on the areal population sizes used to compute the offsets. While the disease counts were recorded during 2015-2016, the population sizes were extracted from the previous census, dating from 2010. Between 2010 and 2015-2016, the population sizes may have increased in some districts, without being reflected in the offsets in this analysis, causing the artificial detection of increased disease risks. Figure \ref{fig:Zika_Out_Highlight} highlights in blue and purple the districts identified as potential outliers and whose sizes have largely increased since 2010, according to more recent aerophotogrammetry flights by the Health Secretariat of the city. The eastern blue districts are pointed out as outliers by all four models in Figure \ref{fig:Zika_Out}. Further investigating these districts would help determine whether they do present outlying disease risks or if they are artificial outliers. An interesting side effect of the proposed model seems to be that by identifying outliers and further investigating the results, the authorities might better understand the population dynamics in the region of interest, in between censuses, and identifying potential issues in the accurate recording of cases.

Therefore, further investigation on the detected districts should be conducted by decision makers and experts to fully comprehend the detected outlying behaviours. Also, it is important to emphasize that some socio-environmental factors that influence the burden and distribution of \textit{Aedes}-borne diseases may be heterogeneous within the districts, our spatial unit of analysis. For example, the same district may have areas with \textit{favelas} (slums) and areas with middle and upper class condominiums. The socio-development index will not capture this intra-district social inequality, and a recent study showed evidence about the presence of socio-economic inequalities in the distribution of dengue, Zika and chikungunya in two Latin American cities \cite{Carabali2020}. Another possibility is the presence of large potential breeding sites, such as dumps and vacant lots.

To conclude, we believe our proposed model to be useful to decision makers. First, the parameters' interpretation eases the use of our model regardless of the data spatial structure. This may help decision makers to create a systematic procedure to analyse data with our proposed model, in which non-informative priors for the parameters could be defined for any spatial structure. Then, the introduction of scaling mixture components improves the recovering of the observed and potentially outlying disease risks, as assessed by the model performance criteria (WAIC and MSE). Finally, these mixture components together with high estimated risk ratios help identify all the potential outlying areas in which interventions may need to be prioritised.

\section*{Data Availability Statement}

The Zika and the population data analysed in this study come from the Brazilian Notifiable Diseases Information System (SINAN - Sistema de Informação de Agravos de Notificação) and the Brazilian Institute of Geography and Statistics (IBGE  - Instituto Brasileiro de Geografia e Estatística), respectively, and are publicly available at the Rio de Janeiro Secretariat of Health website (\url{http://www.rio.rj.gov.br/dlstatic/10112/7079759/4197436/ZIKASE2015.pdf} and \url{http://www.rio.rj.gov.br/dlstatic/10112/10617973/4260330/ZIKASE2016.pdf}, for 2015 and 2016, respectively). Note that SINAN reflects data from the public health system (SUS - Sistema Único de Saúde) only, which does not include data from private hospitals and health plans. The sociodevelopment index data come from the Instituto Pereira Passos and can be found at \url{www.data.rio}.

\bibliographystyle{abbrv}
\bibliography{biblio}%

\appendix

\section{Stan code for the proposed model}
\label{sec:Code}

The stan code used to fit the proposed BYM2-Gamma model in the simulation studies (section \ref{sec:sim_clust}, Appendices \ref{sec:SimFromModel}, \ref{sec:SimFromBLC}, \ref{sec:SimNoOut}, \ref{sec:sim_dist} and \ref{sec:sim_clustdist_Rio}) and in the analysis of the Zika epidemic in Rio de Janeiro (section \ref{sec:zika}) is presented below. 

\begin{lstlisting}[caption=Stan code for the BYM2-Gamma proposed model,
  label=code]
data {
  int<lower=1> N; // Number or areas
  int<lower=1> N_edges; // Total number of neighbours in the region
  int<lower=1> p; // General case where there are p covariates, excluding the intercept
  matrix[N,p] X;
  int<lower=1, upper=N> node1[N_edges]; // vectors of neighbourhood
  int<lower=1, upper=N> node2[N_edges];  // structure
  int<lower=0> y[N];              // Zika counts
  vector<lower=0>[N] log_E;           // offset
  real<lower=0> scaling_factor; // to scale the variance of the latent effects
}

parameters {
  real beta0;            // intercept
  vector[p] beta; // Fixed effects
  real<lower=0> sigma;        // marginal standard deviation
  real<lower=0, upper=1> lambda; // mixing parameter
  vector[N] theta;       // unstructured components
  vector[N] s;         // spatially structured components
  vector<lower=0>[N] kappa; // outlier indicator
  real<lower=0> nu; // parameter included in the prior for each kappa_i
}

transformed parameters {
  vector[N] convolved_re; // complete latent effect
  for(i in 1:N){ convolved_re[i] =  sqrt(1 - lambda) * theta[i] + sqrt(lambda/scaling_factor) * s[i]; }
}

model {
  for(i in 1:N)
   y[i] ~ poisson_log(log_E[i] + beta0 + X[i,]*beta + convolved_re[i] * (sigma/sqrt(kappa[i])) );
   
  target += -0.5 * dot_self(s[node1] - s[node2]); // Prior for the spatially structured components
  sum(s) ~ normal(0, 0.001 * N);  // Soft sum-to-zero constraint to be able to have an intercept
  
  for(j in 1:p){ beta[j] ~ normal(0.0, 10.0); }
  
  beta0 ~ normal(0.0, 10.0);
  theta ~ normal(0.0, 1.0);
  sigma ~ normal(0.0, 1.0); 
  lambda ~ uniform(0.0, 1.0);
  kappa ~ gamma(nu/2.0,nu/2.0);
  nu ~ exponential(1.0/4.0);
}

generated quantities {
  vector[N] mu_log;
  vector[N] lik;
  for(i in 1:N){
    mu_log[i]=log_E[i] + beta0 + X[i,]*beta + sigma*convolved_re[i]/sqrt(kappa[i]);
    lik[i] = exp(poisson_log_lpmf(y[i] | mu_log[i])); // likelihood to compute the WAIC
  }
}
\end{lstlisting}

\section{\textcolor{black}{Convergence diagnostics for the proposed model}}
\label{sec:MCMC_CV}

\textcolor{black}{In this section, we present the trace plots, effective sample sizes and $\widehat{R}$ statistics for a few selected parameters of the two parametrisations of the proposed model, when fitted to the data application in Section \ref{sec:zika}. For the mixture components, $\kappa$'s, we select the ones that produced the best and the worst convergence diagnostics.}

\begin{table}[H]
\footnotesize
    \centering
    \begin{tabular}{c cccc}
         \toprule
         & \multicolumn{2}{c}{\textcolor{black}{BYM-Gamma}} & \multicolumn{2}{c}{\textcolor{black}{BYM2-logCAR}} \\
         & \textcolor{black}{ESS} & \textcolor{black}{$\widehat{R}$} & \textcolor{black}{ESS} & \textcolor{black}{$\widehat{R}$} \\
         \midrule
         \textcolor{black}{$\kappa_{92}$} & \textcolor{black}{1305} & \textcolor{black}{1.000} & \textcolor{black}{1614} & \textcolor{black}{0.999}\\
         \textcolor{black}{$\kappa_{13}$} & \textcolor{black}{2000} & \textcolor{black}{0.999} & \textcolor{black}{2838} & \textcolor{black}{0.999}\\
         \textcolor{black}{$\lambda$} & \textcolor{black}{1958} & \textcolor{black}{1.009} & \textcolor{black}{1211} & \textcolor{black}{1.007}\\
         \textcolor{black}{$\nu$} & \textcolor{black}{2000} & \textcolor{black}{1.000} & \textcolor{black}{1817} & \textcolor{black}{1.000}\\
         \textcolor{black}{$\sigma$} & \textcolor{black}{1912} & \textcolor{black}{1.001} & \textcolor{black}{1987} & \textcolor{black}{1.004}\\
         \bottomrule
    \end{tabular}
    \caption{\footnotesize\textcolor{black}{Effective sample sizes (ESS) and $\widehat{R}$ statistics for some parameters when fitting the two parametrisations of the proposed model to the Zika data. $\kappa_{13}$ and $\kappa_{92}$ were chosen because they produced the best and the worst convergence diagnostics.}}
    \label{tab:Zika_ESS_R}
\end{table}

\begin{figure}[H]
    \centering
    \includegraphics[width=\textwidth]{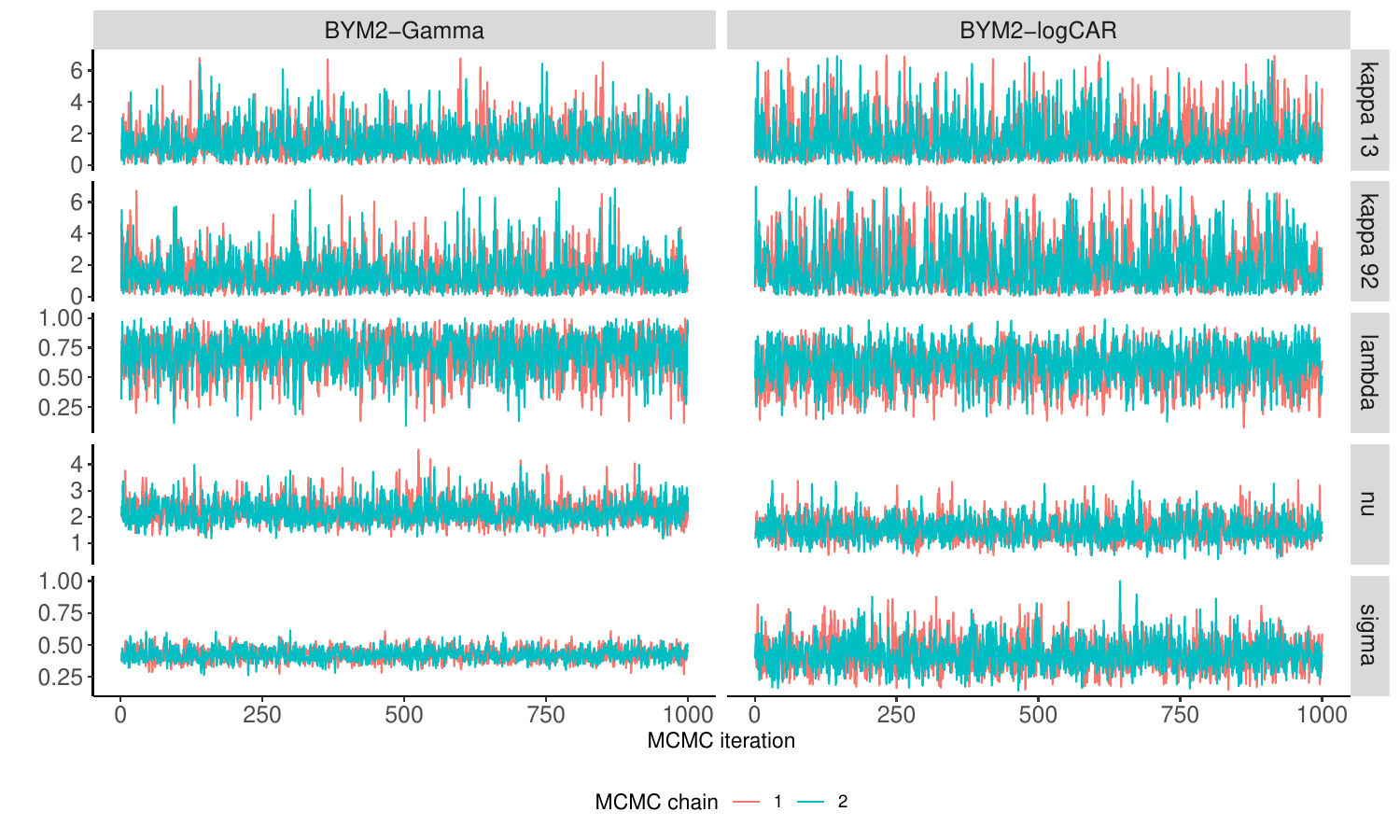}
    \caption{\footnotesize\textcolor{black}{Trace plots for some parameters when fitting the two parametrisations of the proposed model to the Zika data. $\kappa_{13}$ and $\kappa_{92}$ were chosen because they produced the best and the worst convergence diagnostics.}}
    \label{fig:Zika_Traceplots}
\end{figure}

\section{Simulation study: generating data from the proposed BYM2-Gamma model}
\label{sec:SimFromModel}

To assess the proposed BYM2-Gamma model's ability to recover the truth, a simulation study is conducted wherein data are generated from the proposed BYM2-Gamma model. Again, the $n=160$ districts of Rio de Janeiro and their neighbourhood structure are used. The latent effects' unstructured and scaled spatially structured components are generated once: $$\bm{\theta} \sim \mathcal{N}(\bm{0}, \bm{I}), \quad \mbox{and} \quad \bm{u}^\star \sim \mathcal{N}(\bm{0}, \bm{Q}_\star^-),$$ where $\bm{Q}_\star=h(\bm{D}-\bm{W})$, with $h$, the scaling factor, entirely defined by the spatial structure of Rio de Janeiro. An algorithm to generate from the ICAR prior is presented in Chapter 2 of Rue and Held \cite{RueHeld}. The \textcolor{black}{mixing components that induce the marginal heavier tails}, $\bm{\kappa}$, are independently generated once from a $\mathrm{Gamma}(\nu/2,\nu/2)$, with \textcolor{black}{$\nu$ fixed at $4$} to allow for fairly heavy tails. The latent effects are then computed as $$b_i = \left[\sqrt{1-\lambda}\theta_i + \sqrt{\lambda}u_i^\star\right]\times \sigma/\sqrt{\kappa_i}, \ i=1, \dots, n,$$ where $\lambda=0.8$ and $\sigma=0.3$. Finally, 100 replicates of populations of size $n=160$ are generated from the Poisson model $$Y_i \sim \mathcal{P}\left(E_i \exp\left[\beta_0 + b_i\right]\right),$$ with $\beta_0=-0.1$ and the offsets, $[E_1, \dots, E_n]^\top$, taken from the analysis of the Zika counts. \textcolor{black}{Then, models BYM2-Gamma and Congdon are fitted to each of the 100 replicates, using the same inference procedure as in section \ref{sec:sim_clust}. The goal is to check if we recover the true values used to generate the data, and to check if the WAIC is able to distinguish between the proposed model and Congdon's.} 

\begin{figure}[H]
	\centering
	\includegraphics[page=1, width=\textwidth]{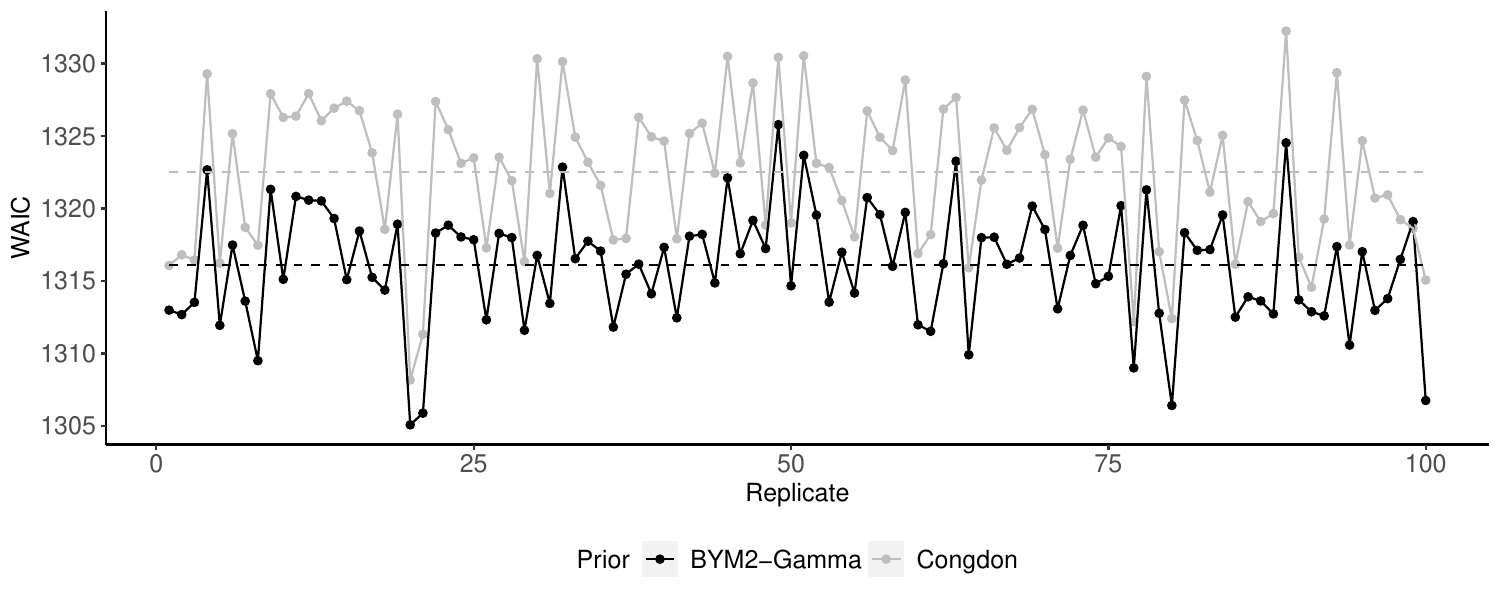}
	\caption{\footnotesize WAIC across the 100 replicates for the proposed BYM2-Gamma model and Congdon's regarding the simulated data from the proposed BYM2-Gamma model. Dashed lines: mean WAIC for each model}
	\label{fig:WAIC_SimFromModel}
\end{figure}

\begin{figure}[H]
	\centering
	\includegraphics[page=1, width=\textwidth]{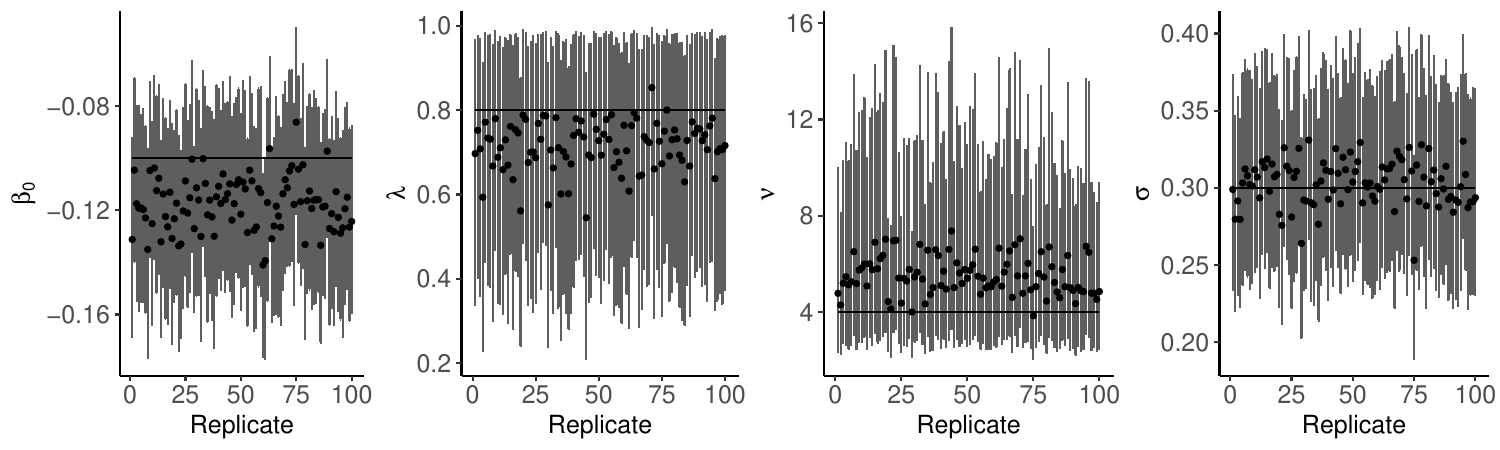}
	\caption{\footnotesize Posterior summaries of the parameters for the proposed BYM2-Gamma model across the 100 replicates regarding the simulated data from the proposed BYM2-Gamma model.\\ Solid circle: posterior mean; Vertical lines: 95\% posterior credible interval; Solid horizontal line: true value. 
	}
	\label{fig:Param_SimFromModel}
\end{figure}

\begin{figure}[H]
	\centering
	\includegraphics[page=1, width=\textwidth]{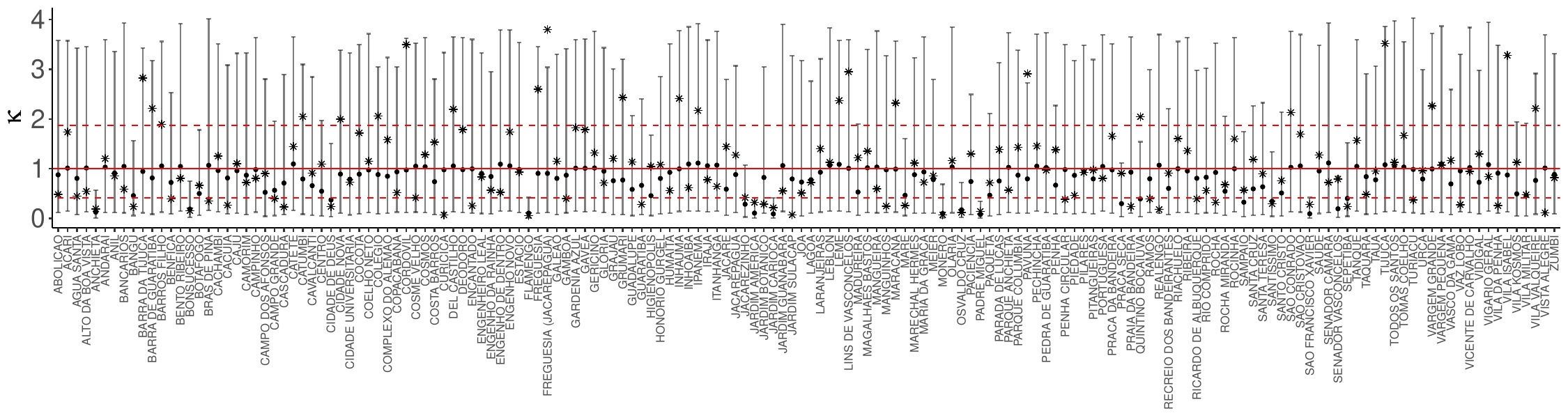}
	\caption{\footnotesize\textcolor{black}{Posterior summaries (mean and 95\% credible interval) of the $\kappa$ parameters across all the districts of Rio de Janeiro for one replicate when fitting the BYM2-Gamma model. The stars correspond to the true generated $\kappa$'s and the red horizontal lines correspond to the prior summary (solid line: prior mean, dashed lines: prior 95\% credible interval).}}
	\label{fig:kappa_SimFromModel}
\end{figure}

\begin{figure}[H]
	\centering
	\includegraphics[page=1, width=\textwidth]{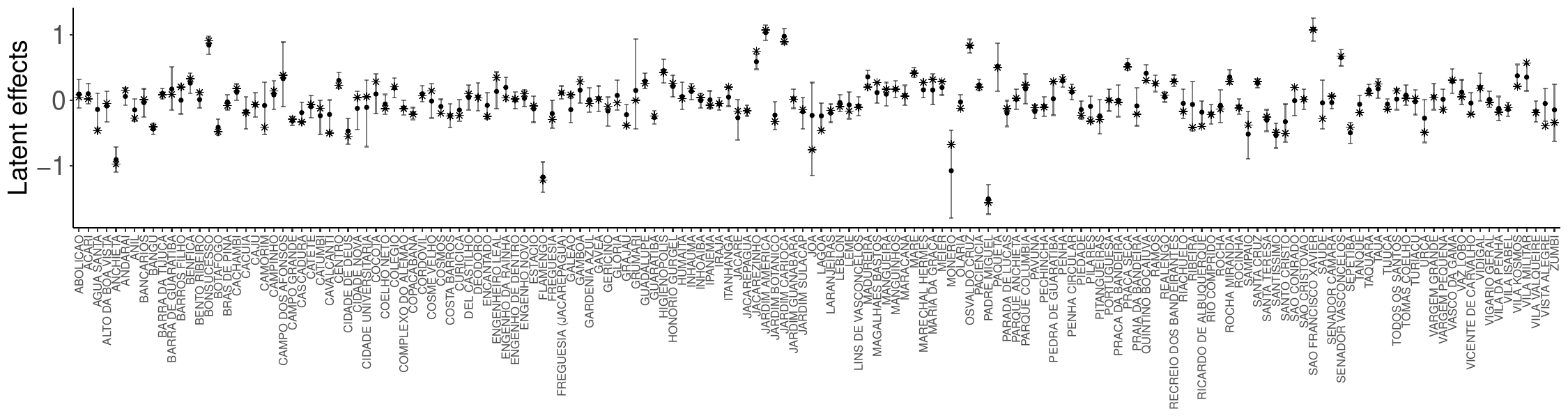}
	\caption{\footnotesize\textcolor{black}{Posterior summaries (mean and 95\% credible interval) of the latent effects across all the districts of Rio de Janeiro for one replicate when fitting the BYM2-Gamma model. The stars correspond to the true generated latent effects.}}
	\label{fig:s_SimFromModel}
\end{figure}

Figure \ref{fig:WAIC_SimFromModel} shows that the WAIC is able to always choose the model that generated the data, namely the BYM2-Gamma model. 
Figure \ref{fig:Param_SimFromModel} presents the posterior summaries obtained from the BYM2-Gamma model across the 100 replicates for the intercept, $\beta_0$, the mixing parameter, $\lambda$, the hyperparameter, $\nu$, and the overall standard deviation, $\sigma$. \textcolor{black}{For all samples, the 95\% posterior credible intervals of all parameters contain the true values used to generate the data.}
The interest lies particularly on the main parameters of the model, such as the outlier indicators, $\bm{\kappa}$. Figure \ref{fig:kappa_SimFromModel} plots the \textcolor{black}{posterior summaries, for one replicate, of the $\kappa$'s across all districts in Rio de Janeiro. \textcolor{black}{Most of the 95\% posterior credible intervals for $\kappa$ contain the true value used to generate the data. Moreover, for those neighbourhoods that have outlying  observations, the estimate for $\kappa$ is quite concentrated around its true value. This suggests that the model is able to point out the neighbourhoods with outlying observations.}
Similarly, the true latent effects, $\bm{b}$, are shown to be recovered by the 95\% posterior credible intervals in Figure \ref{fig:s_SimFromModel}.}

\section{Simulation study: generating data from the proposed BYM2-logCAR model}
\label{sec:SimFromBLC}

We now assess the proposed BYM2-logCAR model's ability to recover the truth. Similar to Appendix \ref{sec:SimFromModel}, a simulation study is conducted wherein data are generated from the proposed BYM2-logCAR model using the $n=160$ districts of Rio de Janeiro. The unstructured and spatially structured components, $\bm{\theta}$ and $\bm{u}^\star$ respectively, are independently generated once, like in Appendix \ref{sec:SimFromModel}. The scaling mixture components, $\bm{\kappa}$, are generated once using the spatial structure as follows: $$\bm{z} \mid \nu_\kappa \sim \mathcal{N}\left(\bm{0}, \nu_\kappa \bm{Q}_{\alpha,\star}^{-1}\right) \quad \mbox{and} \quad \kappa_i = \exp\left(- \frac{\nu_\kappa}{2} + z_i\right), \ i=1, \dots, n,$$
 where $\bm{Q}_{\alpha, \star}=h\bm{Q}_\alpha = h_\alpha[\bm{D}-\alpha\bm{W}]$ is again the valid precision matrix that is scaled by $h_\alpha$, which is computed based on $\bm{D}-\alpha\bm{W}$. We impose $\alpha=0.99$ and define an arbitrary $\nu_\kappa=0.3$ to allow the $\kappa$'s to depart from 1. Like in Appendix \ref{sec:SimFromModel}, the latent effects are then computed as $b_i = \left[\sqrt{1-\lambda}\theta_i + \sqrt{\lambda}u_i^\star\right]\times \sigma/\sqrt{\kappa_i}, \ i=1, \dots, n,$ where $\lambda=0.8$ and $\sigma=0.3$. Finally, 100 replicates of populations of size $n=160$ are generated from the Poisson model, $Y_i \sim \mathcal{P}\left(E_i \exp\left[\beta_0 + b_i\right]\right),$ with $\beta_0=-0.1$ and the offsets, $[E_1, \dots, E_n]^\top$, taken from the analysis of the Zika counts. The proposed BYM2-logCAR model and Congdon's are both fitted on the 100 replicated datasets using the same inference procedure as in section \ref{sec:sim_clust}. 

\begin{figure}[H]
	\centering
	\includegraphics[page=1, width=\textwidth]{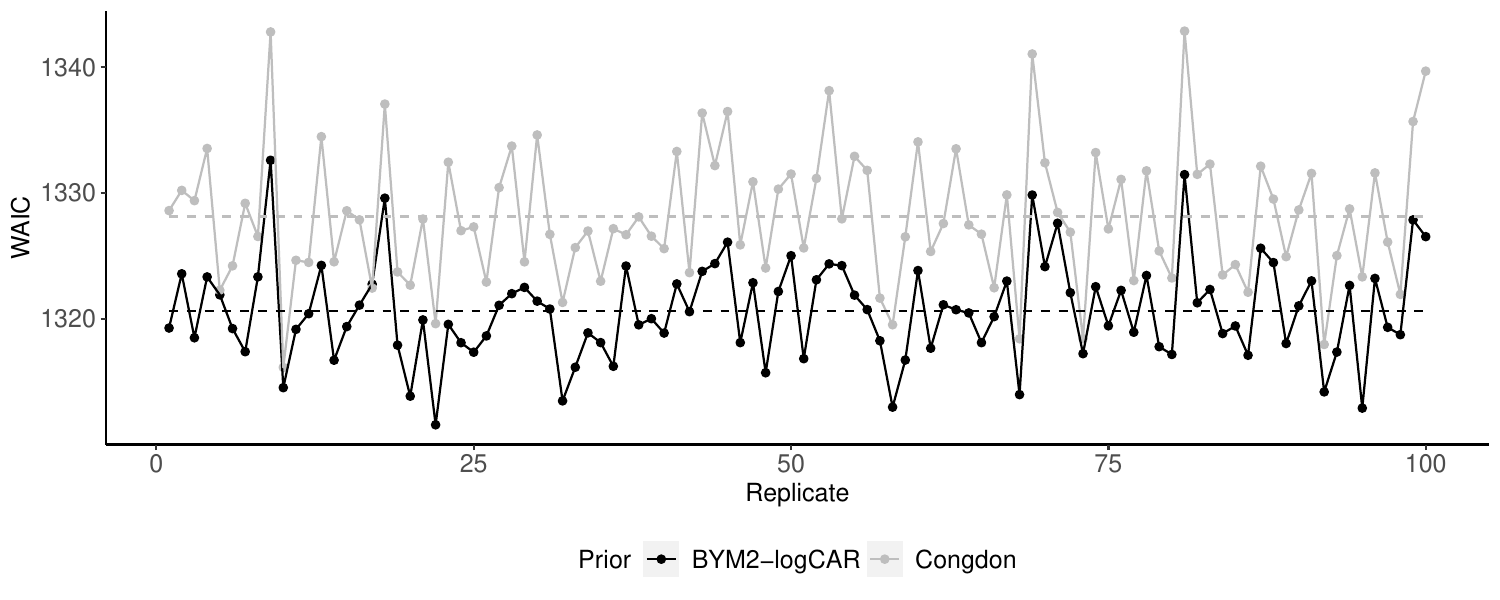}
	\caption{\footnotesize WAIC across the 100 replicates for the proposed BYM2-logCAR model and Congdon's regarding the simulated data from the proposed BYM2-logCAR model. Dashed lines: mean WAIC for each model}
	\label{fig:WAIC_SimFromBLC}
\end{figure}

\begin{figure}[H]
	\centering
	\includegraphics[page=1, width=\textwidth]{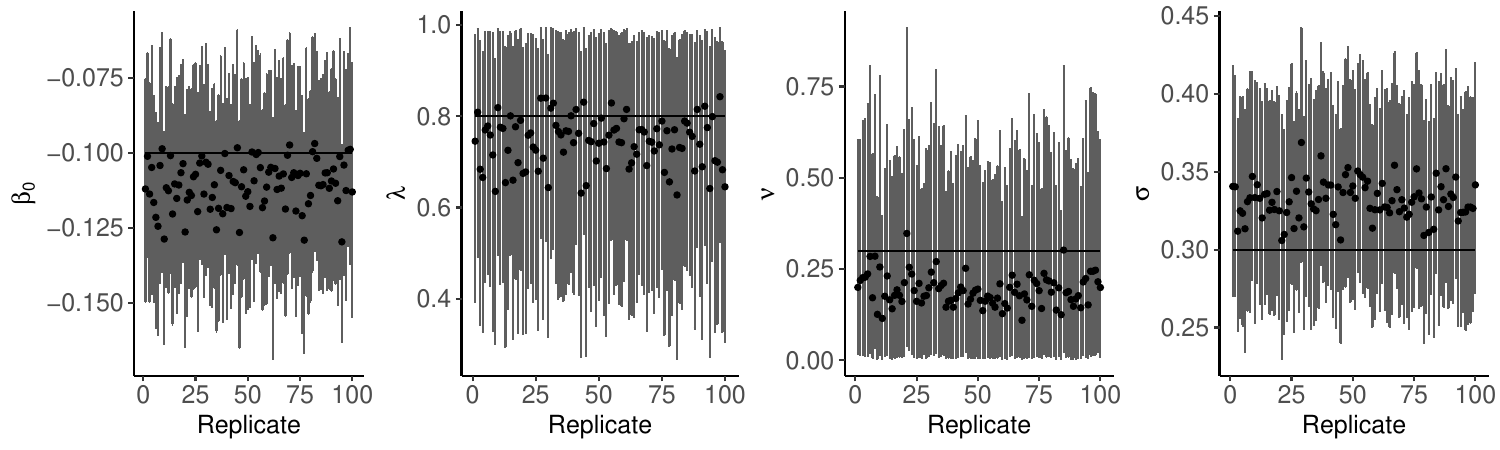}
	\caption{\footnotesize Posterior summaries of the parameters for the proposed BYM2-logCAR model across the 100 replicates regarding the simulated data from the proposed BYM2-logCAR model.\\ Solid circle: posterior mean; Vertical lines: 95\% posterior credible interval; Solid horizontal line: true value. 
	}
	\label{fig:Param_SimFromBLC}
\end{figure}

\begin{figure}[H]
	\centering
	\includegraphics[page=1, width=\textwidth]{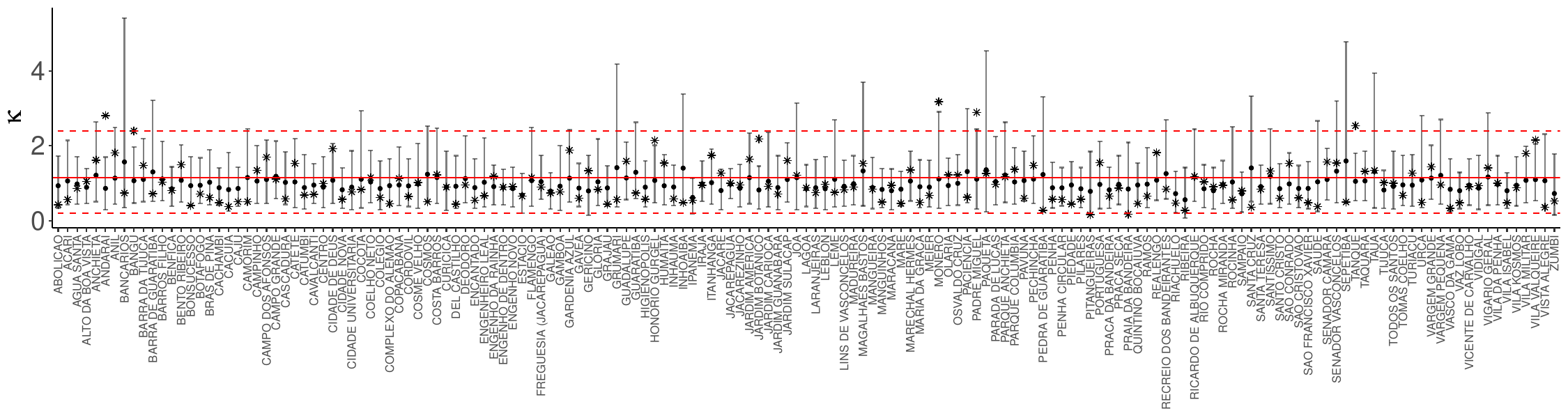}
	\caption{\footnotesize \textcolor{black}{Posterior summaries (mean and 95\% credible interval) of the $\kappa$ parameters across all the districts of Rio de Janeiro for one replicate when fitting the BYM2-logCAR model. The stars correspond to the true generated $\kappa$'s and the red horizontal lines correspond to the prior summary (solid line: prior mean, dashed lines: prior 95\% credible interval).}}
	\label{fig:kappa_SimFromBLC}
\end{figure}

\begin{figure}[H]
	\centering
	\includegraphics[page=1, width=\textwidth]{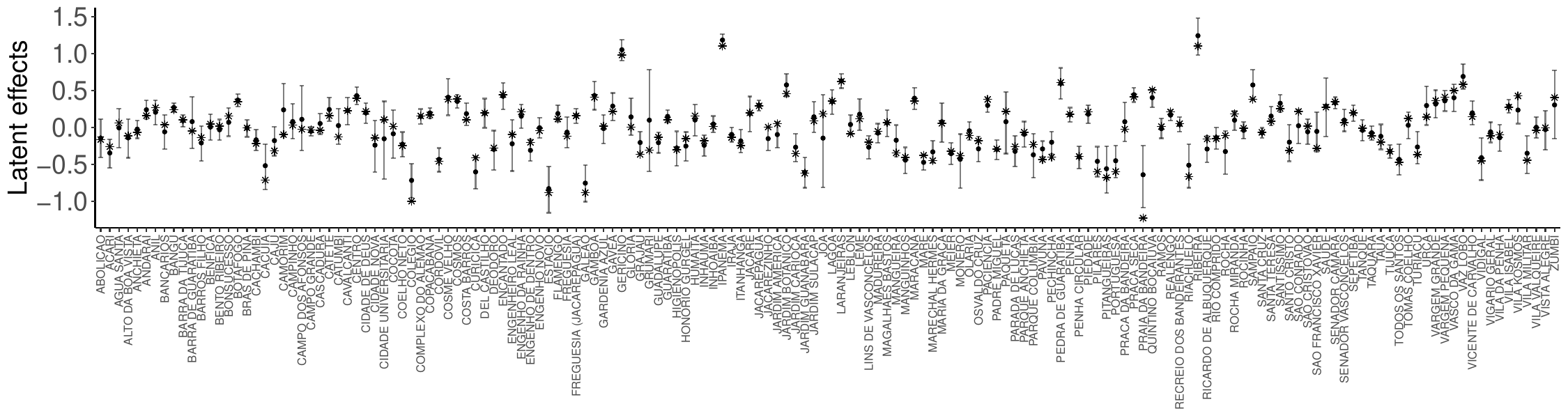}
	\caption{\footnotesize\textcolor{black}{Posterior summaries (mean and 95\% credible interval) of the latent effects across all the districts of Rio de Janeiro for one replicate when fitting the BYM2-logCAR model. The stars correspond to the true generated latent effects.}}
	\label{fig:s_SimFromBLC}
\end{figure}

Figure \ref{fig:WAIC_SimFromBLC} shows that the WAIC always favours the proposed BYM2-logCAR model, which generated the data. Figure \ref{fig:Param_SimFromBLC} shows how well the proposed BYM2-logCAR model is able to recover the true values of the model parameters through the posterior summaries across the 100 replicates for the intercept, $\beta_0$, the mixing parameter, $\lambda$, the hyperparameter, $\nu_\kappa$, and the overall standard deviation, $\sigma$. Across the 100 replicates, the proposed BYM2-logCAR model always captures the truth, as the posterior 95\% credible intervals (vertical lines) always cover the true values of the parameters (solid horizontal lines). 
Regarding the scaling mixture components, $\bm{\kappa}$, Figure \ref{fig:kappa_SimFromBLC} shows the \textcolor{black}{posterior summaries for one replicate and generated values, across all districts. The $\kappa$'s generated following this structured prior seem to vary less than the ones generated from the independent gamma priors in Appendix \ref{sec:SimFromModel}. Therefore, the posterior credible intervals are narrower than the ones from the simulation study presented in Appendix \ref{sec:SimFromModel}. Regardless, the posterior 95\% credible intervals almost always cover the true mixture components. Similarly, the generated latent effects plotted in Figure \ref{fig:s_SimFromBLC} are recovered by the posterior 95\% credible intervals.}

\section{Simulation study: no outlying areas}
\label{sec:SimNoOut}

\begin{figure}[H]
	\centering
	\includegraphics[page=1, width=\textwidth]{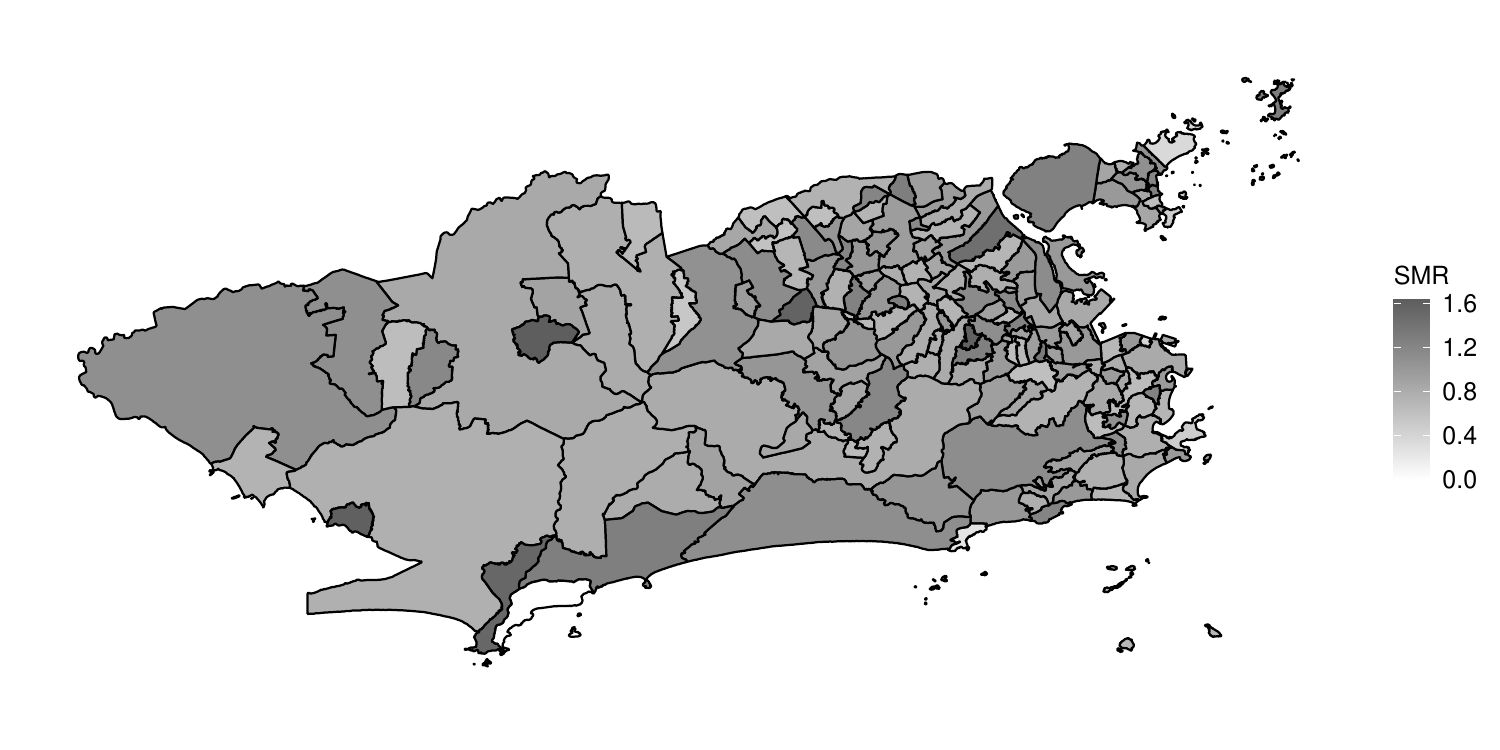}
	\caption{\footnotesize Standardised morbidity ratio for the 50th simulation without outliers.}
	\label{fig:SMR_SimNoOut}
\end{figure}

To confirm that the proposed model does not detect outliers when unnecessary, a simulation study is again conducted on the map of Rio de Janeiro without contaminating any district. Data are generated 100 times as follows: $$Y_i \sim \mathcal{P}\left(E_i \exp[\beta_0 + b_i]\right), \ i=1, \dots, n,$$ with $n=160$, $\beta_0=-0.1$, $\bm{E}=[E_1, \dots, E_n]^\top$ taken from the Zika data analysis. The latent effects, $\bm{b}=[b_1, \dots, b_n]^\top$, are simulated once from a PCAR distribution: $$\bm{b} \sim \mathcal{N}\left(\bm{0}, \sigma_b^2\left[\bm{D}-\alpha\bm{W}\right]^{-1}\right),$$ with $\sigma_b=\sqrt{0.2}$ and $\alpha=0.7$. Figure \ref{fig:SMR_SimNoOut} shows the map of the 50th replicate of the simulated dataset, where no district seems to be an outlier with respect to the whole city. Again, the two parametrisations of the proposed model are compared to Congdon's, using the same prior distributions as described in section \ref{sec:sim_clust}.

\begin{figure}[H]
	\centering
	\includegraphics[page=1, width=\textwidth]{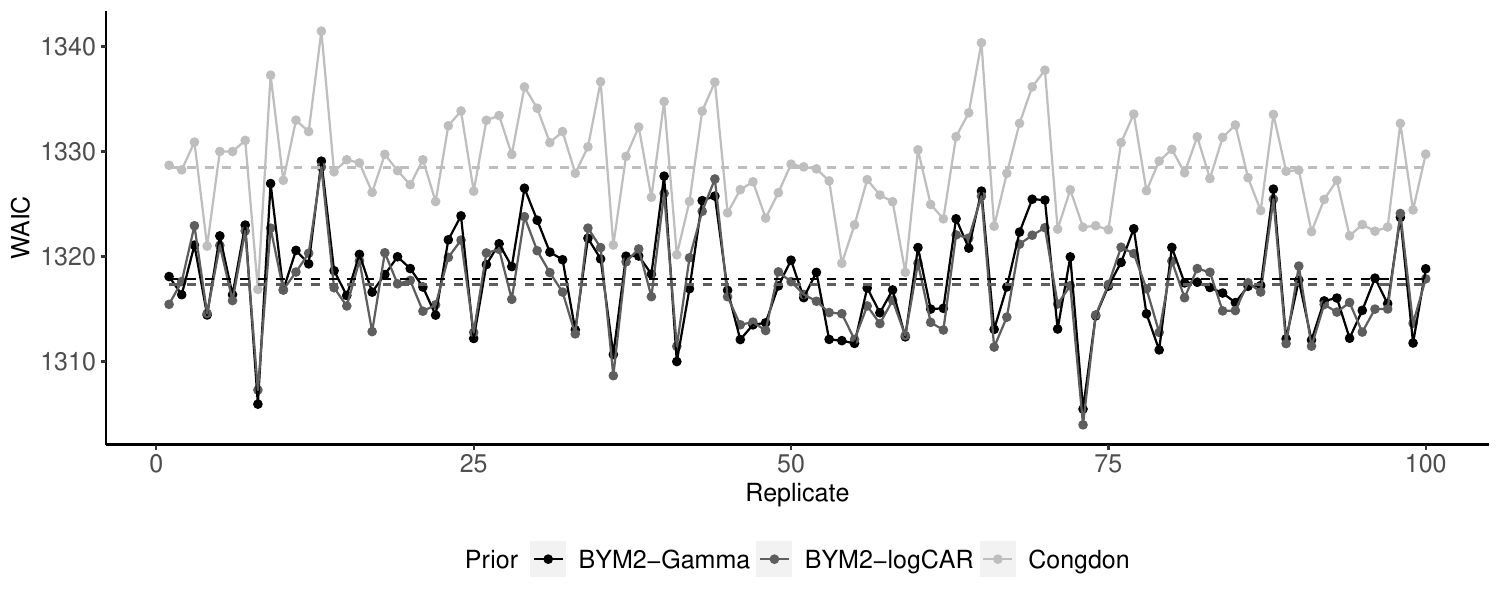}
	\caption{\footnotesize WAIC across the 100 replicates for the proposed models and Congdon's for the simulation without outliers. Dashed lines: mean WAIC for each model}
	\label{fig:WAIC_SimNoOut}
\end{figure}

\begin{figure}[H]
	\centering
	\hspace*{-0.07\textwidth}
	\subfloat[\centering ]{\includegraphics[page=1, width=0.6\textwidth]{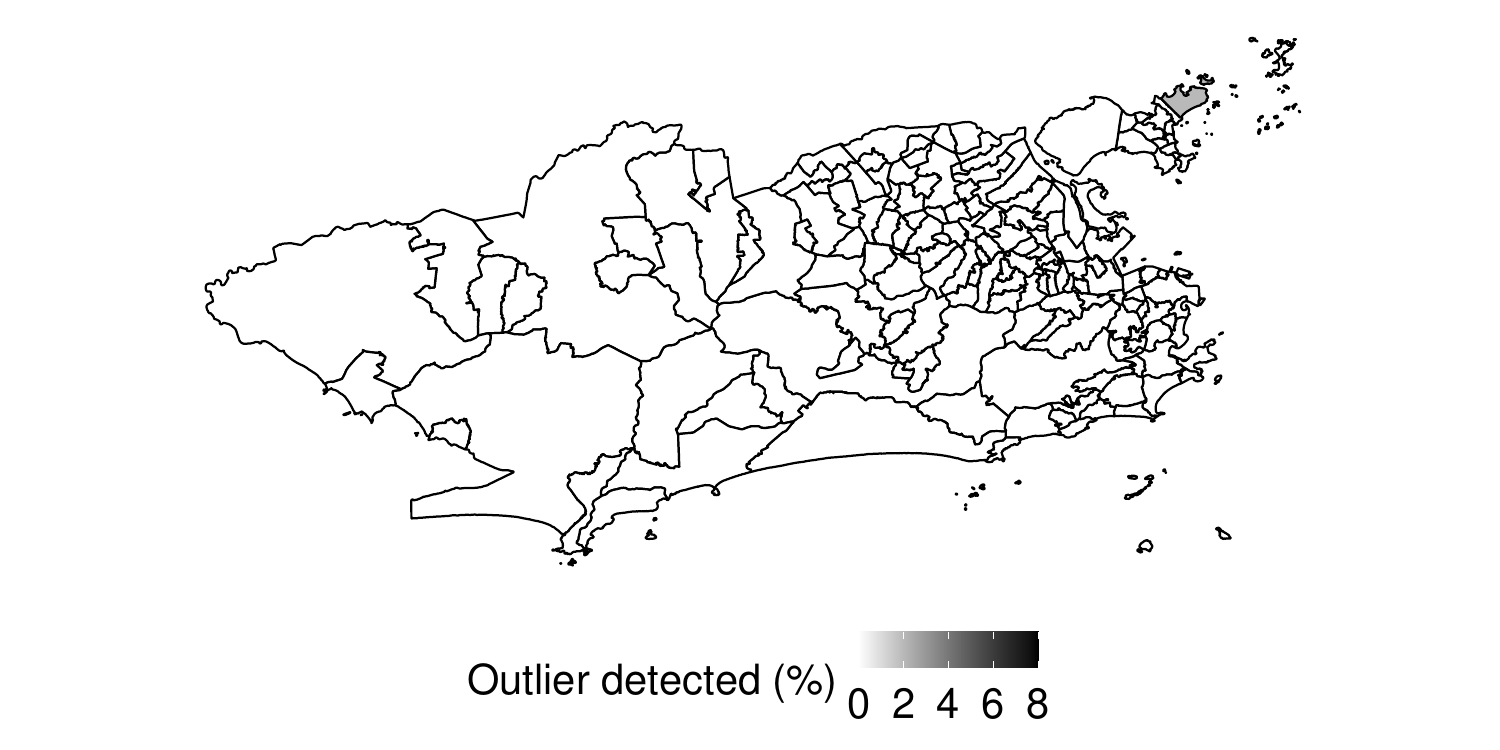}}
	\hspace*{-0.09\textwidth}
	\subfloat[\centering ]{\includegraphics[page=1,width=0.6\textwidth]{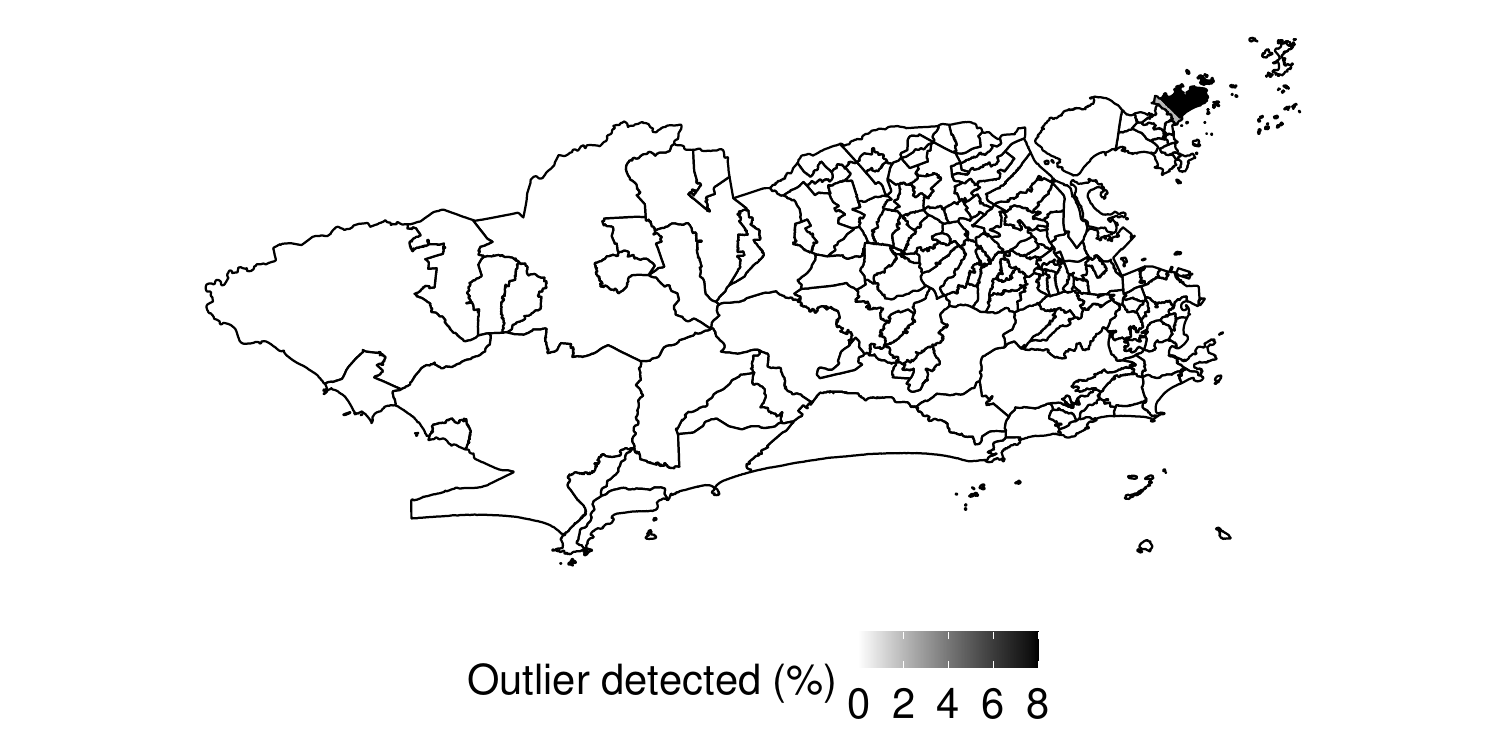}}\\
	\subfloat[\centering ]{\includegraphics[page=1,width=0.6\textwidth]{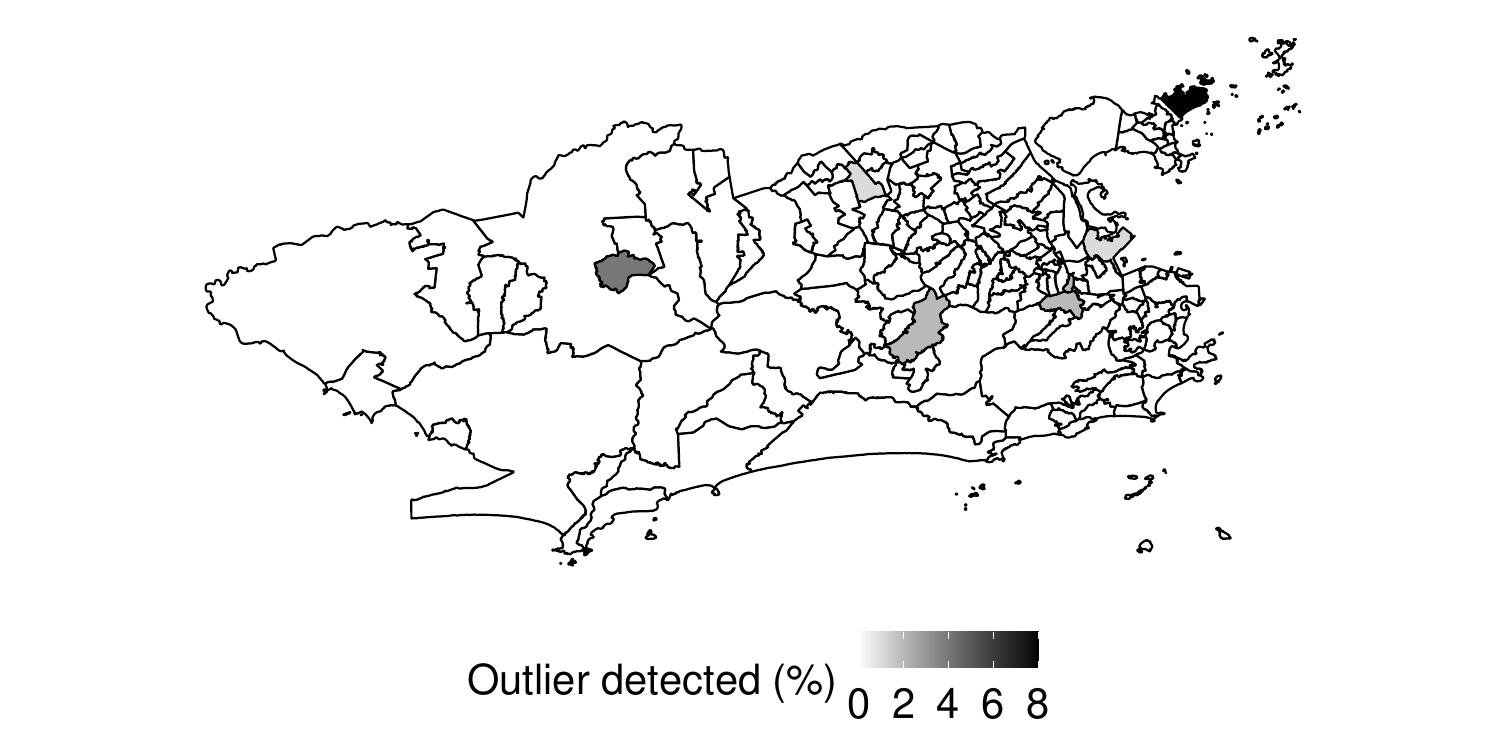}}
	\caption{\footnotesize\centering 
	Maps of the percentages of outliers as indicated by $\kappa_{ur}<1$ across the $r=1, \dots, 100$ replicates, where $\kappa_{ur}$ is the upper bound of the posterior 95\% credible interval of $\kappa$ in the $r$th replicate of the simulated dataset without outliers.
	a) BYM2-Gamma model; 
	b) BYM2-logCAR model;
	c) Congdon's model.}
	\label{fig:Out_SimNoOut}
\end{figure}

In terms of WAIC, the proposed models seem to perform best, as shown in Figure \ref{fig:WAIC_SimNoOut}. For this simulation study, the interest lies particularly in comparing the outliers detections from the two versions of the proposed model and Congdon's. Figure \ref{fig:Out_SimNoOut} presents the districts that are found to be outliers by the BYM2-Gamma proposed model (a), the BYM2-logCAR proposed model (b) and Congdon's (c). The BYM2-Gamma model only identifies one district, Freguesia, to be a potential outlier in 2\% of the replicates. The BYM2-logCAR and Congdon's models on the other hand detect Freguesia up to 8\% of the times, showing more sensitivity to the neighbourhood structure. Congdon's model further identifies 5 districts as potential outliers although no district was contaminated.

\section{\textcolor{black}{Simulation study: distant outliers in France}}
\label{sec:sim_dist}

\begin{figure}[H]
	\centering
	\includegraphics[page=1, width=0.7\textwidth]{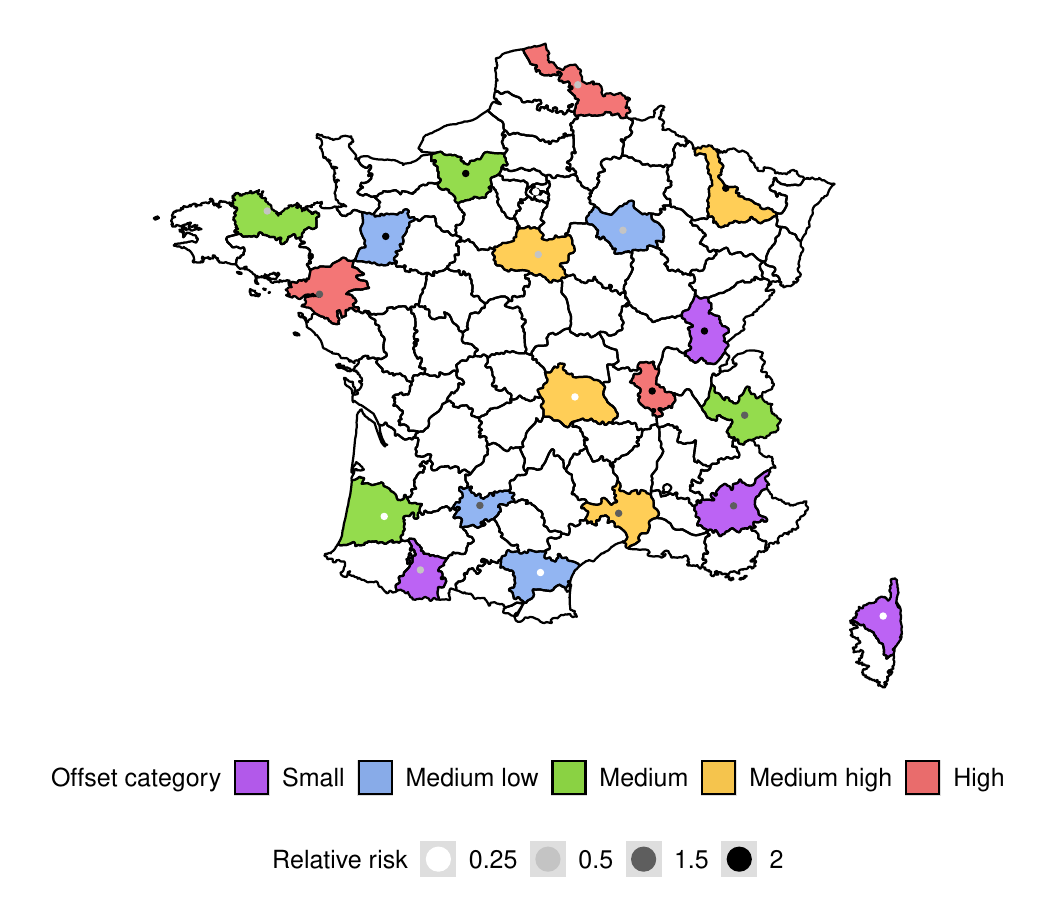}
	\caption{\footnotesize \textcolor{black}{French departments arbitrary chosen to be outliers in simulation study with distant outliers. Colours depict the offset category based on the empirical offset quantiles. The points represent the relative risk set to each outlying district.}}
	\label{fig:Out_Far_France}
\end{figure}

\textcolor{black}{In this simulation study, 20 \textcolor{black}{distant French departments} are contaminated such that outliers are created. \textcolor{black}{Similar to the simulation study presented in Section \ref{sec:sim_clust}}, there are no covariates in this analysis, and all areas are first imposed a relative risk of 1. The same five offset categories are defined. Based on these categories, we select 20 \textcolor{black}{non-neighbouring departments} to be outliers. Four \textcolor{black}{departments} are chosen from each offset category. That is, there are 4 outliers within the smallest offset group, 4 within the second-to-smallest offset group, and so on. Then, within each group of four \textcolor{black}{departments}, the relative risks are contaminated into outliers by setting the relative risks to be equal to $b_{i}=0.25$, $b_{i'}=0.5$, $b_{i''}=1.5$ and $b_{i'''}=2$. The resulting outliers are mapped in Figure \ref{fig:Out_Far_France}, highlighting the offset sizes and imposed relative risks. Again, $R=100$ populations of size $n=96$ are created by generating the number of cases $Y_i \sim \mathcal{P}\left(E_i \exp[b_i]\right)$. The same four models with priors defined in section \ref{sec:sim_clust} are fitted through \texttt{rstan}. After 20,000 iterations with a burn-in period of 10,000 and a thinning factor of 10, the 2 MCMC chains attained convergence as assessed by trace plots, effective sample sizes and $\widehat{R}$ statistics.}

\begin{figure}[H]
	\centering
	\includegraphics[page=1, width=\textwidth]{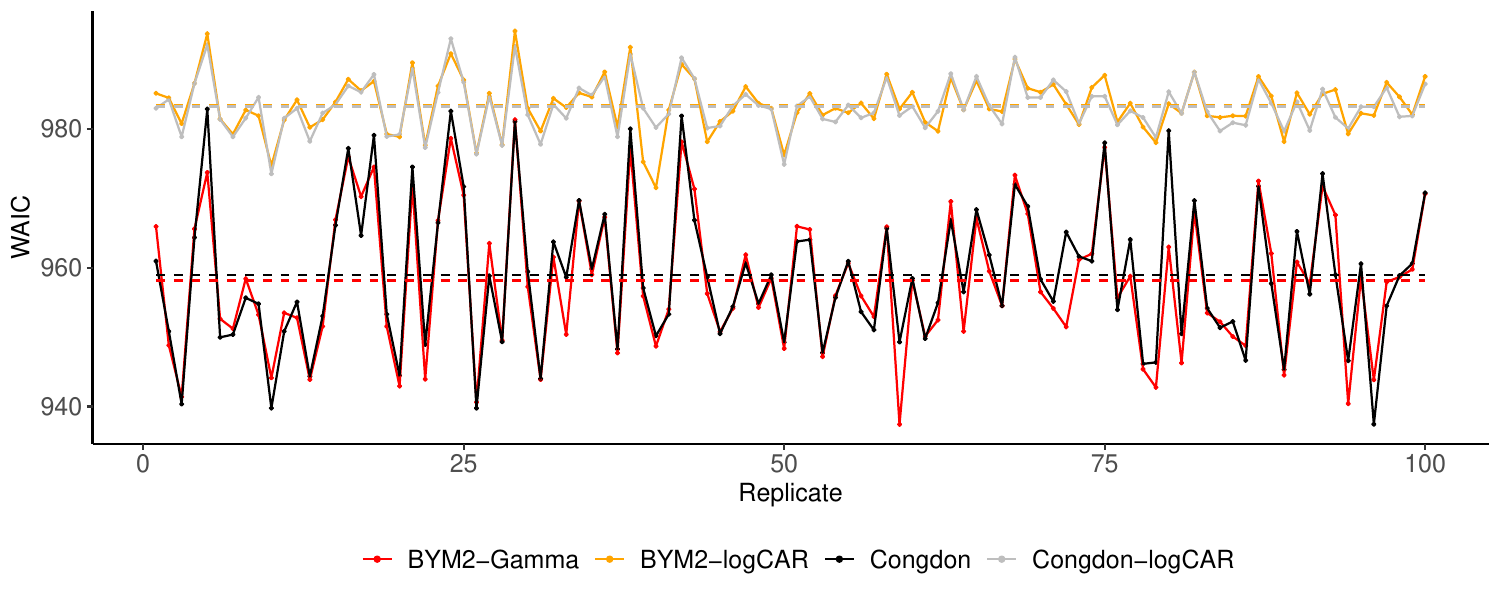}
	\caption{\footnotesize\textcolor{black}{WAIC across the 100 replicates for the proposed models and Congdon's, in the simulation study with distant outliers. Dashed lines: mean WAIC for each model.}}
	\label{fig:WAIC_Far_France}
\end{figure}

\textcolor{black}{In terms of WAIC \cite{WAIC}, for which smaller values are preferred, the proposed BYM2-Gamma model performs \textcolor{black}{similarly to}  Congdon's, as shown in Figure \ref{fig:WAIC_Far_France}. The BYM2-Gamma and original Congdon models always perform better than the models that include spatially structured scaling mixture components. On average, the BYM2-logCAR \textcolor{black}{and Congdon-logCAR models} yield a criterion of \textcolor{black}{983}, while the BYM2-Gamma \textcolor{black}{and Congdon models} present a WAIC of \textcolor{black}{958 and 959, respectively.}}

\begin{figure}[H]
	\centering
	\includegraphics[page=1, width=\textwidth]{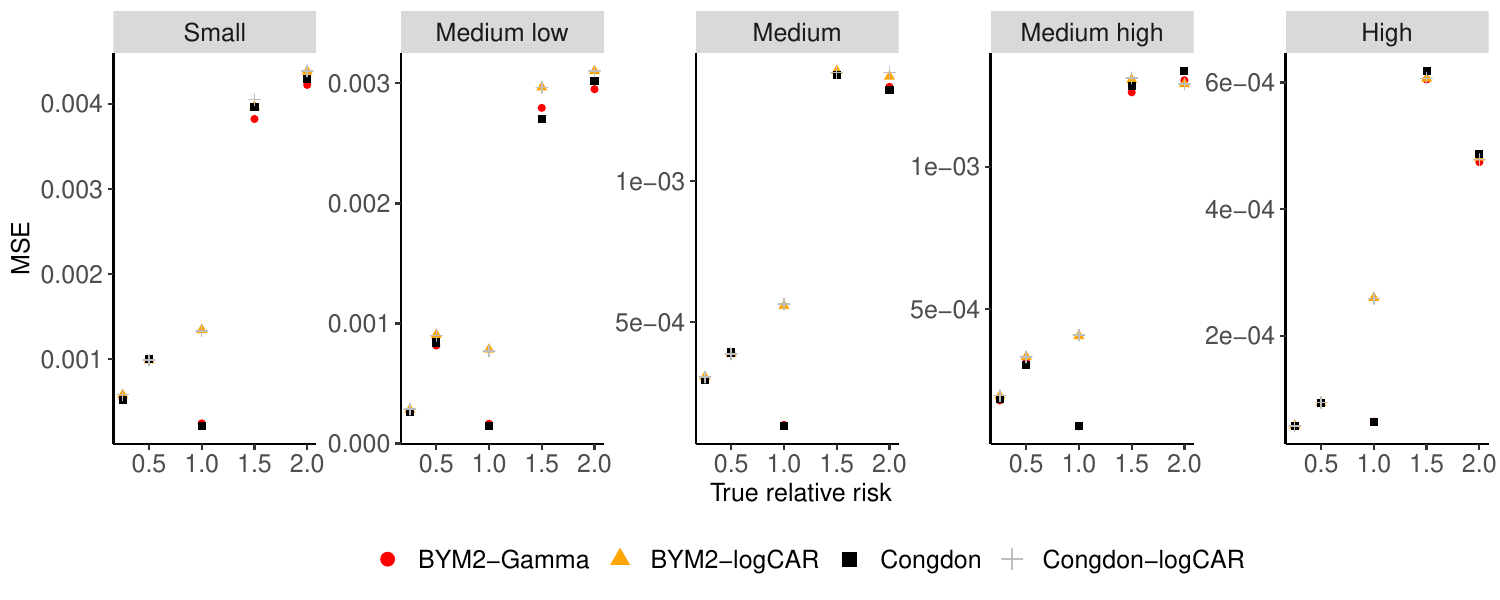}
	\caption{\footnotesize\textcolor{black}{MSE over the 100 replicates for the proposed models and Congdon's according to the true relative risk and the offset size, in the simulation study with distant outliers.}}
	\label{fig:MSE_Far_France}
\end{figure}

\textcolor{black}{The models' performances are also compared in terms of MSE, as shown in Figure \ref{fig:MSE_Far_France}. As expected, all models result in MSEs that are smaller in the areas with large offsets, and MSEs that are larger in the areas with small offsets. Additionally, all models tend to better fit the data in non-outlying areas, that is in the areas with a relative risk of 1. Regarding the outlying areas only, the largest MSEs are observed for extreme risks of 2 whereas the smallest correspond to extreme risks of 0.5. On average over the 100 replicated datasets and across all areas, the MSEs are of \textcolor{black}{0.0010} for the BYM2-Gamma \textcolor{black}{and Congdon} models, and \textcolor{black}{0.0011} for both log-CAR parametrisations.}

\begin{figure}[H]
	\centering
	\includegraphics[page=1, width=\textwidth]{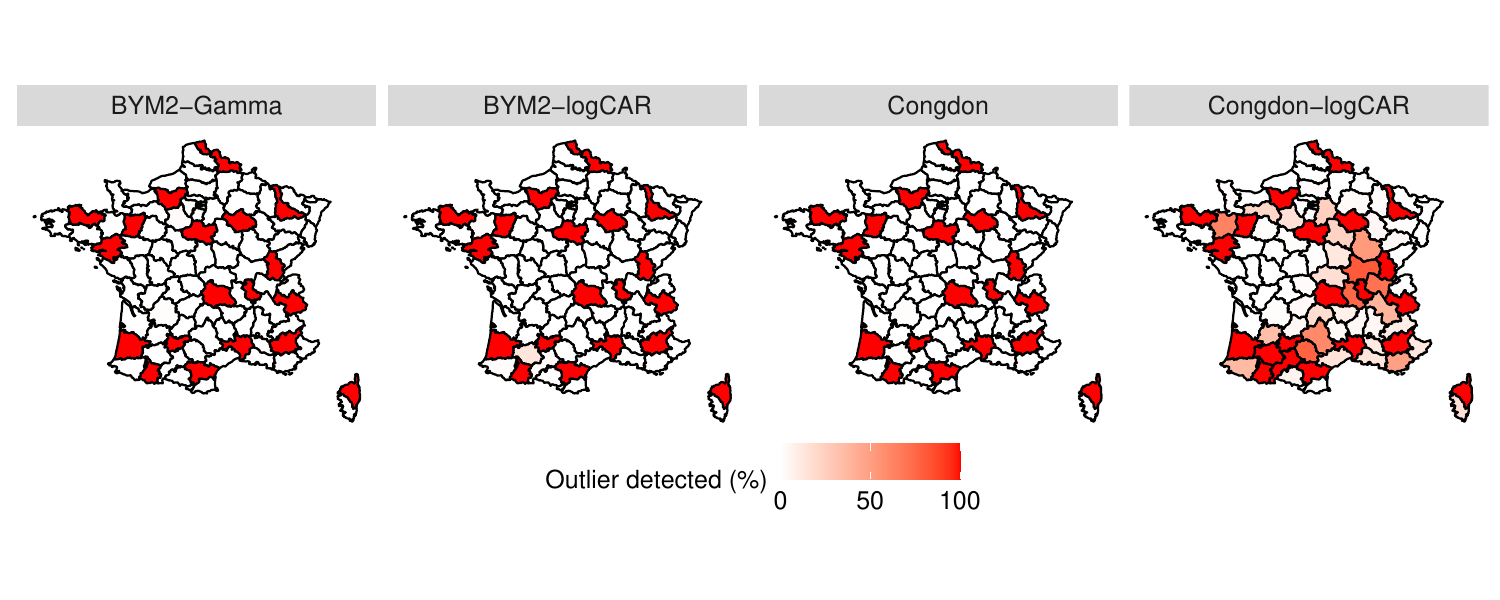}
	\caption{\footnotesize\textcolor{black}{Percentage of times among 100 replicates that the outliers were identified by each model, in the simulation study with distant outliers. The outliers are pointed out when $\kappa_u<1$, where $\kappa_u$ is the upper bound of the posterior 95\% credible interval of $\kappa$.}}
	\label{fig:Out_id_Far_France}
\end{figure}

\textcolor{black}{Regarding the detection of outliers, which is the main focus of this simulation study, Table \ref{tab:Sens_Spe_FarOut_France} shows how often each model accurately detects districts as outliers (sensitivity) and non-outliers (specificity), depending on the offset category. That is, the sensitivity is equal to the percentage of outliers detected among the contaminated districts over the 100 replicates. The specificity is the percentage of districts not identified as outliers among the ones whose true relative risk is equal to 1, over the 100 replicates. \textcolor{black}{The definition for sensitivity and specificity are taken from Richardson et al. \cite{richardson2004interpreting}.} Additionally, Figure \ref{fig:Out_id_Far_France} shows how often each district is detected as a potential outlier by the four models, while indicating the offset sizes. Recall, area $i$ is detected as an outlier when $\kappa_{u,i} < 1$, where $\kappa_{u,i}$ is the upper bound of the 95\% posterior credible interval of $\kappa_i$. Overall, all models are able to find \textcolor{black}{all of} the contaminated districts. \textcolor{black}{Additionally, except for Congdon's model with the logCAR parametrisation, none of the models tend to point out as potential outliers too many of the non-contaminated areas (specificity greater than 99\%).}}

\begin{table}[H]
\footnotesize
    \centering
    \begin{tabular}{cc cccc}
\toprule
& Offset category & BYM2-Gamma & BYM2-logCAR & Congdon & Congdon-logCAR \\
\midrule
\multirow{6}{*}{Sensitivity} & Small & 100.0 & 100.0 & 100.0 & 100.0 \\
& Medium low & 100.0 & 100.0 & 100.0 & 100.0 \\
& Medium & 100.0 & 100.0 & 100.0 & 100.0 \\
& Medium high & 100.0 & 100.0 & 100.0 & 100.0 \\
& High & 100.0 & 100.0 & 100.0 & 100.0 \\
& Overall & 100.0 & 100.0 & 100.0 & 100.0 \\
\addlinespace
\multirow{6}{*}{Specificity} & Small & 99.9 & 99.2 & 100.0 & 88.1 \\
& Medium low & 99.9 & 100.0 & 99.9 & 84.0 \\
& Medium & 99.8 & 100.0 & 99.9 & 88.7 \\
& Medium high & 99.9 & 99.8 & 99.9 & 79.5 \\
& High & 99.9 & 99.8 & 100.0 & 87.1 \\
& Overall & 99.9 & 99.7 & 99.9 & 85.5 \\
\bottomrule
    \end{tabular}
    \caption{\footnotesize\textcolor{black}{Sensitivity and specificity of the outlier detection for each model depending on the offset size, in the simulation study with distant outliers.}}
    \label{tab:Sens_Spe_FarOut_France}
\end{table}

\section{\textcolor{black}{Simulation studies on the map of Rio de Janeiro}}
\label{sec:sim_clustdist_Rio}

In this section, we present the results from simulation studies \textcolor{black}{conducted using the map of Rio de Janeiro} wherein some arbitrary areas are contaminated into outlying areas, to assess the performance of the proposed model in comparison to the one proposed by Congdon\cite{Congdon}. \textcolor{black}{Similar to Section \ref{sec:sim_clust},} the design of the simulation studies is inspired by Richardson et al.\cite{richardson2004interpreting}.
The $n=160$ districts of Rio de Janeiro and their neighbourhood structure are used as the region of study. In the first simulation study (section \ref{sec:sim_dist_Rio}), areas that are far from each other are contaminated into outliers. In the second simulation study (section \ref{sec:sim_clust_Rio}), neighbouring areas are contaminated into outliers. \textcolor{black}{In the third simulation study (section \ref{sec:sim_clustwithx}), neighbouring areas are contaminated and we include a covariate.} In all simulation studies, the goal is to identify the correct districts as outliers.

\subsection{Distant outliers in Rio}
\label{sec:sim_dist_Rio}

\begin{figure}[H]
	\centering
	\includegraphics[page=1, width=\textwidth]{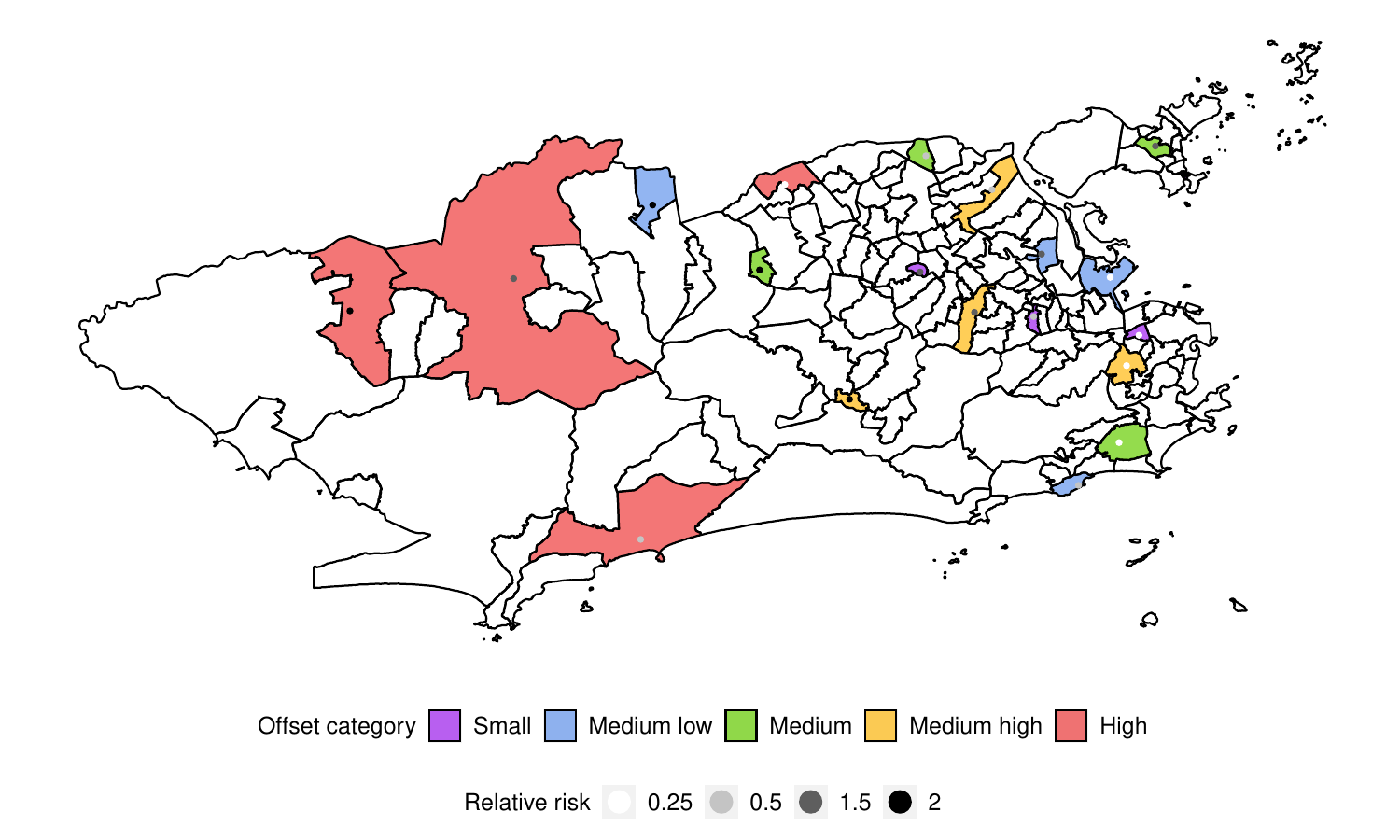}
	\caption{\footnotesize Districts of Rio de Janeiro city arbitrary chosen to be outliers in the first simulation study. Colors depict the offset category based on the empirical offset quantiles. The points represent the relative risk set to each outlying district.}
	\label{fig:Out_Far}
\end{figure}

In the first simulation study, 20 districts are arbitrarily chosen to be outliers. The goal is for our proposed model to accurately identify the outliers. Out of simplicity, there are no covariates included in the generating process nor when fitting the models. First, all $n=160$ latent effects, which correspond to relative risks in this covariate-free simulation study, are set to 1: $b_i=1, \ i=1, \dots, n$. Then, the offsets $[E_1, \dots, E_n]^\top$ are taken from the real data application to Zika counts that is presented in section \ref{sec:zika}. We define five offset categories based on the empirical offset quantiles. The first category corresponds to the smallest offsets and the fifth category, to the largest ones. The categories are termed \textcolor{black}{``}Small" for $E \leq 59.1$, \textcolor{black}{``}Medium low" for $E \in (59.1, 112.4]$, \textcolor{black}{``}Medium" for $E \in (112.4, 177.2]$, \textcolor{black}{``}Medium high" for $E \in (177.2, 267.2)$ and \textcolor{black}{``}High" for $E>267.2$. Based on these categories, we select 20 districts to be outliers. Four districts are chosen from each offset category. That is, there are 4 outliers within the smallest offset group, 4 within the second-to-smallest offset group, and so on. Then, within each group of four districts, the relative risks are contaminated into outliers by setting the relative risks to be equal to $b_{i}=0.25$, $b_{i'}=0.5$, $b_{i''}=1.5$ and $b_{i'''}=2$. Figure \ref{fig:Out_Far} maps the 160 districts of Rio de Janeiro, showing which areas are outliers based on the offset category and the contaminated relative risk. Again, all the white areas have a relative risk of 1. Finally, $R=100$ populations of size $n=160$ are created according to a hierarchical Poisson model. That is, $Y_i \sim \mathcal{P}\left(E_i\exp[b_i]\right)$. The only source of randomness across the 100 replicates comes from the repeated sampling from a Poisson distribution.

Using the two scale mixtures described in section \ref{sec:kappas}, the Congdon model is compared to the proposed model. The first version of the proposed model is denoted BYM2-Gamma and the second, BYM2-logCAR. The original Congdon model is termed Congdon, whereas the one with spatially structured scale mixture components is denoted Congdon-logCAR. For the four models, the intercept is given a quite vague prior: $\beta_0 \sim \mathcal{N}(0, 10^2)$ and the mixing parameter, $\lambda$, is assigned a uniform, $\mathcal{U}(0,1)$, prior distribution. The same $\mathcal{N}_+(0,1)$ prior is considered for $\sigma$, which is a \textit{marginal} standard deviation in the proposed model, while it is a \textit{conditional} standard deviation in Congdon's. Finally, in the BYM2-Gamma and Congdon models, the prior distribution for the $\kappa$'s is described in (\ref{eq:Kappa_Congdon}) with $\nu \sim \mathrm{Exp}(1/4)$. For the BYM2-logCAR and Congdon-logCAR parametrisations, the $\kappa$'s follow \textit{a priori} the distribution in (\ref{eq:kappa_logCAR}) and we set $\nu \sim \mathrm{Exp}(1/0.3)$. 

The models are fitted through the R package {\tt rstan} (Stan Development Team, 2020). For each dataset, the MCMC procedure consists of 2 chains of 20,000 iterations with a 10,000 burn-in period and a thinning factor of 10. Convergence of the chains is assessed through trace plots, effective sample sizes and the $\widehat{R}$ statistic (Gelman et al. \cite{Gelman_Diagnostic}, Vehtari et al. \cite{Rhat_Stan}).

\begin{figure}[H]
	\centering
	\includegraphics[page=1, width=\textwidth]{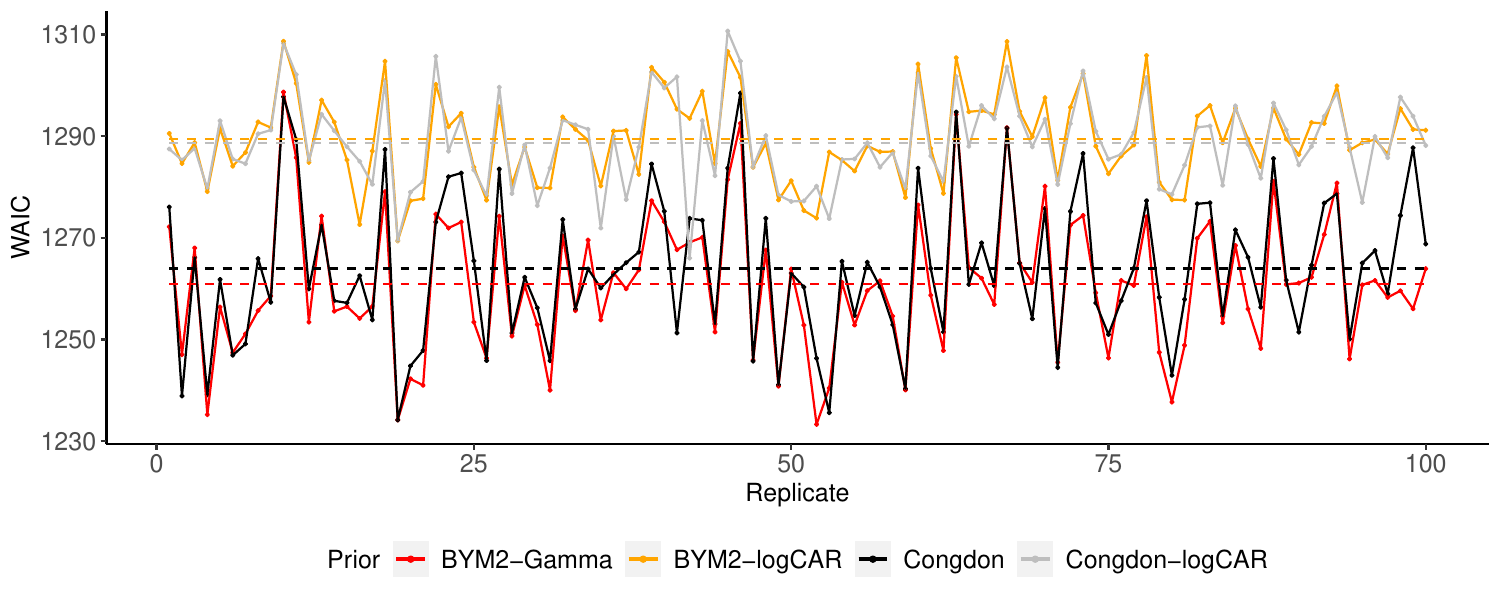}
	\caption{\footnotesize WAIC across the 100 replicates for the proposed models and Congdon's. Dashed lines: mean WAIC for each model.}
	\label{fig:WAIC_Far}
\end{figure}

In terms of WAIC \cite{WAIC}, for which smaller values are preferred, the proposed BYM2-Gamma model performs better than Congdon's, on average, as shown in Figure \ref{fig:WAIC_Far}. The BYM2-Gamma and original Congdon models always perform better than the models that include spatially structured scaling mixture components. On average, the BYM2-logCAR model yields a criterion of 1289.5 versus 1288.6 for the Congdon-logCAR model, while the BYM2-Gamma model presents a WAIC of 1260.9, versus 1263.9 for Congdon's.

\begin{figure}[H]
	\centering
	\includegraphics[page=1, width=\textwidth]{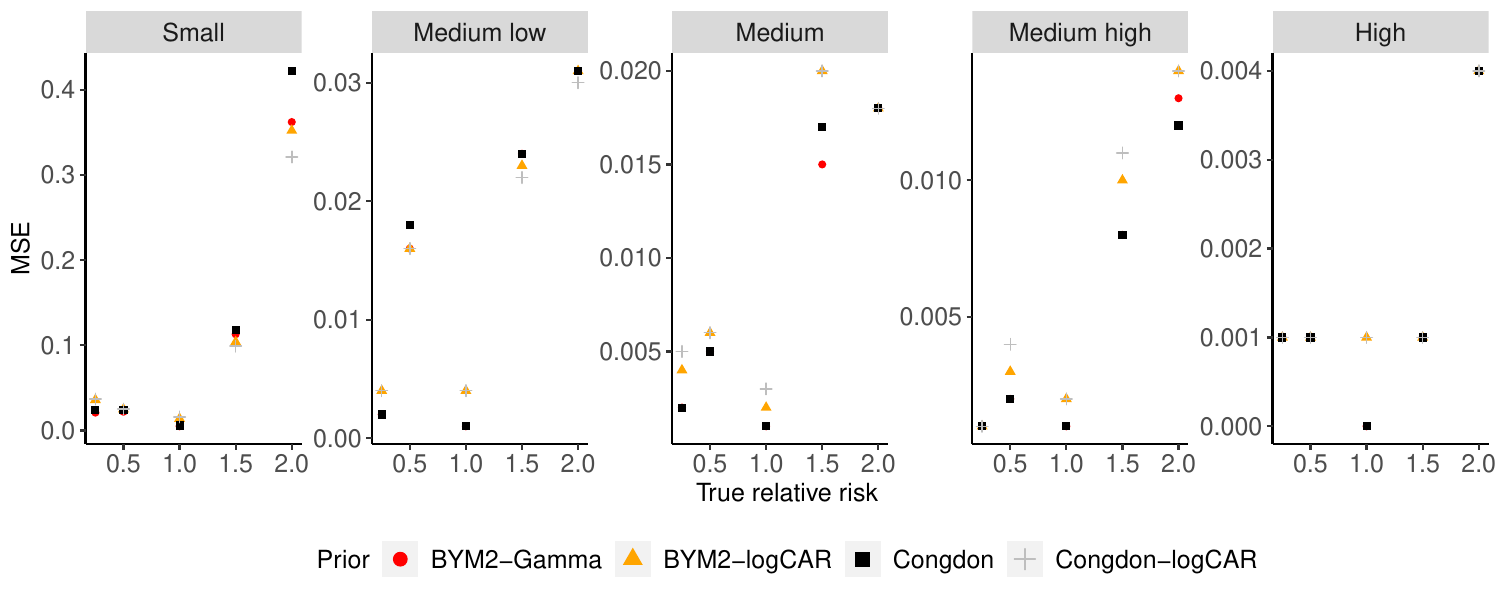}
	\caption{\footnotesize MSE over the 100 replicates for the proposed models and Congdon's according to the true relative risk and the offset size.}
	\label{fig:MSE_Far}
\end{figure}

The models' performances are also compared in terms of MSE, as shown in Figure \ref{fig:MSE_Far}. Again, as expected, all models yield smaller MSEs in areas with larger offsets. Additionally, all models tend to better fit the data in areas that are not outliers, that is in the areas with a relative risk of 1.  
On average over the 100 replicated datasets and across all areas, the MSEs are of 0.005 for the BYM2-Gamma model, 0.006 for Congdon and 0.008 for both log-CAR parametrisations.

\begin{figure}[H]
	\centering
	\includegraphics[page=1, width=\textwidth]{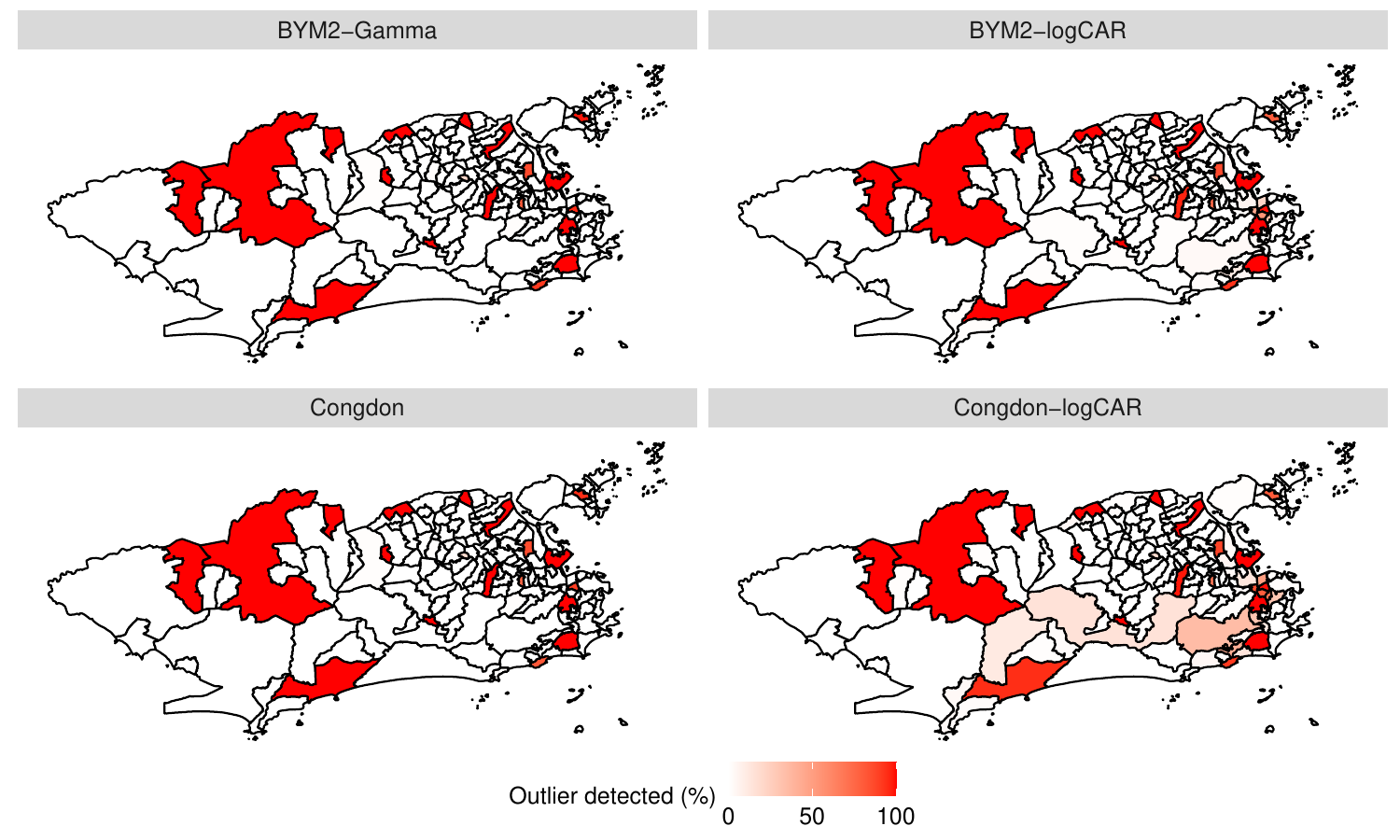}
	\caption{\footnotesize Percentage of times among 100 replicates that the outliers were identified by each model, in the first simulation study. The outliers are pointed out when $\kappa_u<1$, where $\kappa_u$ is the upper bound of the posterior 95\% credible interval of $\kappa$.}
	\label{fig:Out_id_Far}
\end{figure}

Regarding the detection of outliers, which is the main focus of this simulation study, Table \ref{tab:Sens_Spe_FarOut} shows how often each model accurately detects districts as outliers (sensitivity) and non-outliers (specificity), depending on the offset category. That is, the sensitivity is equal to the percentage of outliers detected among the contaminated districts over the 100 replicates. The specificity is the percentage of districts not identified as outliers among the ones whose true relative risk is equal to 1, over the 100 replicates. Additionally, Figure \ref{fig:Out_id_Far} shows how often each district is detected as a potential outlier by the four models, while indicating the offset sizes. Area $i$ is detected as an outlier when $\kappa_{u,i} < 1$, where $\kappa_{u,i}$ is the upper bound of the 95\% posterior credible interval of $\kappa_i$. Overall, all models are able to find the contaminated districts in the four upper offset categories. When the offsets are the smallest, all models detect the outliers only half of the time, with a slight advantage for the proposed models (e.g. sensitivity of 55.5 for BYM2-Gamma versus 50.25 for Congdon). In this simulation study where outliers are distant, the parametrisations with spatially structured scaling mixture components tend to identify slightly more outliers than are truly present in the data (e.g. specificities of 95.4 versus 90.2 for BYM2-logCAR and Congdon-logCAR, respectively). 
\begin{table}[H]
\footnotesize
    \centering
    \begin{tabular}{cc cccc}
\toprule
& Offset category & BYM2-Gamma & BYM2-logCAR & Congdon & Congdon-logCAR \\
\midrule
\multirow{6}{*}{Sensitivity} & Small & 55.50 & 54.00 & 50.25 & 49.50 \\
& Medium low & 94.25 & 94.75 & 91.50 & 94.75 \\
& Medium & 99.50 & 94.75 & 98.50 & 95.00 \\
& Medium high & 100.00 & 99.00 & 100.00 & 99.50 \\
& High & 100.00 & 100.00 & 100.00 & 98.50 \\
& Overall & 89.85 & 88.50 & 88.05 & 87.45 \\
\addlinespace
\multirow{6}{*}{Specificity} & Small & 99.93 & 97.82 & 99.93 & 95.04 \\
& Medium low & 100.00 & 95.39 & 100.00 & 90.21 \\
& Medium & 100.00 & 99.64 & 100.00 & 98.93 \\
& Medium high & 99.93 & 99.57 & 99.93 & 98.11 \\
& High & 99.96 & 99.89 & 99.96 & 99.25 \\
& Overall & 99.96 & 98.46 & 99.96 & 96.31 \\
\bottomrule
    \end{tabular}
    \caption{\footnotesize Sensitivity and specificity of the outlier detection for each model depending on the offset size.}
    \label{tab:Sens_Spe_FarOut}
\end{table}

\subsection{Neighbouring outliers in Rio}
\label{sec:sim_clust_Rio}

\begin{figure}[H]
	\centering
	\includegraphics[page=1, width=\textwidth]{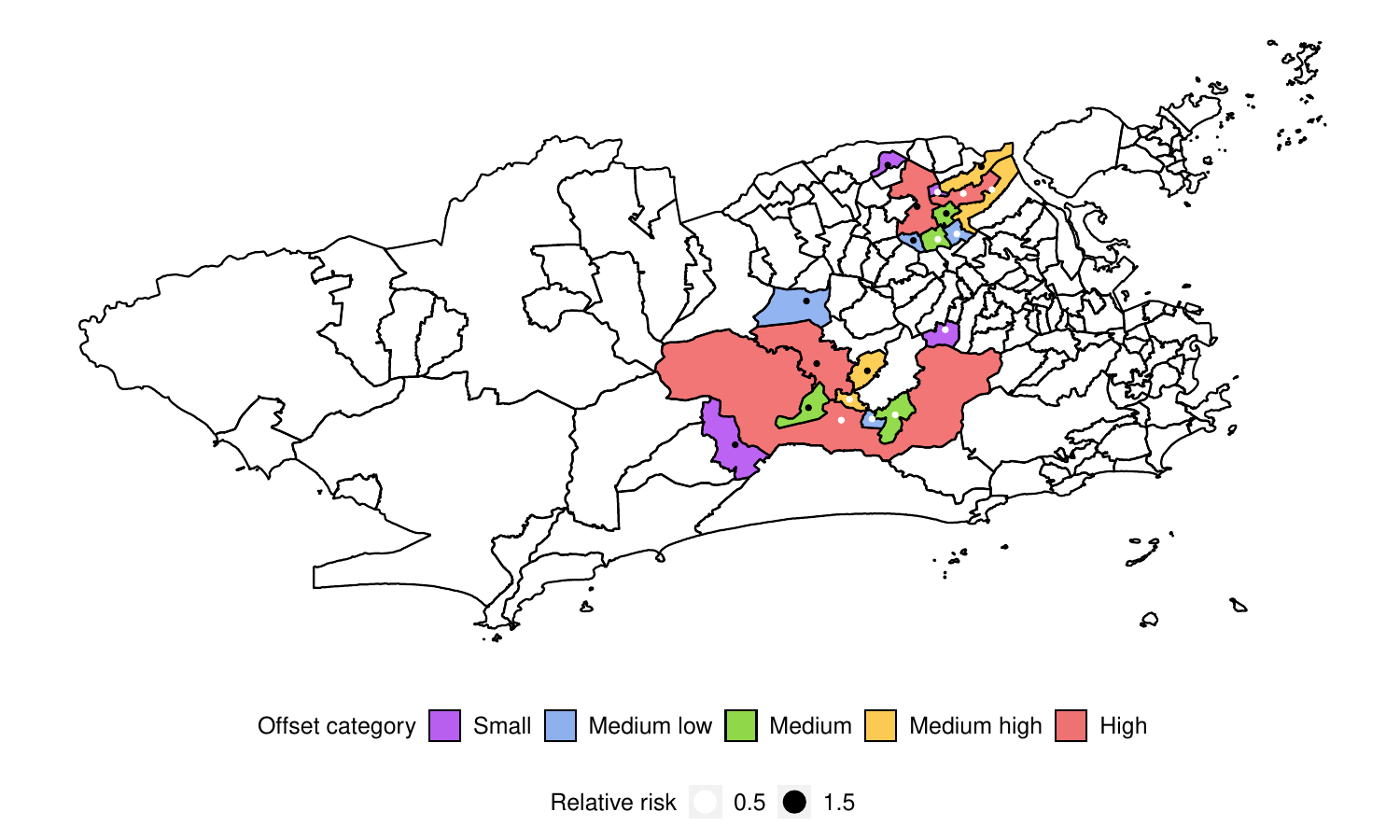}
	\caption{\footnotesize Districts of Rio de Janeiro city arbitrary chosen to be outliers in the second simulation study. Colors depict the offset category based on the empirical offset quantiles. The points represent the relative risk set to each outlying district.}
	\label{fig:Out_Clust}
\end{figure}

In this second simulation study, 20 districts are contaminated such that 2 groups of 10 neighbouring outliers are created. Once again, there are no covariates in this analysis and all areas are first imposed a relative risk of 1. Similarly to section \ref{sec:sim_dist}, the offsets $[E_1, \dots, E_n]^\top$ are taken from the Zika data analysis from section \ref{sec:zika}. Hence, the same five offset categories are defined. Then, 20 districts are selected to be outliers, such that each group of 10 neighbouring outliers contains 2 areas of each offset category. Within each such pair of districts, the relative risks are contaminated into outliers by setting $b_i=0.5$ and $b_{i'}=1.5$. The resulting outliers are mapped in Figure \ref{fig:Out_Clust}, highlighting the offset sizes and imposed relative risks. Again, $R=100$ populations of size $n=160$ are created by generating the number of cases $Y_i \sim \mathcal{P}\left(E_i \exp[b_i]\right)$. The same four models with priors defined in section \ref{sec:sim_dist} are fitted through \texttt{rstan}. After 20,000 iterations with a burn-in period of 10,000 and a thinning factor of 10, the 2 MCMC chains attained convergence as assessed by trace plots, effective sample sizes and $\widehat{R}$ statistics.

\begin{figure}[H]
	\centering
	\includegraphics[page=1, width=\textwidth]{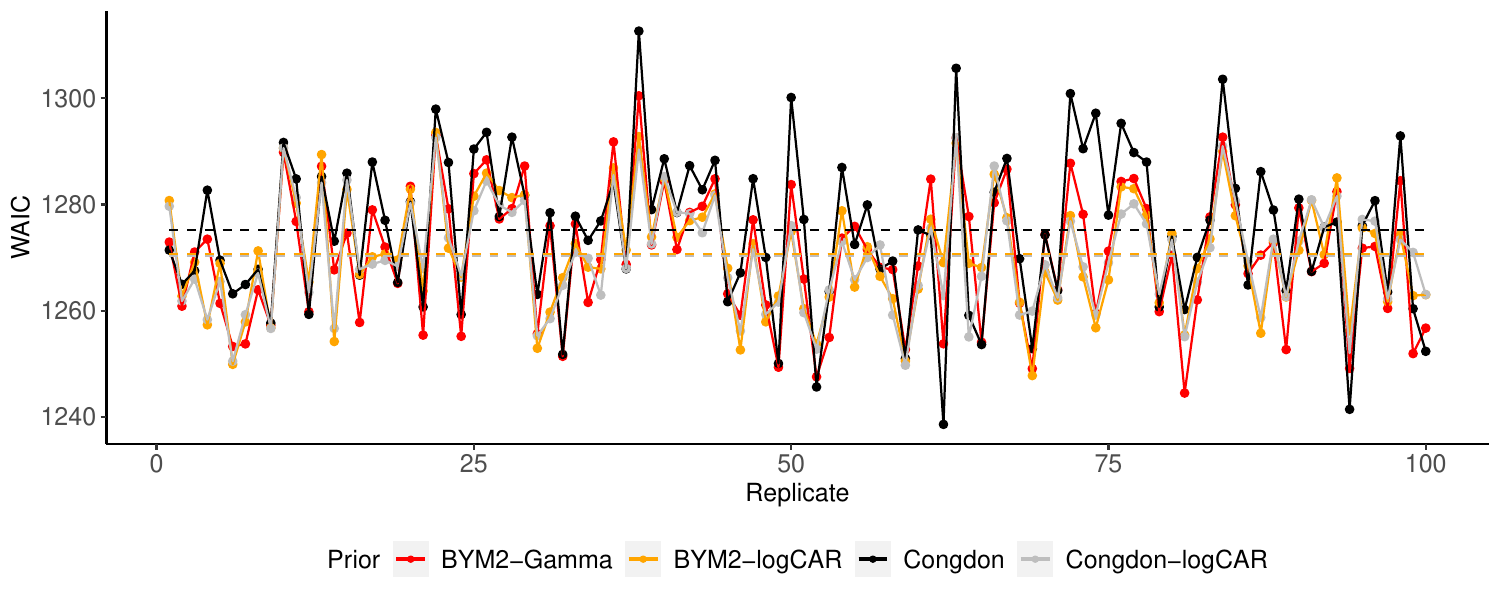}
	\caption{\footnotesize WAIC across the 100 replicates for the proposed models and Congdon's in the presence of neighbouring outliers. Dashed lines: mean WAIC for each model.}
	\label{fig:WAIC_Clust}
\end{figure}

In terms of WAIC, \textcolor{black}{as shown in Figure \ref{fig:WAIC_Clust}}, Congdon performs slightly worse than the other three models, \textcolor{black}{with an average value of 1275, versus 1270 for Congdon-logCAR the two proposals.}

\begin{figure}[H]
	\centering
	\includegraphics[page=1, width=\textwidth]{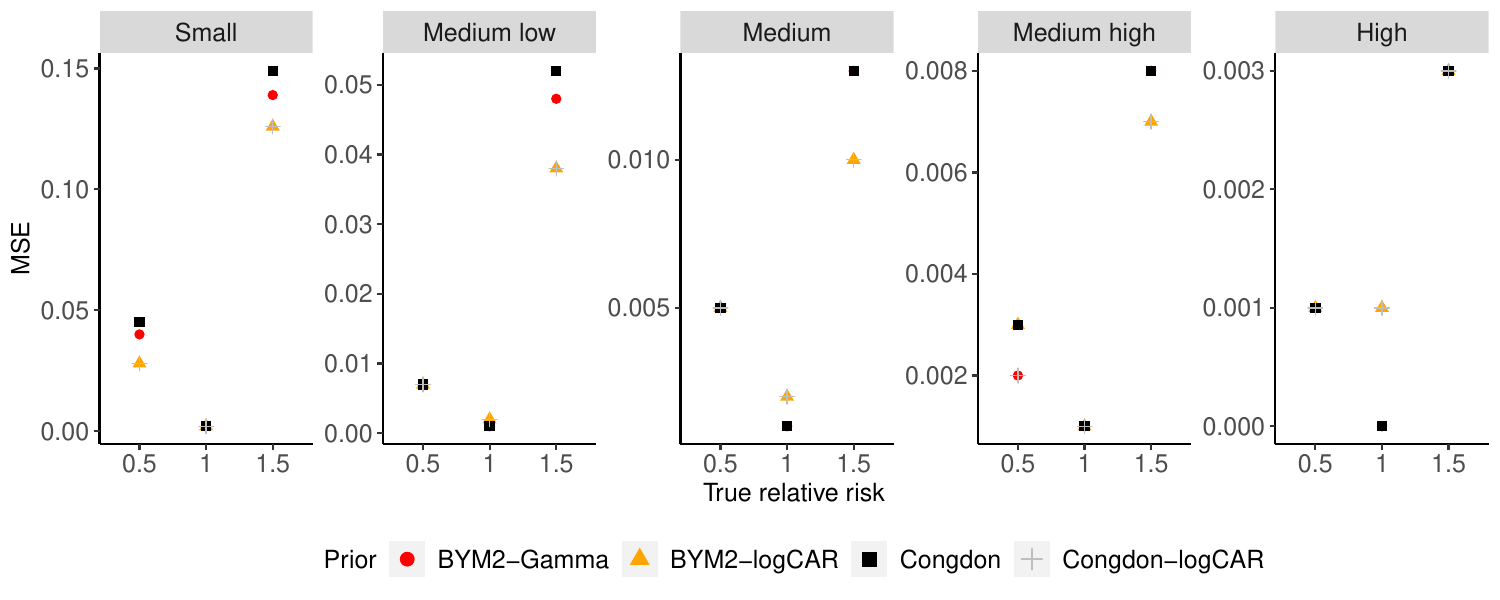}
	\caption{\footnotesize MSE over the 100 replicates for the proposed models and Congdon's according to the true relative risk and the offset size, in the presence of neighbouring outliers.}
	\label{fig:MSE_Clust}
\end{figure}

In terms of MSE, as expected, all models fit better the data in areas with higher offsets than in areas with smaller offsets, as shown in Figure \ref{fig:MSE_Clust}. Again, all models better fit the data in areas that are not outliers, areas with a relative risk of 1. 
Over the 100 replicates and all areas, \textcolor{black}{the four models perform similarly, with an average MSE of 0.004.}

\begin{figure}[H]
	\centering
	\includegraphics[page=1, width=\textwidth]{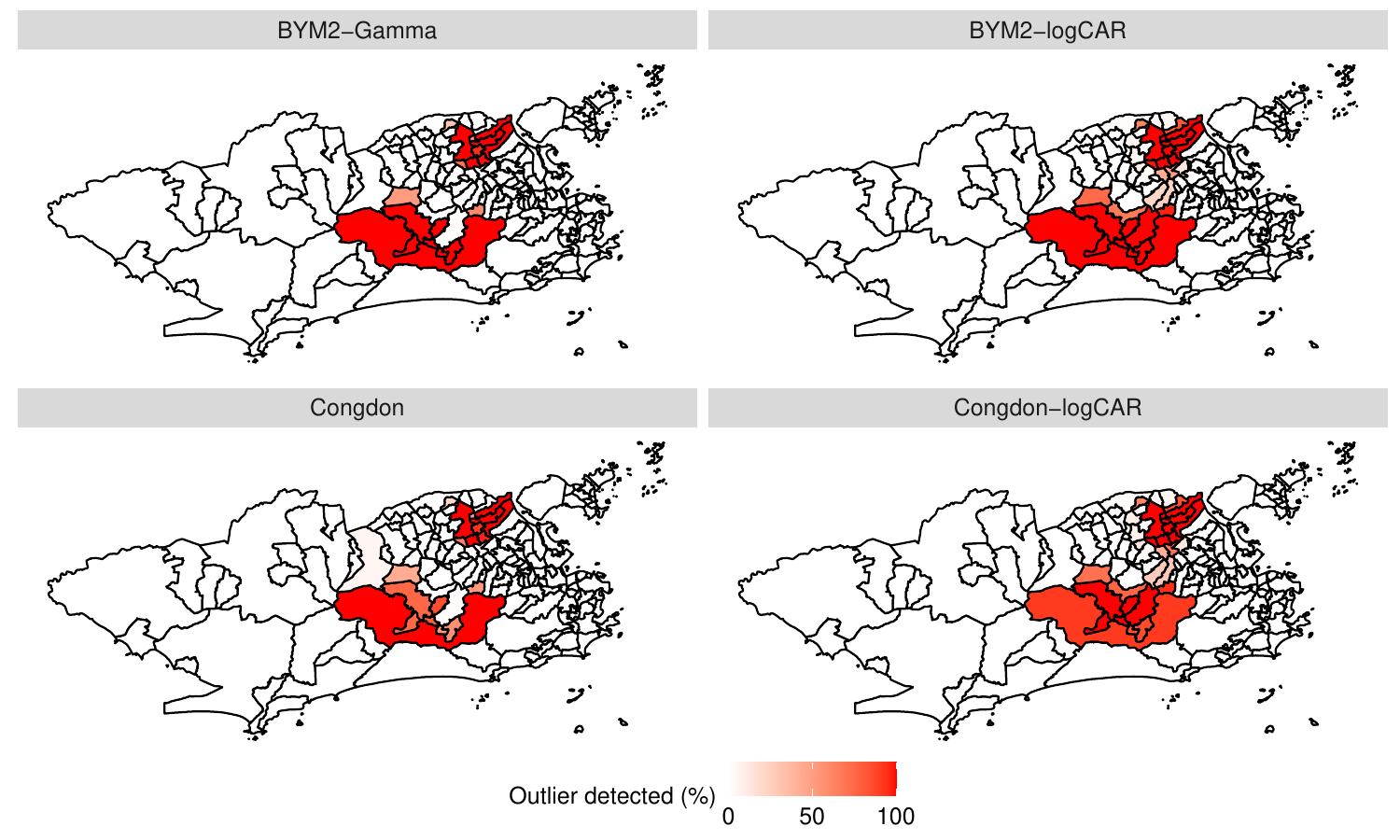}
	\caption{\footnotesize Percentage of times among 100 replicates that the outliers were identified by each model, in the second simulation study. The outliers are pointed out when $\kappa_u<1$, where $\kappa_u$ is the upper bound of the posterior 95\% credible interval of $\kappa$.}
	\label{fig:Out_id_Clust}
\end{figure}

Regarding the detection of outliers, the results are summarised in Table \ref{tab:Sens_Spe_ClustOut} and Figure \ref{fig:Out_id_Clust}. Similarly to the previous simulation study with distant outliers, both models with spatially structured $\kappa$'s tend to identify more outliers than truly present in the data (e.g. overall specificities of 97\% and 96.5\% for BYM2-logCAR and Congdon-logCAR, respectively, versus 99.9\% for both BYM2-Gamma and Congdon). In the smallest offset category, all models often miss the outliers, with a clear advantage for the models with spatially structured $\kappa$'s (e.g. sensitivity of about 30\% for BYM2-Gamma and Congdon versus 64\% for BYM2-logCAR and Congdon-logCAR). Regardless of the offset size, the BYM2-Gamma model performs better than Congdon's in terms of detected outliers. In particular, in the third offset category, the BYM2-Gamma model misses outliers only 1.5\% of the time versus 18.75\% for Congdon's model. 

\begin{table}[H]
\footnotesize
    \centering
    \begin{tabular}{cc cccc}
\toprule
& Offset category & BYM2-Gamma & BYM2-logCAR & Congdon & Congdon-logCAR \\
\midrule
\multirow{6}{*}{Sensitivity} & Small & 35.25 & 64.25 & 30.50 & 64.00 \\
& Medium low & 80.25 & 93.00 & 64.25 & 89.00\\
& Medium & 98.50 & 100.00 & 81.25 & 96.25 \\
& Medium high & 100.00 & 100.00 & 91.00 & 100.00 \\
& High & 100.00 & 100.00 & 93.75 & 97.75 \\
& Overall & 82.80 & 91.45 & 72.15 & 89.40 \\
\addlinespace
\multirow{6}{*}{Specificity} & Small & 100.00 & 100.00 & 99.93 & 100.00 \\
& Medium low & 99.96 & 98.07 & 99.96 & 98.00 \\
& Medium & 99.89 & 93.79 & 99.93 & 92.50 \\
& Medium high & 99.96 & 96.50 & 100.00 & 95.79 \\
& High & 99.96 & 96.36 & 99.71 & 96.21 \\
& Overall & 99.96 & 96.96 & 99.91 & 96.51 \\
\bottomrule
    \end{tabular}
    \caption{\footnotesize Sensitivity and specificity of the outlier detection, in the presence of neighbouring outliers, for each model depending on the offset size.}
    \label{tab:Sens_Spe_ClustOut}
\end{table}

\subsection{Neighbouring outliers with a covariate in Rio}
\label{sec:sim_clustwithx}

\begin{figure}[H]
    \centering
    \subfloat[\centering ]{\includegraphics[page=1, width=0.5\textwidth]{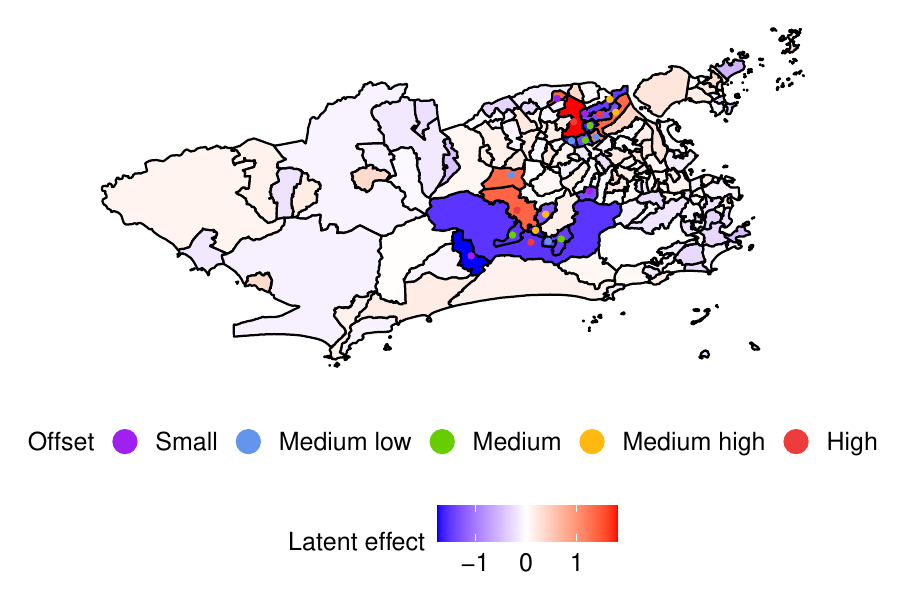}}
	\subfloat[\centering ]{\includegraphics[page=1,width=0.5\textwidth]{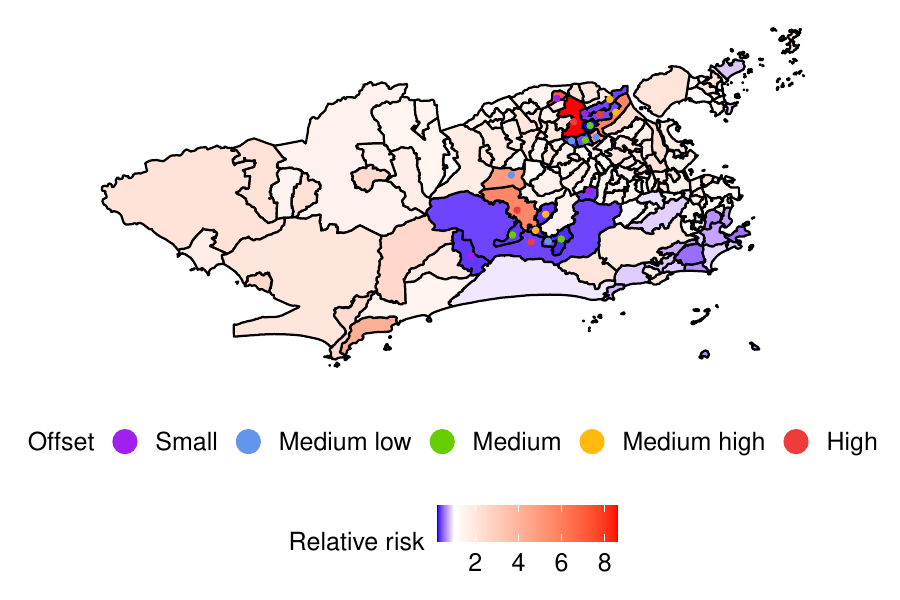}}
    \caption{\footnotesize Rio de Janeiro maps of the latent effects (a) and relative risks (b) after contamination, in the third simulation study. The coloured points depict the offset category based on the empirical offset quantiles.}
    \label{fig:MapLatRisk_ClustWithX}
\end{figure}

\textcolor{black}{In this third simulation study, the same offset categories and 2 groups of 10 neighbouring outliers as in section \ref{sec:sim_clust_Rio} are chosen. Again, the goal is to identify the outlying areas. First, all $n=160$ latent effects are generated following a proper CAR (PCAR) distribution: $\bm{b} \sim \mathcal{N}\left(\bm{0}, \sigma^2\left[\bm{D}-\alpha\bm{W}\right]^{-1}\right),$ where the matrices $\bm{W}$ and $\bm{D}$ are computed as defined in Section \ref{sec:Litt}, using the neighbourhood structure of Rio de Janeiro. We set $\sigma^2=0.1$ and $\alpha=0.99$ such that the proper spatial distribution is close to an ICAR distribution. Following Section \ref{sec:sim_clust_Rio}, four districts are chosen from each offset category and their generated latent effects are contaminated as $b_i^{contam} = b_i + e_i,$ with $e_i \sim \mathcal{U}\left(2\max(|b_{(1)}|, |b_{(n)}|), 3\max(|b_{(1)}|, |b_{(n)}|)\right),$ where $b_{(1)}$ and $b_{(n)}$ denote the minimum and maximum generated latent effects, respectively. Figure \ref{fig:MapLatRisk_ClustWithX} (a) maps the resulting 160 latent effects, showing which areas are outliers based on the offset category. Finally, $R=100$ populations of size $n=160$ are created according to the hierarchical Poisson model $Y_i \sim \mathcal{P}\left(E_i\exp[\beta_0 + \beta x_i + b_i]\right)$, where $\beta_0 = 2.5$, $\beta = -3.5$ and the covariate $x$ is the development index taken from the real data application to Zika counts presented in Section \ref{sec:zika}. The resulting relative risks are mapped in Figure \ref{fig:MapLatRisk_ClustWithX} (b), showing the outlying areas based on the offset category. Once again, the same four models are fitted through \texttt{rstan} and convergence of the 2 MCMC chains was attained after 20,000 iterations with a burn-in period of 10,000 and a thinning factor of 10.}

\begin{figure}[H]
	\centering
	\includegraphics[page=1, width=\textwidth]{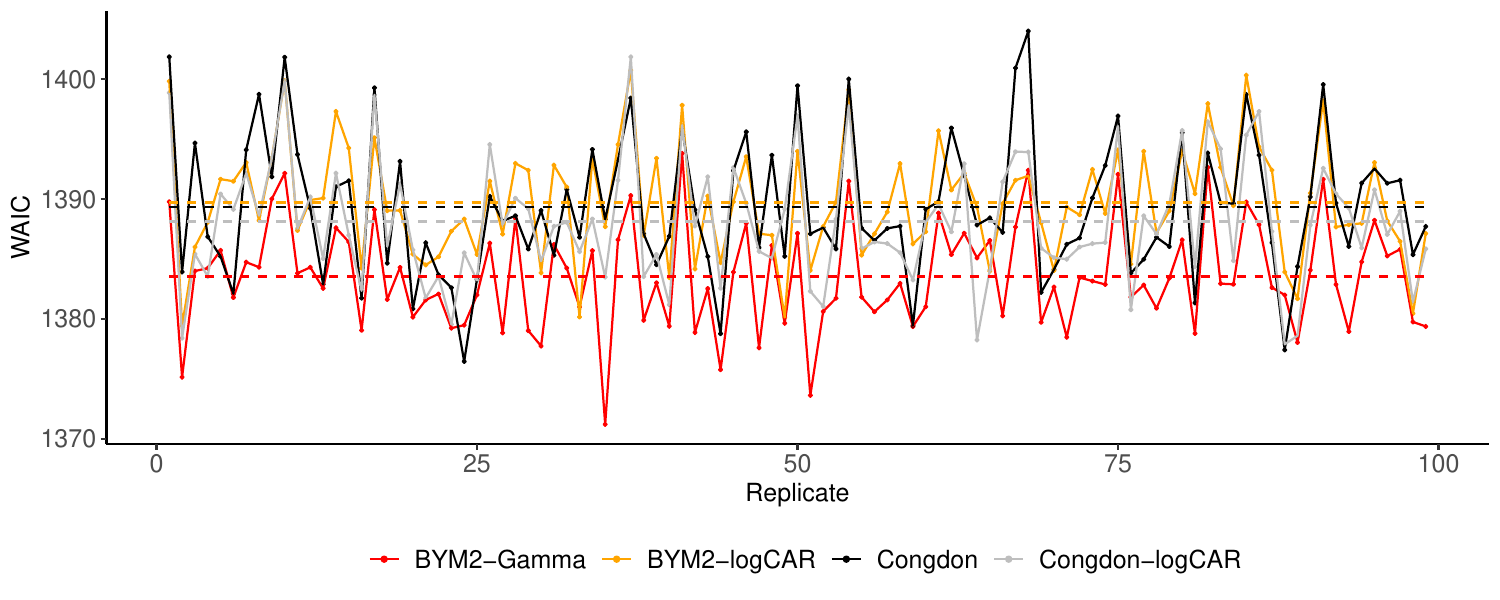}
	\caption{\footnotesize WAIC across the 100 replicates for the proposed models and Congdon's, in the third simulation study. Dashed lines: mean WAIC for each model.}
	\label{fig:WAIC_ClustWithX}
\end{figure}

\textcolor{black}{In terms of WAIC, the proposed BYM2-Gamma model performed the best, with a mean WAIC of 1383 over the 100 replicates. As shown in Figure \ref{fig:WAIC_ClustWithX}, the other three models' performances are similar to each other, with average values of 1389 (Congdon), 1390 (BYM2-logCAR) and 1388 (Congdon-logCAR).}

\begin{figure}[H]
	\centering
	\includegraphics[page=1, width=\textwidth]{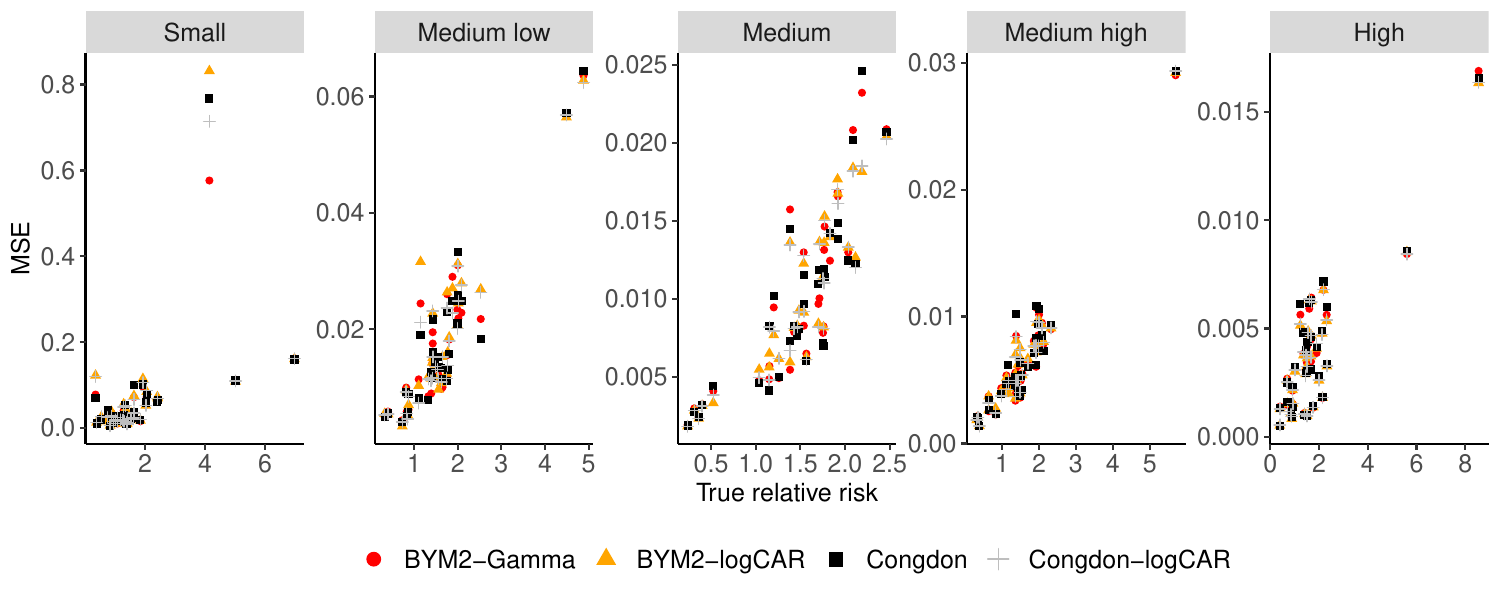}
	\caption{\footnotesize MSE over the 100 replicates for the proposed models and Congdon's according to the true relative risk and the offset size, in the third simulation study.}
	\label{fig:MSE_ClustWithX}
\end{figure}

\textcolor{black}{Figure \ref{fig:MSE_ClustWithX} shows each model's MSE for every districts across the different offset categories. The four models yield again smaller MSEs in districts with relative risks closer to 1, regardless of the offset size. Additionally, regardless of the relative risk size, all models reach smaller MSEs values for larger offset values. Overall, the proposed BYM2-Gamma model performed better with a mean MSE of 0.0189, versus 0.0212, 0.0204 and 0.0202, for the proposed BYM2-logCAR, Congdon and Congdon-logCAR models, respectively.}

\begin{figure}[H]
	\centering
	\includegraphics[page=1, width=\textwidth]{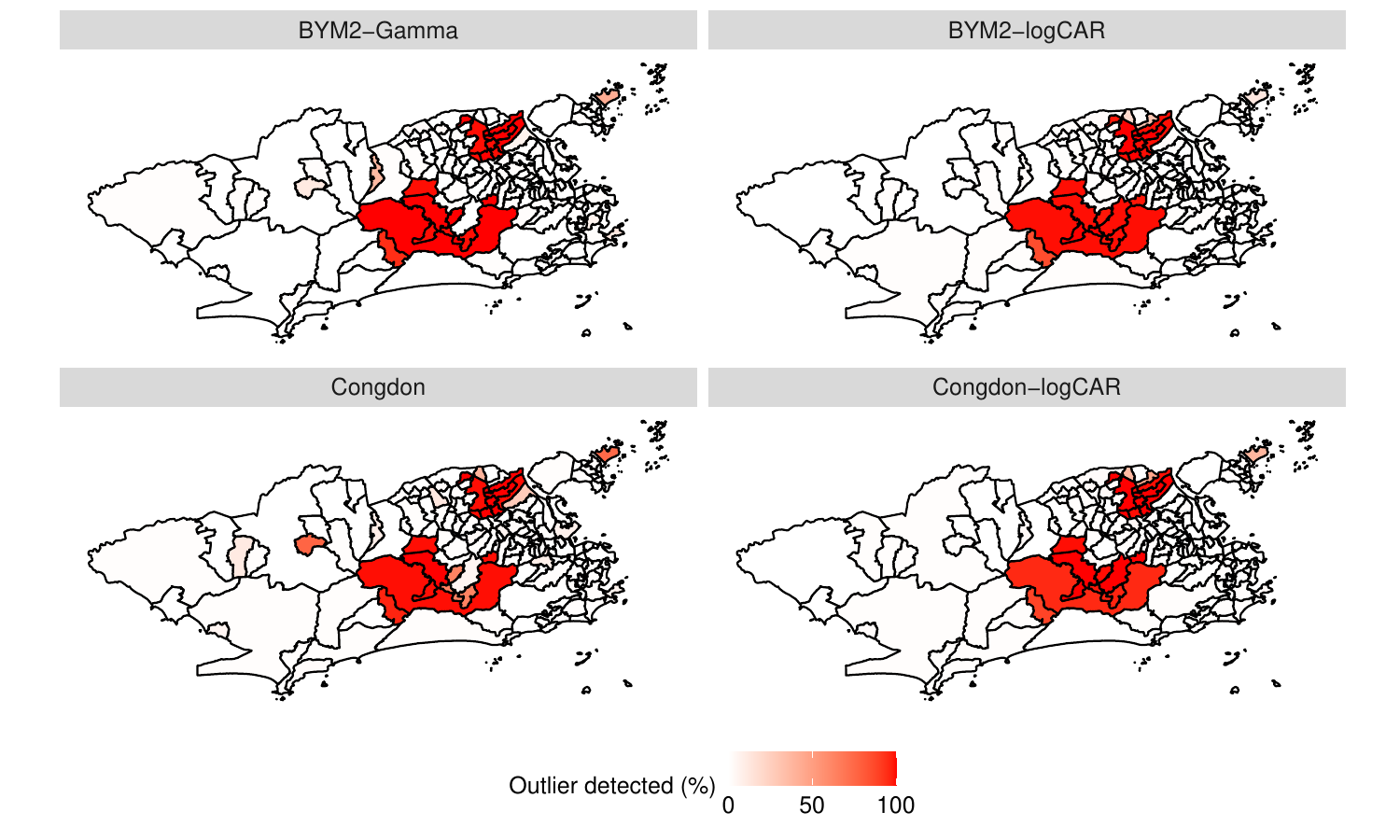}
	\caption{\footnotesize Percentage of times among 100 replicates that the outliers were identified by each model, in the third simulation study. The outliers are pointed out when $\kappa_u<1$, where $\kappa_u$ is the upper bound of the posterior 95\% credible interval of $\kappa$.}
	\label{fig:Out_id_ClustWithX}
\end{figure}

\textcolor{black}{Table \ref{tab:Sens_Spe_ClustWithX} shows the sensitivities and specificities of outlier identification produced by each model across the five offset categories. The proposed BYM2-Gamma model performs better in both identifying the correct outliers, and not pointing out the non-contaminated areas. Overall, Congdon's model misses some outlying districts 8\% of the time, and up to 19\% of the time, in the fourth offset category. The proposed spatially structured prior for the mixture components improved Congdon's model performance, where Congdon-logCAR only misses 2\% of the contaminated areas, overall. Additionally, Congdon's model tends to capture more outliers than were contaminated, like the western and eastern non-contaminated districts that are detected 75\% of the time, as shown in Figure \ref{fig:Out_id_ClustWithX}.}

\begin{table}[H]
\footnotesize
	\centering
	\begin{tabular}{cc cccc}
		\toprule
		& Offset category & BYM2-Gamma & BYM2-logCAR & Congdon & Congdon-logCAR  \\
		\midrule
		\multirow{6}{*}{Sensitivity} & Small &98.5     &     96.0  &  99.2     &        97.0 \\
		& Medium low & 100.0         &   99.2 &   90.4 &           99.0  \\
		& Medium & 100.0        &    99.8   & 90.9       &      99.0  \\
		& Medium high & 100.0        &    99.5 &   81.8        &     99.5  \\
		& High &  100.0      &      99.8  & 100.0     &          99.0  \\
		& Overall & 99.7 & 98.8 & 92.5 & 98.7 \\
		\addlinespace
		\multirow{6}{*}{Specificity} & Small & 99.3     &     99.9  &  99.1      &       99.9 \\
		& Medium low & 98.4     &     99.5  &  96.9       &      98.5 \\
		& Medium &  99.4      &    98.2   & 95.6       &      96.8  \\
		& Medium high & 99.8     &     99.9    &99.4     &        99.9  \\
		& High & 98.3      &    96.3 &   97.9       &      96.3    \\
		& Overall & 99.0 & 98.8 & 97.8 & 98.3 \\
		\bottomrule
	\end{tabular}
	\caption{\footnotesize Sensitivity and specificity of the outlier detection for each model depending on the offset size, in the third simulation study.}
 \label{tab:Sens_Spe_ClustWithX}
\end{table}

\end{document}